\documentclass[onecolumn,aps,preprintnumbers,showkeys,nofootinbib]{revtex4}

\usepackage{bm}
\usepackage{natbib}
\usepackage{graphicx}
\usepackage{amsmath,titletoc}

\newcommand{\ie}{{\it i.e.~}}

\newcommand{\beq}{\begin{equation}}
\newcommand{\eeq}{\end{equation}}
\newcommand{\bdm}{\begin{displaymath}}
\newcommand{\edm}{\end{displaymath}}
\newcommand{\bea}{\begin{eqnarray}}
\newcommand{\eea}{\end{eqnarray}}
\newcommand{\bt}{\begin{tabular}}
\newcommand{\et}{\end{tabular}}
\newcommand{\xv}{{\bf x}}
\newcommand{\kv}{{\bf k}}

\newcommand{\qv}{{\bf q}}

\newcommand{\intq}{\int\!\!d^3 q}

\def\d{\delta}
\def\D{\Delta}
\def\Ms{\, h^{-1} \, {\rm M}_{\odot}}
\def\Mpc{\, h^{-1} \, {\rm Mpc}}

\def\cGpc{\, h^{-3} \, {\rm Gpc}^3}
\def\kMpc{\, h \, {\rm Mpc}^{-1}}
\def\icMpc{\, h^3 \, {\rm Mpc}^{-3}}
\def\fNL{f_{NL}}

\def\la{\langle}

\def\O{\mathcal O}
\def\kall{k_1,k_2,k_3}

\begin{document}

\title{The halo bispectrum in N-body simulations with non-Gaussian initial conditions}

\author{Emiliano Sefusatti}
\email{emiliano.sefusatti@cea.fr}
\affiliation{Institut de Physique Th\'eorique, CEA/DSM/IPhT, Unit\'e de Recherche Associ\'ee au CNRS, CEA/Saclay, F-91191, Gif-sur-Yvette, C\'edex, France}

\author{Mart\'in Crocce}
\email{martincrocce@gmail.com}
\affiliation{Institut de Ci\`encies de l'Espai, IEEC-CSIC, Campus UAB, Facultat de Ci\`encies, Torre C5 par-2, Barcelona 08193, Spain }

\author{Vincent Desjacques}
\email{Vincent.Desjacques@unige.ch}
\affiliation{Universit\'e de Gen\`eve and Center for Astroparticle Physics, 24 Quai Ernest Ansermet, 1211 Gen\`eve 4, Switzerland}
\affiliation{Institute for Theoretical Physics, Universit\"at Z\"urich, Winterthurerstrasse 190, CH-8057 Z\"urich, Switzerland}

\begin{abstract}
We present measurements of the bispectrum of dark matter halos in numerical simulations with non-Gaussian initial conditions of the local type. We show, in the first place, that the overall effect of primordial non-Gaussianity on the halo bispectrum is larger than on the halo power spectrum when all measurable configurations are taken into account. We then compare our measurements with a tree-level perturbative prediction finding good agreement at large scale when the constant Gaussian bias parameter, both linear and quadratic, and their constant non-Gaussian corrections are fitted for. The best-fit values of the Gaussian bias factors and their non-Gaussian, scale-independent corrections are in qualitative agreement with the peak-background split expectations. In particular, we show that the effect of non-Gaussian initial conditions on squeezed configurations is fairly large (up to $30\%$ for $\fNL=100$ at redshift $z=0.5$) and results from contributions of similar amplitude induced by the initial matter bispectrum, scale-dependent bias corrections as well as from nonlinear matter bispectrum corrections. We show, in addition, that effects at second order in $\fNL$ are irrelevant for the range of values allowed by CMB and galaxy power spectrum measurements, at least on the scales probed by our simulations. Finally, we present a Fisher matrix analysis to assess the possibility of constraining primordial non-Gaussianity with future measurements of the galaxy bispectrum. We find that a survey with a volume of about $10\cGpc$ at mean redshift $z\simeq 1$ could provide an error on $\fNL$ of the order of a few. This shows the relevance of a joint analysis of galaxy power spectrum and bispectrum in future redshift surveys.
\end{abstract}


\maketitle

\section{Introduction}

Recent years have witnessed an intense activity related to the possibility, offered by future observations, to detect a non-Gaussian component in the primordial density perturbations. In fact, the detection of a non-vanishing primordial bispectrum would have profound implications for our understanding of the inflationary mechanisms, possibly ruling out the simplest model of canonical, single-field, slow roll inflation \citep{KomatsuEtal2009A}.

On the theoretical side, primordial non-Gaussianity (PNG) has been recognized as a relevant ``prediction'' of early Universe models which has the potential to shed light on inflaton interactions or multiple fields scenarios (see \citep{Chen2010, ByrnesChoi2010, Barnaby2010} for recent reviews and \citep{CheungEtal2008, SenatoreZaldarriaga2010} for an effective field theory approach). On the observational side, the Planck satellite \citep{PLANCK2006} will soon improve significantly the constraints on non-Gaussian parameters from the Cosmic Microwave Background (CMB) bispectrum measurements currently provided by the WMAP mission \citep{KomatsuEtal2011}. 

While the CMB clearly represents the most direct window on the primeval perturbations, the effect of non-Gaussian initial conditions on the large scale bias of halos and galaxies discovered by \citep{DalalEtal2008} has turned the galaxy power spectrum into a powerful probe of PNG of the local kind \citep{SalopekBond1990}, or, more generically but to a lesser degree, of an initial bispectrum assuming large values in the limit of squeezed triangular configurations such as in the folded model \citep{ChenEastherLim2007, HolmanTolley2008, MeerburgVanDerSchaarCorasaniti2009} or quasi-single field models \citep{ChenWang2010A, ChenWang2010B}. This discovery has been followed by a number of studies investigating this effect in details \citep{MatarreseVerde2008, SlosarEtal2008, TaruyaKoyamaMatsubara2008, AfshordiTolley2008, McDonald2008, DesjacquesSeljakIliev2009, Valageas2009, GrossiEtal2009, PillepichPorcianiHahn2010, GiannantonioPorciani2010, TseliakhovicHirataSlosar2010, BeckerHutererKadota2010, SchmidtKamionkowsky2010, WagnerVerde2011,  ShanderaDalalHuterer2011, CyrRacineSchmidt2011, ScoccimarroEtal2011, DesjacquesJeongSchmidt2011B, DesjacquesJeongSchmidt2011A} and has motivated several different groups to runs large suites of N-body simulations seeded with non-Gaussian initial conditions \citep{DesjacquesSeljakIliev2009, PillepichPorcianiHahn2010, NishimichiEtal2010, WagnerVerdeBoubeker2010,  WagnerVerde2011, LoVerdeSmith2011, ShanderaDalalHuterer2011, ScoccimarroEtal2011}. Remarkably, observations of the large scale structure can now place limits on the local non-Gaussian parameter $\fNL$ similar to those from the CMB \citep{SlosarEtal2008, XiaEtal2010B, XiaEtal2011}, while even better constraints are expected from upcoming large-volume galaxy redshift surveys \citep{FedeliMoscardiniMatarrese2009, CarboneMenaVerde2010,  CunhaHutererDore2010,  SartorisEtal2010,  FedeliEtal2011, GiannantonioEtal2011}, especially if the stochasticity induced by sampling variance and shot noise can be suppressed \citep{Seljak2009, Slosar2009, HamausSeljakDesjacques2011, YooEtal2011}.

The interest in large-scale structure observations as a probe of the initial conditions has, however, a much longer history. In fact, it has long been expected that the main effect of PNG on the large scale mass and galaxy distribution consists in an additional contribution to the matter bispectrum, linearly evolved to present time by gravitational instability (see \citep{LiguoriEtal2010} and references therein). Focusing on this effect, \citep{ScoccimarroSefusattiZaldarriaga2004, SefusattiKomatsu2007} have shown that future galaxy surveys will be able to provide constraints on the non-Gaussian parameters which surpass the best CMB limits. These results follow from the fact that future three-dimensional redshift surveys will provide a larger number of observable modes than two-dimensional CMB observations. Moreover, the primordial contribution to the matter bispectrum is to the large scale structure what the CMB bispectrum is to the temperature anisotropies. Hence, the impact of non-Gaussian initial conditions of the local type on the galaxy power spectrum {\em should not} overshadow the sensitivity of the galaxy bispectrum to PNG. In fact, the latter is much more sensitive to the amplitude and configuration shape of the three-point function of primordial curvature perturbations. Therefore, while the galaxy power spectrum might be competitive with the CMB bispectrum as far as local non-Gaussianity is concerned, this does not happen for other models of non-Gaussianity. In more general terms, a complete study of large-scale structure data will naturally involve the {\em combined} analysis of power spectrum and bispectrum measurements (see, for example, \citep{SefusattiScoccimarro2005,SefusattiEtal2006} in the context of cosmological parameter determination). This work represents a first step in the direction of such an analysis for non-Gaussian initial conditions.   

In this work, we focus on the local model of primordial non-Gaussianity \citep{GanguiEtal1994, VerdeEtal2000, KomatsuSpergel2001}. In this specific case we expect, for the halo bispectrum the combination of effects due to scale-dependent corrections to the both linear and quadratic halo bias and the effect of the primordial bispectrum component. First attempts at incorporating large-scale bias corrections in a description of the halo or galaxy three-point function can be found in \citep{Sefusatti2009, JeongKomatsu2009B}. These works relied however on a local model for galaxy bias or high-peak statistics leading to equivalent results. It is now clear that such a prescription does not correctly describe two-point statistics (see, for instance, \citep{ScoccimarroEtal2011, DesjacquesJeongSchmidt2011B}). A simple description that takes into account nonlocal corrections in terms of a multivariate bias expansion is proposed by \citep{GiannantonioPorciani2010}, developing the earlier results of \citep{SlosarEtal2008, TaruyaKoyamaMatsubara2008, AfshordiTolley2008, McDonald2008, DesjacquesSeljakIliev2009} to include nonlinear bias contributions.  They also provide a partial expression for the halo bispectrum including a few relevant terms. A complete tree-level expression for the halo bispectrum based on the multivariate halo bias expansion is derived instead in \citep{BaldaufSeljakSenatore2011}. This is essentially the model that we compare to numerical simulations in this work. 

Previous measurements of the halo bispectrum in numerical simulations with local non-Gaussian initial conditions have been presented so far, to the best of our knowledge, only in \citep{NishimichiEtal2010}. This preliminary work employed simulations with large values for the non-Gaussian parameter $\fNL$ ($\fNL\gg 100$), pointing-out a peculiar dependence on $\fNL^2$, not present in the same configurations of the matter bispectrum. These results focused on the effects of non-Gaussianity on squeezed configurations and qualitatively tested the scale dependence of linear and quadratic corrections in $\fNL$, together with their dependence on redshift and halo mass threshold. 

We consider N-body simulations with local non-Gaussian initial conditions corresponding to $\fNL=\pm 100$, that is, characterized by a relatively small departure from Gaussianity, as suggested by CMB observations \citep{KomatsuEtal2009A}. We improve over the preliminary results of \citep{NishimichiEtal2010} by presenting systematic measurements of {\em all} triangular configurations shapes from large down to mildly nonlinear scales, along the lines of our previous work on non-Gaussian effects on the matter bispectrum \citep{SefusattiCrocceDesjacques2010}. We consider a low and high mass halo sample and a unique output redshift at $z=0.5$ (which is of the order of the median redshift of forthcoming redshift surveys). This allows us, in the first place, to provide an estimate of the signal generated by local PNG and compare it to that induced in the halo power spectrum. For all triangular configurations, we compare our measurements to the model of \citep{BaldaufSeljakSenatore2011}, where some of the scale-independent bias factors are fitted to the data rather than derived using the peak-background split prescriptions. A comparison of best-fit values for the bias parameters with the peak-background split predictions is also presented. 

This paper is organized as follows. In Section \ref{sec:nonlocal}, we review the multivariate bias expansion proposed by \citep{GiannantonioPorciani2010} and, in Section \ref{sec:model}, we spell out the tree-level expression for the halo bispectrum derived by \citep{BaldaufSeljakSenatore2011} together with the analogous expression for the matter-matter-halo cross-bispectrum. In Section \ref{sec:simulations}, we describe the numerical simulations employed in our work and, in Section \ref{sec:signal}, we estimate the cumulative signal-to-noise for the non-Gaussian effects for both the power spectrum and bispectrum. In Section \ref{sec:results}, we perform a detailed comparison between the model and the measurements of the halo and matter-matter-halo bispectrum, and discuss the values of the best-fit bias parameters in light of the peak-background split expectations. We also perform a simple Fisher matrix analysis based on the halo bispectrum model, in an attempt to provide a first estimate of the ability of the halo bispectrum to constrain PNG. We present our conclusions in Section \ref{sec:conclusions}.

\section{Large-scale bias, $\fNL$, and the peak-background split}
\label{sec:nonlocal}

In this section, we derive an expression for the Eulerian halo overdensity $\d_h$ as function of the nonlinear and non-Gaussian matter overdensity $\d$ {\em and} of the initial, \ie linear, curvature perturbations $\phi$ including corrections at second order in $\d$ and proportional to the product $\phi\,\d$. In doing so we follow the approach of \citep{GiannantonioPorciani2010} and \citep{BaldaufSeljakSenatore2011} and assume a {\em multivariate} bias expansion of the halo overdensity. The expression for $\d_h$ is obtained by applying the peak-background argument along the lines of \cite{SlosarEtal2008} and \cite{GiannantonioPorciani2010}. Note that we will not take into account the additional non-Gaussian bias corrections computed by \citep{DesjacquesJeongSchmidt2011A, DesjacquesJeongSchmidt2011B} and \citep{ScoccimarroEtal2011} since they are negligible for the local model with constant $\fNL$.

Throughout this paper we will assume non-Gaussian initial conditions of the local kind and work consistently at linear order in the non-Gaussian $\fNL$ parameter. This is justified in light of the limits provided by CMB \citep{KomatsuEtal2011} and LSS \citep{SlosarEtal2008,  XiaEtal2010, XiaEtal2010B, XiaEtal2011} observations of a local $\fNL$: typically $|\fNL| \lesssim 100$ at 99\% CL. As we will see in Section~\ref{sec:results}, this is a good approximation even for the comparison with our simulations, which assume $\fNL=\pm 100$. 

We shall work with a local expansion of the form \citep{GiannantonioPorciani2010, BaldaufSeljakSenatore2011}
\beq\label{eq:dhpos}
\d_{h} \simeq b_{10}\,\d+b_{01}\,\phi_{0}+\frac{1}{2}\,b_{20}\,\d^2+b_{11}\,\phi_0\,\d+\O(\fNL^2)
\eeq
where $\d$ still stands for the {\em nonlinear} and {\em non-Gaussian} matter density contrast, while $\phi_0$ represents the {\em linear} and {\em Gaussian} curvature perturbations. This expression will be used to compute the leading, ``tree-level'' contribution to the halo bispectrum whereas, for matter and matter-curvature correlators, we will consider additional perturbative corrections.

\subsection{Lagrangian bias}

For local quadratic non-Gaussianity, the Bardeen's curvature perturbation in the matter dominated era is given by 
\beq
\Phi(\xv)=\phi(\xv)+\fNL\phi(\xv)^2\,,
\eeq
where $\phi$ is a Gaussian field. In the peak-background split framework, we can separate the perturbations into their long-wavelength and short-wavelength piece, $\phi_0$ and $\phi_1$, and thus obtain
\beq\label{eq:Phi}
\Phi(\xv)=\phi_0(\xv)+\fNL\phi_0(\xv)^2+\left[1+2\fNL\phi_0(\xv)\right]\phi_1(\xv)+\fNL\phi_1(\xv)^2 +{\rm const.}
\eeq
In this expansion, the most relevant term is $(1+2\fNL\phi_0)\phi_1$ since, in a region where $\phi_0$ takes some constant value, it can be interpreted as a local, scale-dependent rescaling of the amplitude of short-wavelength fluctuations. As shown in \citep{SlosarEtal2008}, the second and fourth terms can be ignored as far as one is interested in the $k$-dependent bias correction. At linear order, the Fourier modes of the density perturbations $\d$ are related to those of the curvature perturbations $\Phi$ through the linearized Poisson equation 
\beq\label{eq:Poisson}
\d(k,z)=M(k,z)\Phi(k),
\eeq
where 
\beq
M(k,z)\equiv\frac23\frac{k^2T(k)D(z)}{\Omega_mH_0^2}\,,
\eeq
with $T(k)$ representing the matter transfer function and $D(z)$ the linear growth factor, while $\Omega_m$ and $H_0$ are the present time relative matter density and the Hubble parameter, respectively. 

Taking $\phi_0$ constant and convolving the left- and right-hand side of Eq.(\ref{eq:Phi}) with $M$, we can see that the effect of non-Gaussian initial conditions can be interpreted as a local, $\fNL$-dependent modulation of the r.m.s. amplitude $\sigma_1$ of the short-wavelength density perturbations (for a more rigorous derivation, see \citep{SchmidtKamionkowsky2010, DesjacquesJeongSchmidt2011B})
\beq\label{eq:delta1}
\sigma_1 \to \sigma_1 (1+2\fNL\phi_0)\,,
\eeq
Following \citep{GiannantonioPorciani2010, BaldaufSeljakSenatore2011}, we define the Lagrangian bias parameters $b_{ij}^L(M_1,\d_1)$ from the expansion of the Lagrangian halo density field $\d_h^L$ in the {\em large scale, linear} matter density $\d_0$ and curvature perturbations $\phi_0$,
\bea
\d_h^L(M_1,z_1|M_0,z_0) & \equiv & \frac{{\mathcal N}(M_1,z_1|M_0,z_0)}{n(M_1,z_1)V_0}-1=\sum_{i,j=0}^{\infty}\frac{1}{i!j!}b_{ij}^L(M_1,\d_1)\d_0^i\phi_0^j
\nonumber\\
& = & b_{10}^L\,\d_0+b_{01}^L\,\phi_0+\frac{1}{2}\,b_{20}^L\,\d_0^2+b_{11}^L\,\d_0\,\phi_0+\dots
\eea
Assuming universality, the bias parameters can be derived from the shape of the unconditional mass function $f(\nu)$ alone, provided that we substitute the variable $\nu\equiv\d_c/\sigma_1$ with
\beq\label{eq:nu10}
\nu_{10}\simeq\frac{\d_1-\d_0}{(1+2\fNL\phi_0)\sigma_1},
\eeq 
so that, in particular
\bea
b_{10}^L & = & \frac{1}{f(\nu_{10})}\left.\frac{\partial f(\nu_{10})}{\partial \d_0}\right|_{\d_0,\phi_0=0}\,,\label{eq:biasLGa}\\
b_{20}^L & = & \frac{1}{f(\nu_{10})}\left.\frac{\partial^2 f(\nu_{10})}{\partial \d_0^2}\right|_{\d_0,\phi_0=0}\,.\label{eq:biasLGb}
\eea
Notice that the halo mass function itself depends on the non-Gaussian parameter $\fNL$. If we factorize the effect of non-Gaussianity on the mass function as
\beq
f(\nu)=f_G(\nu)\,R_{NG}(\nu)\,,
\eeq
where $f_G(\nu)$ is the halo mass function for Gaussian initial conditions and $R_{NG}$ represents the relative effect of PNG, we can split the scale-independent bias parameters $b_{10}^L$ and $b_{20}^L$ into
\bea
b_{10}^L & = & b_{10,G}^L+\Delta b_{10,NG}^L\,,\\
b_{20}^L & = & b_{20,G}^L+\Delta b_{20,NG}^L\,,
\eea
where the Gaussian components $b_{i0,G}$ are obtained from the Eq.s~(\ref{eq:biasLGa}) and (\ref{eq:biasLGb}) in terms of the Gaussian mass function $f_G$, while the non-Gaussian corrections are given by
\bea
\Delta b_{10,NG}^L & = & \frac{1}{R_{NG}(\nu_{10})}\left.\frac{\partial R_{NG}(\nu_{10})}{\partial \d_0}\right|_{\d_0,\phi_0=0}\,,\label{eq:dngbiasL1}\\
\Delta b_{20,NG}^L & = & \frac{1}{R_{NG}(\nu_{10})}\left.\frac{\partial^2 R_{NG}(\nu_{10})}{\partial \d_0^2}\right|_{\d_0,\phi_0=0}+2\,b_{10,G}^L\,\Delta b_{10,NG}^L\,.\label{eq:dngbiasL2}
\eea

As shown in \citep{GiannantonioPorciani2010}, all the $b_{ij}$ with $j\ne0$ can be written in terms of the $b_{i0}$, so that in particular
\bea
b_{01}^L & = & 2\fNL\d_c\,b_{10}^L\,,\label{eq:b01Lb10}\\
b_{11}^L & = & 2\fNL\left(\d_c\,b_{20}^L-b_{10}^L\right)\,.\label{eq:b11Lb20b10}
\eea
One should keep in mind that these relations are strictly valid for a universal mass function. For non-universal mass functions, the non-Gaussian bias corrections should be computed through a direct evaluation of derivatives of the halo mass function with respect to mass \citep{ScoccimarroEtal2011}.

\subsection{Eulerian bias}

The Eulerian halo density $\d_h$ can be expressed in terms of $\d_h^L$ as \citep{MoWhite1996}
\beq\label{eq:dhEulerianExp}
\d_h(\d,\phi_0)=\d+(1+\d)\d_h^L(\d_0,\phi_0)\,,
\eeq
where we notice now the additional dependence of $\d_h^L$ (and $\d_h$) on the linear and Gaussian curvature perturbation $\phi_0$.
To obtain an expression of $\d_h$ as a function of the {\em nonlinear} matter overdensity $\d$, we need to express the {\em linear} density perturbations $\d_0$ as a function of $\d$. This is usually done assuming the expansion $\d_0=\sum_{i=1}^{\infty}a_i\d^i=a_1\d+a_2\d^2+ ...$ derived in the spherical collapse approximation. In this case, the series coefficients correspond to spherical averages of the kernel of the perturbative expansion, with, in particular, $a_1=1$ and $a_2=-17/21$\footnote{One can in principle consider the more general case where the relation between $\d_0$ and $\d$ is given, in Fourier space, by the perturbative expansion
\bdm
\d_{0,\kv}=\d_{\kv}-\int d^3q_1\,d^3q_2\,\d_D(\kv-\qv_{12})\,F_2(\qv_1,\qv_2)\,\d_{\qv_1}\,\d_{\qv_2}+\dots\,.
\edm
This can have relevant consequences on the correction to the quadratic bias due to PNG, when the Gaussian, linear perturbation $\d_0$ in term $b_{01}\phi_{0,\kv}=b_{01}\d_{0,\kv}/M(k)$ is replaced by its the nonlinear and non-Gaussian counterpart $\d$ as suggested by the results of \citep{ScoccimarroEtal2011}. We will not further discuss this rather thorny issue, leaving it for future work.}.

On inserting Eq.~(\ref{eq:dhpos}) into Eq.~(\ref{eq:dhEulerianExp}), we can express the Eulerian bias parameters in terms of the Lagrangian ones \citep{GiannantonioPorciani2010, BaldaufSeljakSenatore2011}\footnote{Our expression for $b_{11}$ is different from the same expression in \citep{GiannantonioPorciani2010}, where $b_{11} = a_1b_{11}^L/2+b_{01}^L$.}
\bea
b_{10} & = & 1+b_{10}^L\,, \label{eq:biasEUa}\\
b_{01} & = & b_{01}^L\,,\label{eq:biasEUb}  \\
b_{20} & = & b_{20}^L+2\left(a_1+a_2\right)b_{10}^L\,,\label{eq:biasEUc} \\
b_{11} & = & a_1b_{11}^L+b_{01}^L\,.\label{eq:biasEUd} 
\eea
As in the Lagrangian case, Eq.~(\ref{eq:b11Lb20b10}) and (\ref{eq:b01Lb10}), the two extra parameters induced by PNG can be written as a function of $b_{10}$ and $b_{20}$
\bea
\label{eq:b01Eb10}
b_{01} & = & 2\fNL\d_c\,(b_{10}-1)\,,
\\
\label{eq:b11Eb20b10}
b_{11} & = & 2\fNL\left[\d_c\,b_{20}+\left(\frac{13}{21}\d_c-1\right)\left(b_{10}-1\right)\right]\,,
\eea
where we replaced the coefficients $a_1$ and $a_2$ of the spherical collapse expansion with their numerical values. We note that these additional terms vanish for an unbiased population ($b_{10}=1$, $b_{20}=0$). 

In our calculations, we will assume the above expressions for $b_{01}$ and $b_{11}$ to depend only on the Gaussian component of $b_{10}$ and $b_{20}$. In fact, in order to explicitly account for all the corrections linear in $\fNL$, we split $b_{10}$ and $b_{20}$ into their Gaussian and non-Gaussian pieces:
\bea
b_{10} & = & b_{10,G}+\D b_{10,NG}\,,\\
b_{20} & = & b_{20,G}+\D b_{20,NG}\,,
\eea
with
\bea
\D b_{10,NG} & = & \D b_{10,NG}^L\,,\\
\D b_{20,NG} & = & \D b_{20,NG}^L+\frac{8}{21}\D b_{10,NG}^L\,.
\eea
Taking into account Eq.s~(\ref{eq:dngbiasL1}), (\ref{eq:dngbiasL2}) and (\ref{eq:nu10}) we have 
\bea
\D b_{10,NG} & = & -\frac{1}{\d_c}\frac{\nu}{R_{NG}}\frac{\partial R_{NG}}{\partial\nu}\,,\label{eq:dngbias1}\\
\D b_{20,NG} & = & \frac{\nu^2}{\d_c^2}\frac{1}{R_{NG}}\frac{\partial^2 R_{NG}}{\partial\nu^2}-2\frac{\nu}{\d_c}\left(b_{10,G}-\frac{17}{21}\right)\frac{\nu}{R_{NG}}\frac{\partial R_{NG}}{\partial\nu}\,.\label{eq:dngbias2}
\eea
where the expression for $\D b_{10,NG}$ has been derived in \citep{DesjacquesSeljakIliev2009}.

\section{The model}
\label{sec:model}

As already mentioned, we will not consider any loop-correction to the halo power spectrum and bispectrum induced by the nonlinear local bias expansion. Here, we will focus on the leading-order, tree-level expressions for the matter-matter-halo cross-bispectrum and for the halo bispectrum. They will be compared to the simulation results in section~\ref{sec:results}. For comparison, we will also consider the matter-halo cross-power spectrum and the halo power spectrum, already studied in numerical simulations in \citep{DalalEtal2008, DesjacquesSeljakIliev2009, GrossiEtal2009, PillepichPorcianiHahn2010, WagnerVerde2011, ScoccimarroEtal2011}.  

The starting point is Eq.~(\ref{eq:dhpos}) which, in Fourier space, up to quadratic corrections in the density and curvature fields reads
\bea
\d_h(\kv) & \simeq & b_{10}\d(\kv)+b_{01}\phi_0(\kv)+\frac12\,b_{20}\int d^3k_1 d^3k_2\,\d_D(\kv-\kv_{12})\,\d(\kv_1)\,\d(\kv_2)\nonumber\\
& & +\,b_{11}\int d^3k_1 d^3k_2\,\d_D(\kv-\kv_{12})\,\phi_0(\kv_1)\,\d(\kv_2)+\O(\fNL^2)\,.
\eea
Here and henceforth we denote $\kv_{i_1\dots i_n}\equiv\kv_{i_1}+\dots+\kv_{i_n}$.
To highlight the ``scale-dependent'' corrections to halo bias factors, it is convenient to rewrite this expansion in terms of the fully nonlinear and non-Gaussian matter density $\d$ and of the Gaussian component of the {\em linear} matter density $\d_0$. In the following we will assume that $\d_0$ corresponds to the Gaussian component. We then have
\bea\label{eq:dhexp}
\d_h(\kv) & \simeq & b_{10}\d(\kv)+c_{01}(k)\d_0(\kv)+\frac12\,b_{20}\int d^3k_1 d^3k_2\,\d_D(\kv-\kv_{12})\,\d(\kv_1)\,\d(\kv_2)\nonumber\\
& & +\frac12\int d^3k_1 d^3k_2\,\d_D(\kv-\kv_{12})\,\left[c_{11}(k_1)\,\d_0(\kv_1)\,\d(\kv_2)+(k_1\leftrightarrow k_2)\right]\O(\fNL^2)\,,
\eea
where we have defined 
\bea
c_{01}(k) & \equiv & \frac{b_{01}}{M(k)}\,,\\
c_{11}(k) & \equiv & \frac{b_{11}}{M(k)}\,.
\eea
$c_{01}(k)$ clearly corresponds to the usual non-Gaussian scale-dependent correction to linear bias, whereas $c_{11}(k)$ is the analogous contribution to the quadratic bias.

\subsection{The matter-halo and halo power spectrum}

To calculate the two-point correlators, we will neglect the contribution of loop corrections induced by nonlinear bias and, therefore, limit ourselves to the two linear terms of Eq.~(\ref{eq:dhexp}).

The leading order contribution to the halo-matter cross-power spectrum is then simply given by
\beq
P_{\d h}(k)=b_{10}\, P_\d(k)+c_{01}(k)\, P_{\d\d_0}(k)\,,
\eeq
while for the halo power spectrum we have
\beq
P_h(k)=b_{10}^2\, P_\d(k)+2\, b_{10}\, c_{01}(k)\, P_{\d\d_0}(k)\,.
\eeq
Both expressions simply correspond to the usual halo bias correction of \citep{DalalEtal2008} plus the scale-independent correction introduced by \citep{DesjacquesSeljakIliev2009}. 
Here, $P_{\d}$ is the nonlinear matter power spectrum whereas $P_{\d\d_0}$ is defined as $\la\d_{\kv_1}\d_{0,G,{\kv_2}}\rangle\equiv \d_D(\kv_{12})P_{\d\d_0}(k_1)$. We will compute these quantities in standard perturbation theory including loop corrections corresponding to ${\mathcal O}(\d_0^6)$ in the linear density field. We refer the reader to Appendix~\ref{app:PT} for details of the PT expressions.  

At nonlinear level, both $P_\d$ and $P_{\d\d_0}$ receive correction due to the non-Gaussian initial conditions and, therefore, depend on $\fNL$. As we will do later for the three-point matter correlators, we separate the Gaussian from the non-Gaussian contribution and write $P_{\d}=P_{\d,G}+\D P_{\d,NG}$ and $P_{\d\d_0}=P_{\d\d_0,G}+\D P_{\d\d_0,NG}$. The non-Gaussian corrections $\D P_{\d,NG}$ and $\D P_{\d\d_0,NG}$ will be evaluated at linear order in $\fNL$. 

Similarly, we can distinguish the Gaussian and non-Gaussian components in the matter-halo and halo power spectrum. For Gaussian initial conditions, the matter-halo cross-power spectrum reduces to 
\beq\label{eq:PmhG}
P_{\d h,G}(k)=b_{10,G}\, P_{\d,G}(k)\,,
\eeq 
while the non-Gaussian correction is given by
\beq\label{eq:dPmhNG}
\Delta P_{\d h,NG}(k)=b_{10,G}\, \Delta P_{\d,NG}(k)+\Delta b_{10,NG}\, P_{\d,G}(k)+c_{01}(k)\,P_{\d\d_0,G}+\O(\fNL^2)\,.
\eeq
For the halo power spectrum we have instead
\beq\label{eq:PhG}
P_{h,G}(k)=b_{10}^2\, P_\d(k)\,,
\eeq
with the non-Gaussian correction given by
\beq\label{eq:dPhNG}
\Delta P_{h,NG}(k)=b_{10,G}^2\, \Delta P_{\d,NG}(k)+2\,b_{10,G}\,\Delta b_{10,NG} P_{\d,G}(k)+2\,b_{10,G}\,c_{01}(k)\,P_{\d\d_0,G}+\O(\fNL^2)\,.
\eeq
This distinction between Gaussian and non-Gaussian contribution will be crucial in the analysis of Section~\ref{sec:results} as we will compare them separately with the simulations results.

\subsection{The matter-matter-halo and halo bispectrum}

We define the cross matter-matter-halo bispectrum $B_{\d\d h}$ as
\beq
\la\d(\kv_1)\d(\kv_2)\d_h(\kv_3)\rangle\equiv \d_D(\kv_{123})\,B_{\d\d h}(k_1,k_2;k_3)\,,
\eeq
where we assume that the third wavenumber $k_3$ to represent the halo overdensity variable. At leading order in the local bias expansion, the expression for the matter-matter-halo cross-bispectrum is given by
\bea\label{eq:Bmmh}
B_{\d\d h}(k_1,k_2;k_3) & = & b_{10}\, B_\d(k_1,k_2,k_3)+c_{01}(k_3)\, B_{\d\d\d_0}(k_1,k_2;k_3)+b_{20}\, P_\d(k_1)P_\d(k_2)\nonumber\\
& & +\frac12\left[c_{11}(k_1)\, P_{\d\d_0}(k_1)P_\d(k_2)+(k_1\leftrightarrow k_2)\right]+\O(\fNL^2)\,,
\eea
where, in addition to the matter power spectra $P_{\d}$ and $P_{\d\d_0}$ already introduced, $B_\d$ represents the nonlinear matter bispectrum while $B_{\d\d\d_0}$, defined as $\la\d(\kv_1)\d(\kv_2)\d_0(\kv_3)\rangle\equiv \d_D(\kv_{123})\,B_{\d\d\d_0}(k_1,k_2;k_3)$ is the cross-bispectrum between two nonlinear matter density fields $\d$ and one linear (and Gaussian) mass overdensity $\d_0$. Again the last argument, $k_3$, refers to the $\d_0$ variable. For Gaussian initial conditions $B_{\d\d h}$ reduces to 
\beq\label{eq:BmmhG}
B_{\d\d h,G}(k_1,k_2;k_3) = b_{10,G}\, B_{\d,G}(k_1,k_2,k_3)+b_{20,G}\, P_{\d,G}(k_1)P_{\d,G}(k_2)\,,
\eeq
while the non-Gaussian correction explicitly is
\bea\label{eq:dBmmhNG}
\Delta B_{\d\d h,NG}(k_1,k_2;k_3) & = & b_{10,G}\, \Delta B_{\d,NG}(k_1,k_2,k_3)+\Delta b_{10,NG}\,B_{\d,G}(k_1,k_2,k_3)+c_{01}(k_3)\, B_{\d\d\d_0,G}(k_1,k_2;k_3)\nonumber\\
& & +b_{20,G}\, \left[\Delta P_{\d,NG}(k_1)P_{\d,G}(k_2)+(k_1\leftrightarrow k_2)\right]
\nonumber\\ & & 
+\D b_{20,NG}\, \left[P_{\d,G}(k_1)P_{\d,G}(k_2)+(k_1\leftrightarrow k_2)\right]\nonumber\\
& & +\frac12\left[c_{11}(k_1)\, P_{\d\d_0,G}(k_1)P_{\d,G}(k_2)+(k_1\leftrightarrow k_2)\right]+\O(\fNL^2)\,.
\eea
For the tree-level halo bispectrum, we obtain 
\bea
B_{h}(k_1,k_2,k_3) & = & b_{10}^3\, B_\d(k_1,k_2,k_3)+b_{10}^2\left[c_{01}(k_3)B_{\d\d\d_0}(k_1,k_2;k_3)+2\,{\rm perm.}\right]\nonumber\\
& & +b_{10}^2\,b_{20}\left[ P_\d(k_1)P_\d(k_2)+2~{\rm perm.}\right]\nonumber\\
& & +\frac12\,b_{10}\,b_{20}\left[c_{01}(k_1) P_{\d\d_0}(k_1)P_\d(k_2)+c_{01}(k_2) P_{\d\d_0}(k_2)P_\d(k_1)+2~{\rm perm.}\right]\nonumber\\
& & +\frac12\, b_{10}^2\,\left[c_{11}(k_1)\,P_{\d\d_0}(k_1)\,P_\d(k_2)+c_{11}(k_2)\,P_{\d\d_0}(k_2)\,P_\d(k_1)+2~{\rm perm.}\right]+\O(\fNL^2)\,.
\eea
For Gaussian initial conditions, the halo bispectrum is
\beq\label{eq:BhG}
B_{h,G}(k_1,k_2,k_3) =  b_{10,G}^3\, B_{\d,G}(k_1,k_2,k_3)+b_{10,G}^2\,b_{20,G}\left[ P_{\d,G}(k_1)P_{\d,G}(k_2)+2~{\rm perm.}\right]+\O(\fNL^2)\,.
\eeq
while the non-Gaussian correction is given by
\bea\label{eq:dBhNG}
\Delta B_{h,NG}(k_1,k_2,k_3) & = & b_{10,G}^3\, \Delta B_{\d,NG}(k_1,k_2,k_3) +3\,b_{10,G}^2\,\Delta b_{10,NG}\,B_{\d,G}(k_1,k_2,k_3)\nonumber\\
&  & +b_{10}^2\left[c_{01}(k_3)B_{\d\d\d_0}(k_1,k_2;k_3)+2\,{\rm perm.}\right]\nonumber\\
& & +b_{10,G}\,\left(2\,\D b_{10,NG}\,b_{20,G}+b_{10,G}\,\D b_{20,NG}\right)\left[P_{\d,G}(k_1)P_{\d,G}(k_2)+2~{\rm perm.}\right]\nonumber\\
& & +b_{10,G}^2\,b_{20,G}\,\left[\D P_{\d,NG}(k_1)P_{\d,G}(k_2)+\D P_{\d,NG}(k_2)P_{\d,G}(k_1)+2~{\rm perm.}\right]\nonumber\\
& & +\frac12\,b_{10,G}\,b_{20,G}\left[c_{01}(k_1) P_{\d\d_0,G}(k_1)P_{\d,G}(k_2)+c_{01}(k_2) P_{\d\d_0,G}(k_2)P_{\d,G}(k_1)+2~{\rm perm.}\right]\nonumber\\
& & +\frac12\, b_{10,G}^2\,\left[c_{11}(k_1)\,P_{\d\d_0,G}(k_1)\,P_{\d,G}(k_2)+c_{11}(k_2)\,P_{\d\d_0,G}(k_2)\,P_{\d,G}(k_1)+2~{\rm perm.}\right]\nonumber\\
& &+\O(\fNL^2)\,.
\eea
We refer again the reader to Appendix~\ref{app:PT} for the details of the evaluation of the matter correlators, including the bispectra $B_{\d}$ and $B_{\d\d\d_0}$, in perturbation theory.

\section{Simulations and Halo samples}
\label{sec:simulations}

To measure the effect of non-Gaussian initial conditions of the local quadratic type on the halo bispectrum, we use a series of eight twelve sets of three N-body simulations of the $\Lambda$CDM cosmology, each of which has $\fNL=0,\pm 100$ \citep{DesjacquesSeljakIliev2009}. The same Gaussian random seed field $\phi$ is employed in each triplet of runs so as to minimize the sampling variance. These simulations evolve 1024$^3$ dark matter particle in a cubical box of size 1600$\Mpc$. The force resolution is 0.04 times the mean inter-particle distance. The (dimensionless) power spectrum of the Gaussian part $\phi(\xv)$ of the Bardeen potential is the usual power-law $\Delta_\phi^2(k)\equiv  k^3 P_\phi(k)/(2\pi^2)=A_\phi (k/k_0)^{n_s-1}$. The spectral index is $n_s=0.96$ and the normalization of the Gaussian curvature perturbations is $A_\phi=7.96\times 10^{-10}$ at the pivot point $k_0=0.02$Mpc$^{-1}$, close to the best-fitting values inferred from CMB measurements \citep{KomatsuEtal2009B}. We will consider a single output at redshift $z=0.509$.

Friends-of-friends (FoF) halos were extracted using a linking length of 0.2 times the mean inter-particle distance. We will present results for a low-mass bin defined by $8.8\times 10^{12}\Ms<M<1.6\times 10^{13}\Ms$, and a high-mass bin given by $M>1.6\times 10^{13}\Ms$. The mass thresholds has been chosen in order to have the same halo number density of about $\bar{n}_h=1.8\times 10^{-4}\icMpc$ for both mass bins. As we will see, the linear bias for the low mass sample is about $1.6$ while for the high mass sample we have about $2.3$. In addition the choice provides two halo populations characterized by quadratic bias parameters of different sign: negative for low mass halos and positive for high mass halos. This has a direct implication for the sign of non-Gaussian corrections due to nonlinear bias. We finally remark that the shot noise contribution to the power spectrum is comparable to the halo power spectrum itself at about $0.14\kMpc$ and $0.2\kMpc$ for the low and high mass bin respectively. In all our results we present the halo power spectrum $P_h$ and the halo bispectrum $B_h$ corrected for shot-noise, while we ignore such correction for cross-correlations between matter and halo density fields.

\section{Signal-to-noise}
\label{sec:signal}

To assess the effect of non-Gaussianity on the halo and matter correlation functions, it is useful to compare the cumulative signal-to-noise for various statistics. In the case of the the power spectrum, this quantity is defined as
\beq\label{eq:PStoN}
\left(\frac{S}{N}\right)^2_{P}=\sum_{k = k_f}^{k_{max}}\frac{\left[P_{m,NG}(k)-P_{m,G}(k)\right]^2}{{\rm Var}[P_{m,G}(k)]}\,,
\eeq 
where $P_{m,NG}$ and $P_{m,G}$ represent the matter power spectrum for non-Gaussian and Gaussian initial conditions, respectively, while ${\rm Var}[P_{m,G}]$ is the variance of the mass power spectrum in the Gaussian case. Given a surveyed volume $V\propto L^3$, the sum runs over all wavenumbers $k$ in steps of the fundamental mode $k_f=2\pi/L$. Similarly, the signal-to-noise for the bispectrum is
\beq\label{eq:BStoN}
\left(\frac{S}{N}\right)^2_{B}=\sum_{k_1\le k_2\le k_3 = k_f}^{k_{max}}\frac{\left[B_{m,NG}(k_1,k_2,k_3)-B_{m,G}(k_1,k_2,k_3)\right]^2}{{\rm Var}[B_{m,G}(k_1,k_2,k_3)]}\,.
\eeq 
Here the sum is over all triangular configurations, \ie all the triplets $k_1$, $k_2$ and $k_3$ forming a closed triangle. For simplicity, we only include in both cases the variance of the correlators while, in principle, the complete covariance between $k$-bins or bispectrum triangles should be considered. Since this calculation is for illustrative purposes only, we just note that the effects of covariance can significantly reduce the signal-to-noise for both the power spectrum and bispectrum, but they marginally affect the comparison between the two correlators at large and mildly non-linear scales (see, for instance, \citep{SefusattiScoccimarro2005, SefusattiEtal2006} for a realistic estimate of the effects of covariance on the galaxy power spectrum and bispectrum). 

\begin{figure}[!t]
\begin{center}
{\includegraphics[width=0.48\textwidth]{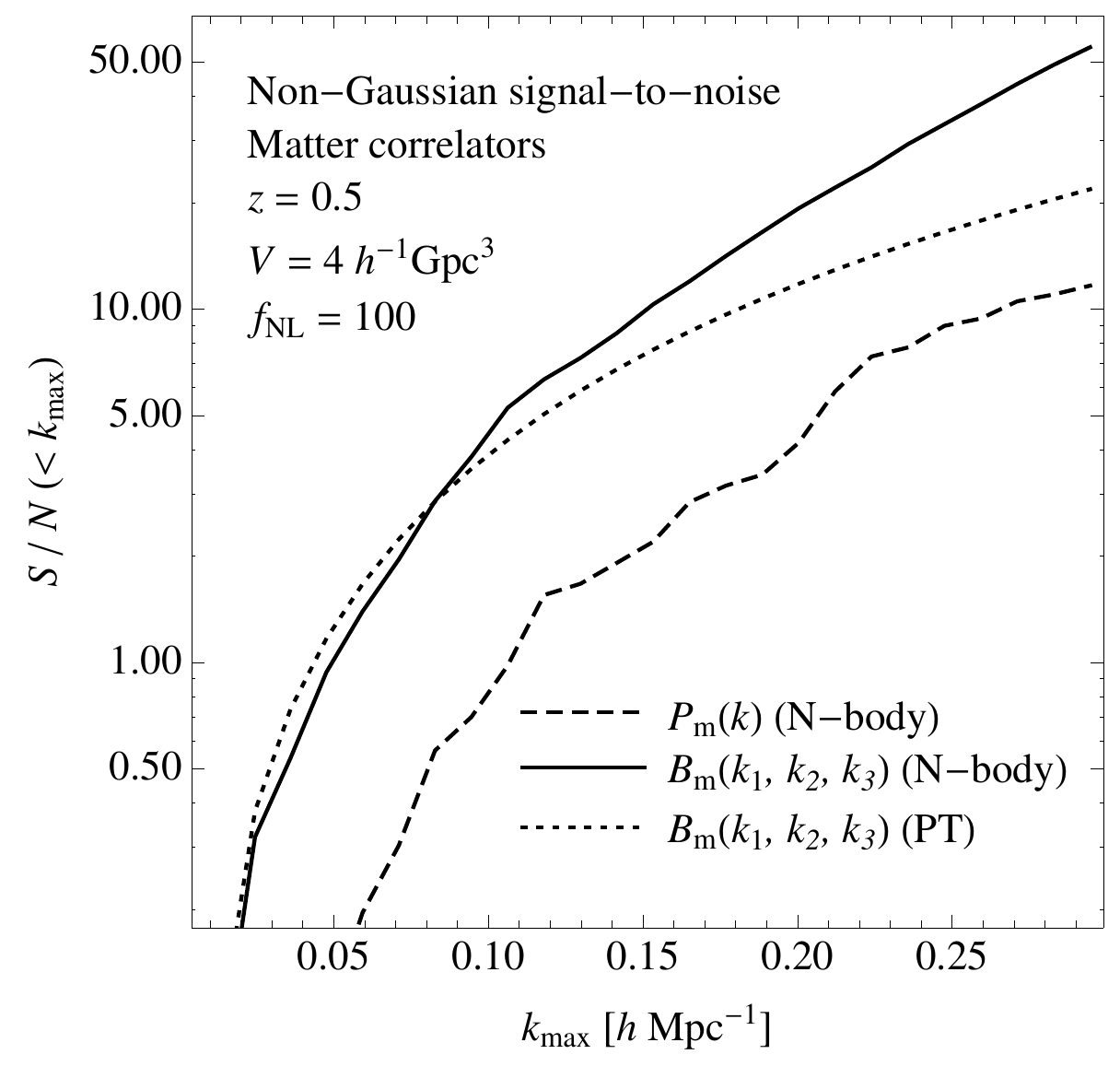}}
{\includegraphics[width=0.48\textwidth]{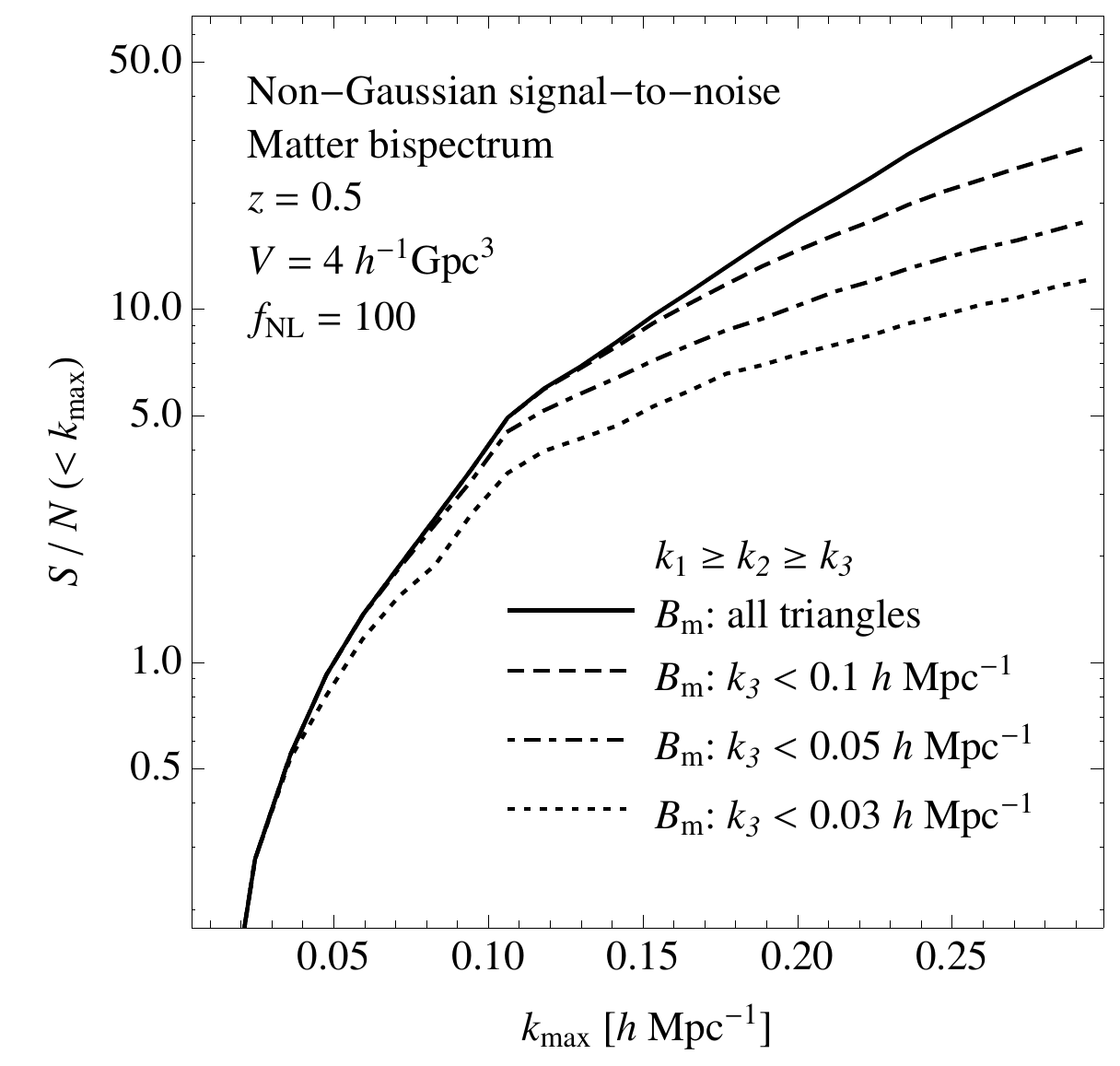}}
\caption{{\em Left panel}: Cumulative signal-to-noise for the effect of non-Gaussian initial conditions on the matter power spectrum and bispectrum as defined in Eq.~(\ref{eq:PStoN}) and Eq.~(\ref{eq:BStoN}) as a function of the largest wavenumber included ($k_{max}$). Continuous and dashed curves indicate the signal-to-noise for the matter bispectrum and power spectrum, respectively, when the correlators and their variance are determined from the simulations. The dotted line correspond to the tree-level prediction for the matter bispectrum and its variance in Eulerian PT. {\em Right panel:} Cumulative signal-to-noise for the matter bispectrum with the sum in Eq.~(\ref{eq:BStoN}) restricted to triangles with $k_3<0.03\kMpc$ ({\em dotted curve}), $k_3<0.05\kMpc$ ({\em dot-dashed curve}) and $k_3<0.1\kMpc$ ({\em dashed curve}) compared to the case where all triangles are included ({\em continuous curve}).}
\label{fig:StoNmatter}
\end{center}
\end{figure}
In the left panel of Fig.~\ref{fig:StoNmatter} we show the cumulative signal-to-noise for the matter power spectrum ({\em dashed curve}) and bispectrum ({\em continuous curve}) as measured from the simulations. The dotted curve represents the predicted signal-to-noise for the matter bispectrum, assuming the correction induced by non-Gaussian initial conditions is given by the linearly evolved initial bispectrum $B_0$ and the bispectrum variance is given by its leading Gaussian component \citep{ScoccimarroEtal1998, ScoccimarroSefusattiZaldarriaga2004}
\beq
{\rm Var}[B(k_1,k_2,k_3)]=\frac{s_B}{8\pi^2 k_1 k_2 k_3}P(k_1)P(k_2)P(k_3)\,,
\eeq
with $s_B=6$, $2$ or $1$ for equilateral, isosceles or scalene triangles respectively. The theoretical prediction agrees reasonably well with the measured signal at large scales. On the other hand, the excess in the measured cumulative signal-to-noise at small scales, which are affected as well by a larger variance, is due to the impact of non-Gaussian initial conditions on the nonlinear evolution of the matter bispectrum. Most importantly, Fig.~\ref{fig:StoNmatter} clearly shows that the effect of PNG in the matter bispectrum is larger by roughly a factor of $4$ than in the matter power spectrum, where PNG only enters at the nonlinear level. This is true on mildly nonlinear scales, but it is possible that this behavior extends to smaller scales as well \citep{SmithDesjacquesMarian2010}. In the right panel of Fig.~\ref{fig:StoNmatter}, we show the cumulative signal-to-noise for the matter bispectrum restricted to different classes of triangular configurations. Specifically, the sum in Eq.~(\ref{eq:BStoN}) runs over triangles with $k_3<0.03\kMpc$ ({\em dotted curve}), $k_3<0.05\kMpc$ ({\em dot-dashed curve}) and $k_3<0.1\kMpc$ ({\em dashed curve}). The dotted curve, for instance, corresponds mainly to squeezed triangles, with one mode (corresponding to $k_3$) deeply in the linear regime and the two others (for which $k_1$, $k_2\ge k_3$) gradually probing nonlinear scales as $k_{max}$ increases. Despite the fact that squeezed configurations provide a significant contribution to the signal-to-noise as is expected for local non-Gaussianity, the signal is fairly equally distributed over all configurations. For instance, for $k_{\max}=0.2\kMpc$, triangles with $k_3<0.03\kMpc$ account for one-third of the total signal solely. 

\begin{figure}[!t]
\begin{center}
{\includegraphics[width=0.48\textwidth]{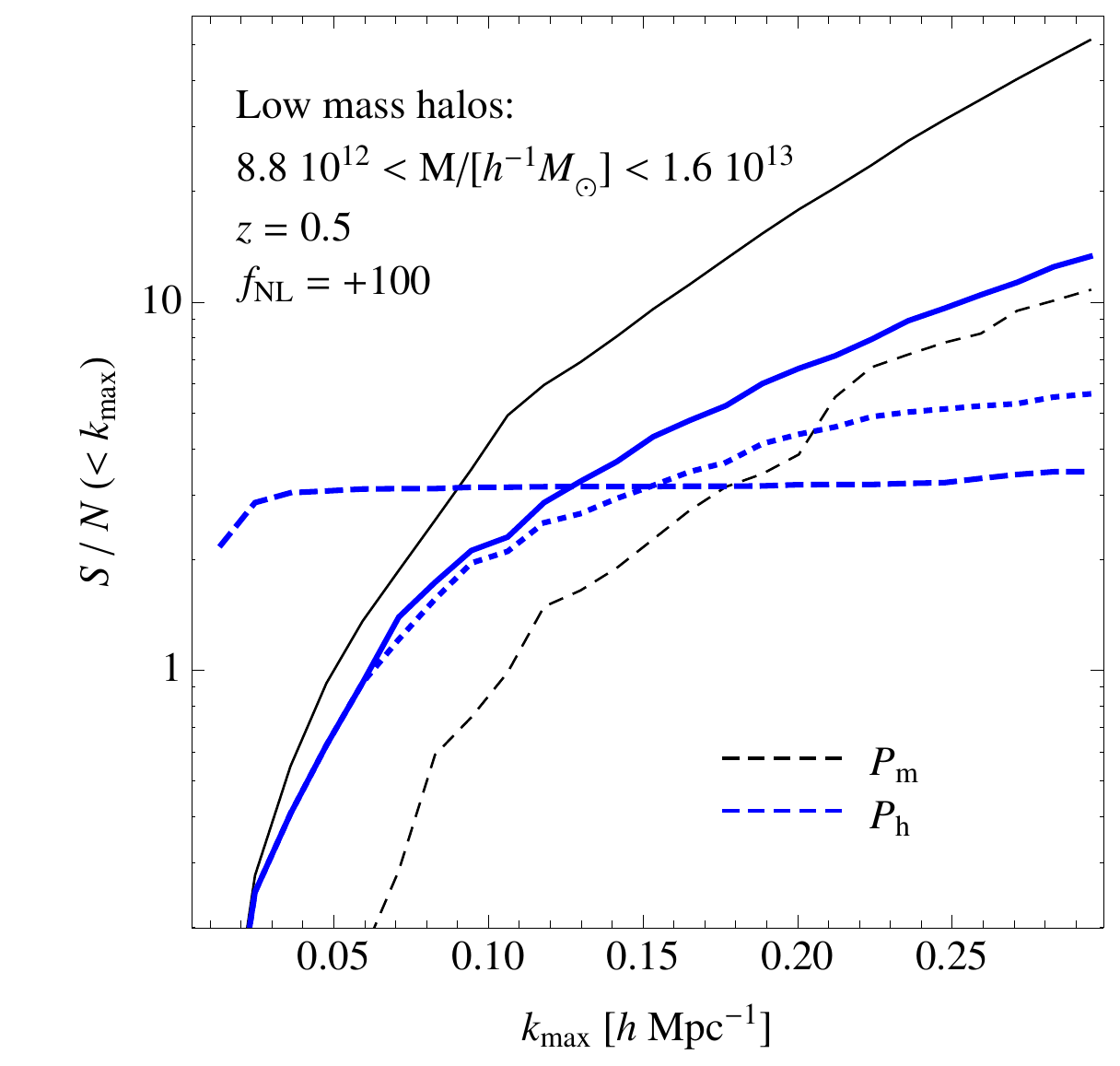}}
{\includegraphics[width=0.48\textwidth]{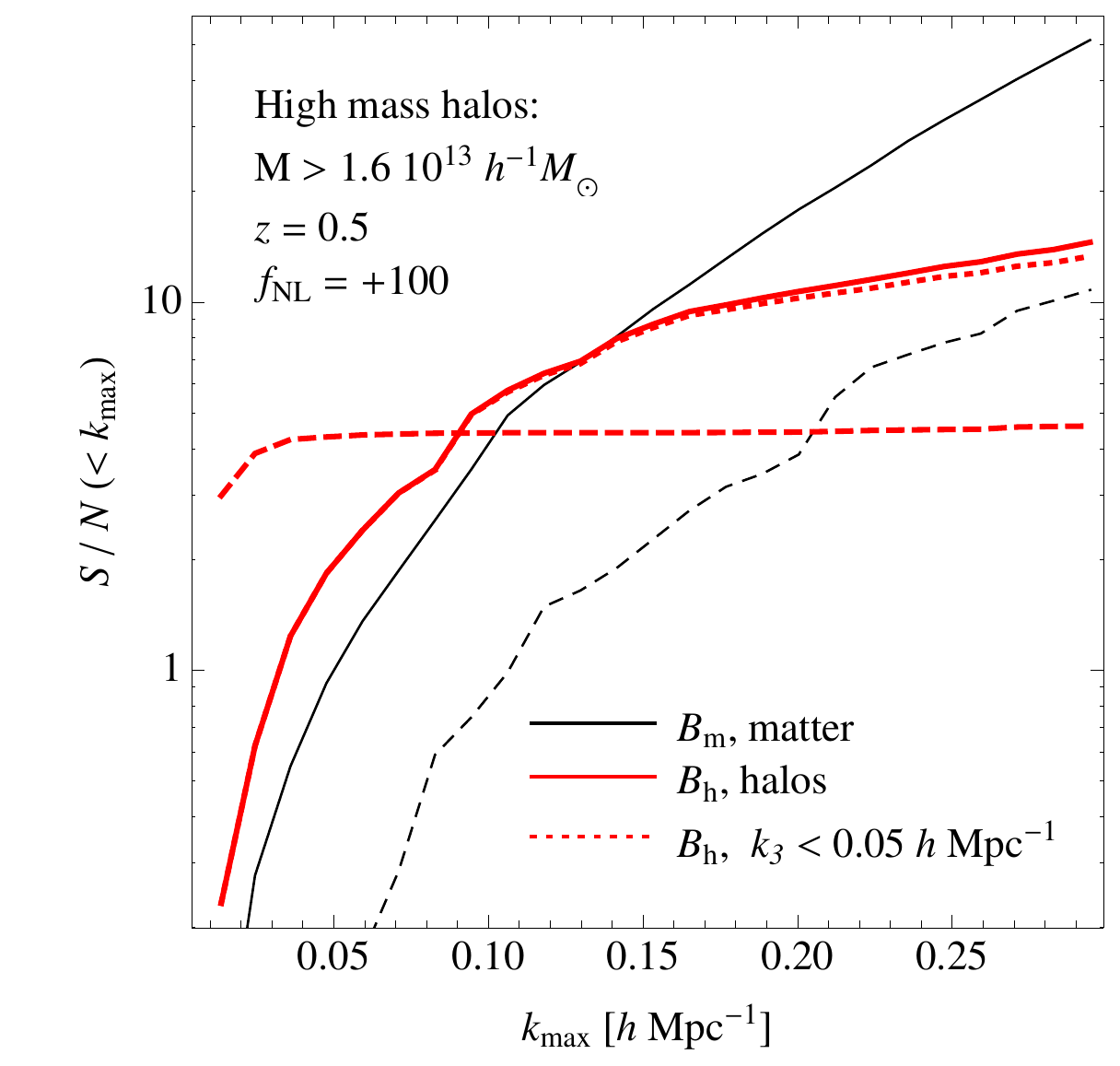}}
\caption{Cumulative signal-to-noise for the effect of non-Gaussian initial conditions as defined in Eq.s~\eqref{eq:PStoN} and \eqref{eq:BStoN} but evaluated respectively for the halo power spectrum $P_h$ ({\em dashed curves}) and halo bispectrum $B_h$ ({\em continuous curves}), as a function of the small scale included ($k_{max}$). Also shown are the signal-to-noise for the halo bispectrum restricted to triangles with $k_3<0.03\kMpc$ ({\em dotted curves}) and, for comparison, the signal-to-noise for the matter power spectrum and bispectrum ({\em respectively dashed and continuous black, thin curves}). Left panels shows the results for low mass halos ({\em blue curves}) while right panels show the results for the high mass halos ({\em red curves}). Inset labels are common to both panels.}
\label{fig:StoNhalos}
\end{center}
\end{figure}
This picture changes substantially when we consider halo correlation functions. For the halo power spectrum, the scale-dependent correction to the linear bias dominates the signal at large scales. In Fig.~\ref{fig:StoNhalos} we plot the cumulative signal-to-noise defined in Eq.s~\eqref{eq:PStoN} and \eqref{eq:BStoN} as estimated from measurements of the halo power spectrum $P_h$ ({\em dashed curves}) and bispectrum $B_h$ ({\em continuous curves}), respectively. Results are shown for the low mass ({\em blue curves, left panel}) and the high mass halo sample ({\em red curves, right panel}) assuming $\fNL=100$. In addition, we show the signal-to-noise for the halo bispectrum restricted to triangles with $k_3<0.03\kMpc$ ({\em dotted curves}) and, for comparison, overlay the signal-to-noise for the matter power spectrum and bispectrum ({\em respectively dashed and continuous black, thin curves}) reproduced from Fig.~\ref{fig:StoNmatter}. 

It is clear from all the cases considered here that the distribution of the signal as a function of scale differs significantly between the power spectrum and the bispectrum. In the power spectrum case, the main effect of non-Gaussian initial conditions is the scale-dependent correction to the linear bias. Since it is largest at the largest scales, the cumulative signal reaches its maximum already at $k\lesssim 0.05\kMpc$, nonlinear corrections at smaller scales adding little signal. For the bispectrum, the cumulative signal-to-noise has a different dependence on the maximum wavenumber $k_{max}$ mainly because it is a sum over all possible triangular configurations, whose number grows as $k_{max}^3$ \citep{SefusattiScoccimarro2005}. At large scales, the signal is suppressed relative to that of the power spectrum since there are only a few measurable triangles characterized by a large variance. For larger values of $k_{max}$ the number of triangles grows considerably. As a consequence, the effect of PNG becomes larger in the bispectrum at relatively small $k_{max}\sim 0.1\kMpc$ for the halo samples considered. The fact that the matter bispectrum includes the initial components linearly extrapolated as well as relatively larger nonlinear corrections contributes to enhance this effect, although, as we will see, the overall signal in the halo bispectrum is characterized as well by significant cancellations of individual terms on the r.h.s. of Eq.~(\ref{eq:dBhNG}).

It is interesting to notice how the distribution of the signal among different triangles differs between halos and mass. As mentioned above, the dotted curves correspond to triangles with one side being less than $0.03\kMpc$, \ie mostly squeezed configurations. For halos, this subset of triangles accounts for a significantly larger fraction of the total signal-to-noise. As we will see later, this is related to the fact that the primordial component of the matter bispectrum and the effects of PNG on linear and quadratic bias both peak on squeezed configurations. At first sight, this is good news as one may think that most of the information on {\em local} non-Gaussian initial conditions can be extracted from a limited number of squeezed configurations. However, one should bear in mind that such triangles are expected to be highly correlated as they share at least one wavenumber. In this regards, we emphasize that, even though the cumulative signal-to-noise shown in Fig.~\ref{fig:StoNmatter} and \ref{fig:StoNhalos} is obtained from measurements in N-body simulation, it does not include the {\em covariance} among different triangular configurations. We expect it to be quite significant for triangles sharing one or more sides, particularly at large scales \citep{SefusattiScoccimarro2005, GaztanagaScoccimarro2005, SefusattiEtal2006}.

\section{Results}
\label{sec:results}

\subsection{Measurements and analysis}
\label{ssec:Analysis}

In this section we present measurements of the cross matter-matter-halo bispectrum $B_{mmh}$ and the halo bispectrum $B_h$ in the simulations described in Section~\ref{sec:simulations}. To facilitate the comparison with recently published results, we show as well measurements of the cross matter-halo power spectrum and halo power spectrum. These measurements provide a first assessment of the model developed in Section~\ref{sec:model}. The comparison between N-body measurements and predictions will involve fitting for several constant bias parameters such as the Gaussian component of the linear and quadratic bias $b_{10,G}$ and $b_{20,G}$ and their scale-independent, non-Gaussian corrections $\D b_{10,NG}$ and $\D b_{20,NG}$. The best fit values will be compared with expectations from the peak-background split theory in Section~\ref{sec:bias}.

For each of the four correlators, we will consider three specific measurements:
\begin{enumerate}
\item {\bf the Gaussian component}, \ie the measurements of the correlators in the simulations with Gaussian initial conditions, that is $P_{mh,G}$, $P_{h,G}$, $B_{mmh,G}$ and $B_{h,G}$;
\item {\bf the non-Gaussian correction}, \ie the {\em difference} between the non-Gaussian ($\fNL=+100$) and the Gaussian measurements,
\bdm
\D C_{NG}\equiv C_{NG}(\fNL=+100)-C_{G}(\fNL=0)\,,
\edm
where $C$ stands for any correlator such as $P_{mh}$, $P_m$, $B_{mmh}$ and $B_h$; this difference is {\em first} computed for each realization separately and {\em then} averaged over the available realizations, thereby reducing the scatter as each Gaussian/non-Gaussian simulations pair is obtained from the same Gaussian seeds; 
\item {\bf the $\O(\fNL^2)$ component}, obtained as the mean over all realizations of the combination 
\bdm
[C_{NG}(\fNL=+100)+C_{NG}(\fNL=-100)-2\,C_{NG}(\fNL=0)]/2=\O(\fNL^2)\,,
\edm 
measured for each $\fNL=0$, $\pm 100$ triplet; this quantity is sensitive to any non-Gaussian correction beyond linear order in the nonlinear parameter and, therefore, is a measure of the amplitude of the corrections quadratic in $\fNL$, which are neglected in our model.
\end{enumerate}
In other words, we do not compare the model directly to the measurements of each correlator, but we analyze separately the Gaussian and non-Gaussian components. The prediction for the non-Gaussian correction are easily obtained from the perturbative expressions of our model, Eq.s~(\ref{eq:dPmhNG}), (\ref{eq:dPhNG}), (\ref{eq:dBmmhNG}) and (\ref{eq:dBhNG}).

We are interested, in the first place, in the behavior of the halo correlators at large scales and in the ability of the tree-level approximation for the bias expansion to capture the main effects of non-Gaussian initial conditions on the matter-matter-halo and halo bispectra. In particular we want to verify if the {\em functional form} of the different components in this approximation present the correct dependence on scale {\em and} on the triangle shape. In other words, we want to establish if the terms contributing to $B_{mmh}$ and $B_h$, Eq.s~(\ref{eq:dBmmhNG}) and (\ref{eq:dBhNG}) respectively, can provide an accurate model for all triangular configurations at large scales.  

As mentioned above, we will treat the (scale-independent) Gaussian bias factors $b_{10,G}$ and $b_{20,G}$ and their non-Gaussian corrections $\D b_{10,NG}$ and $\D b_{20,NG}$ as free parameters. While these parameters control the amplitude of several terms in the expressions of the halo correlators, our choice does not limit significantly the predictivity of the model. Firstly, we assume that the terms generating the scale-dependent bias corrections in the halo density expansion Eq.~(\ref{eq:dhexp}), like $c_{01}(k)$ and $c_{11}(k)$, are fully determined by $b_{10,G}$ and $b_{20,G}$ (which can be measured from the Gaussian realizations). Secondly, the fit involves {\em all} the triangular configurations down to a certain scale or, equivalently, a large number of degrees of freedom. In our model, varying those four free parameters (as we will see shortly, there are in practice less than four) can affect the shape dependence of the halo bispectra only to a limited extent. 

The fitting procedure of the halo power spectrum {\em and} bispectrum follows these general steps:
\begin{enumerate}
\item the {\em Gaussian linear bias parameter} $b_{10,G}$ is obtained from the Gaussian halo power spectrum $P_{h,G}$;
\item the {\em non-Gaussian scale-independent correction to the linear bias} $\Delta b_{10,NG}$ is determined from the non-Gaussian correction to the halo power spectrum, $\Delta P_{h,NG}=P_{h,NG}-P_{h,G}$, {\em assuming} the best-fit value of $b_{10,G}$ obtained from $P_{h,G}$;
\item the {\em Gaussian quadratic bias} $b_{20,G}$ is determined from the Gaussian halo bispectrum $B_{h,G}$ upon setting $b_{10,G}$ to its best-fit value from step 1.
\item the {\em non-Gaussian scale-independent correction to the quadratic bias} $\Delta b_{20,NG}$ is computed from the non-Gaussian correction to the halo bispectrum, $\Delta B_{h,NG}=B_{h,NG}-B_{h,G}$, {\em assuming} the best-fit values of $b_{10,G}$, $\Delta b_{10,NG}$ and $b_{20,G}$ from steps 1, 2 and 3, respectively.
\end{enumerate} 
Analogously, we independently fit the model of the cross matter-halo power spectrum $P_{mh}$ and matter-matter-halo bispectrum $B_{mmh}$. 

In this way, the fit to each of the measured quantities $P_{h,G}$, $\D P_{h,NG}$, $B_{h,G}$ and $\D B_{h,NG}$ effectively is a one-parameter fit. Clearly, we could have adopted other fitting procedures where, for instance, all the parameters are obtained from measurements of the bispectrum alone. Keeping the model as predictive as possible is the main motivation of our choice. A comparison between the best fit bias parameters and the values predicted from the peak-background split justifies {\em a posteriori} our approach. These aspects will be discussed in details in Section~\ref{sec:bias}.

Since we can only expect the tree-level approximation from the bias expansion to be valid at large scales, we restrict the fits to wavenumbers $k\le 0.07\kMpc$ for both the power spectra and the three sides of the triangular bispectrum configurations. This choice notwithstanding, we will show in the figures measurements of the power spectrum and bispectrum up to $0.2\kMpc$, together with the extrapolation of the theoretical model. Therefore, one should keep in mind that the predictions shown are not obtained from a fit to all the data points displayed in the figure. 

In addition, it is important to note that {\em the fits to the bispectra}, although limited to large scales,  {\em assume all measurable configurations} up to the aforementioned $k_{max}=0.07\kMpc$. The bispectra are measured for triangle sides which are multiples of $\Delta k\equiv 3 k_f=0.012\kMpc$, where $k_f\equiv 2\pi/L\simeq 0.004\kMpc$ is the fundamental frequency of the box. It follows that, for the halo bispectrum, we measure $597$ triangular configurations characterized by $k_3\le k_2\le k_1<0.2\kMpc$ and $k_3\ge k_1-k_2$, which ensure no double counting and enforce the triangle constraint. However, only $43$ out of the $597$ triangle configurations, for which $k_3\le k_2\le k_1<0.07\kMpc$, are eventually considered for the fit of the bias parameters. For the matter-matter-halo bispectrum, the variable $k_3$ is corresponds to the halo density wavemode. Therefore, its value is allowed to vary in the range $ k_1-k_2\le k_3\le k_1+k_2$ since, in this case, values of $k_3$ greater than $k_2$ do not correspond to double counting. The total number of configurations measured for the matter-matter-halo  is $1,549$, among which $93$ are used for the fits. While in figures \ref{fig:bmmhGsq} to \ref{fig:bhGeq} we will present results for specific bispectrum configurations only, we stress once more that we are {\em not} simply fitting the triangles shown at any given time but perform instead a global fit to all large scales triangles of any shape.

\subsection{Power spectra}

\begin{figure}[!p]
\begin{center}
\begin{center}{\bf Matter-halo power spectrum, $P_{\d h}(k)$}\end{center}\vspace{0.2cm}
{\includegraphics[width=0.48\textwidth]{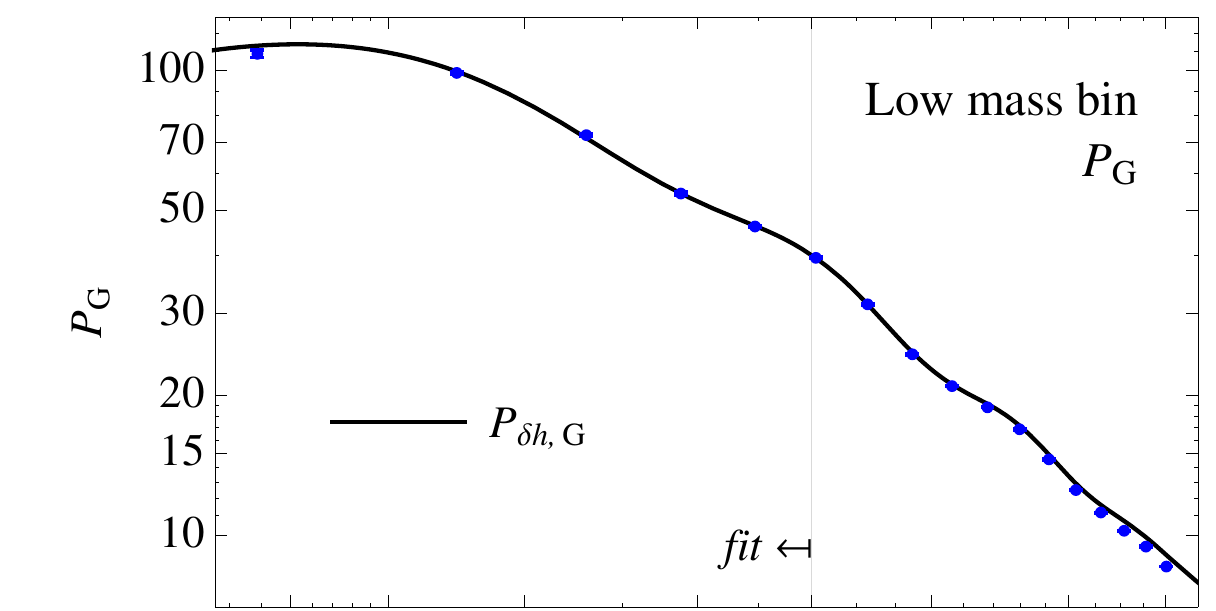}}
{\includegraphics[width=0.48\textwidth]{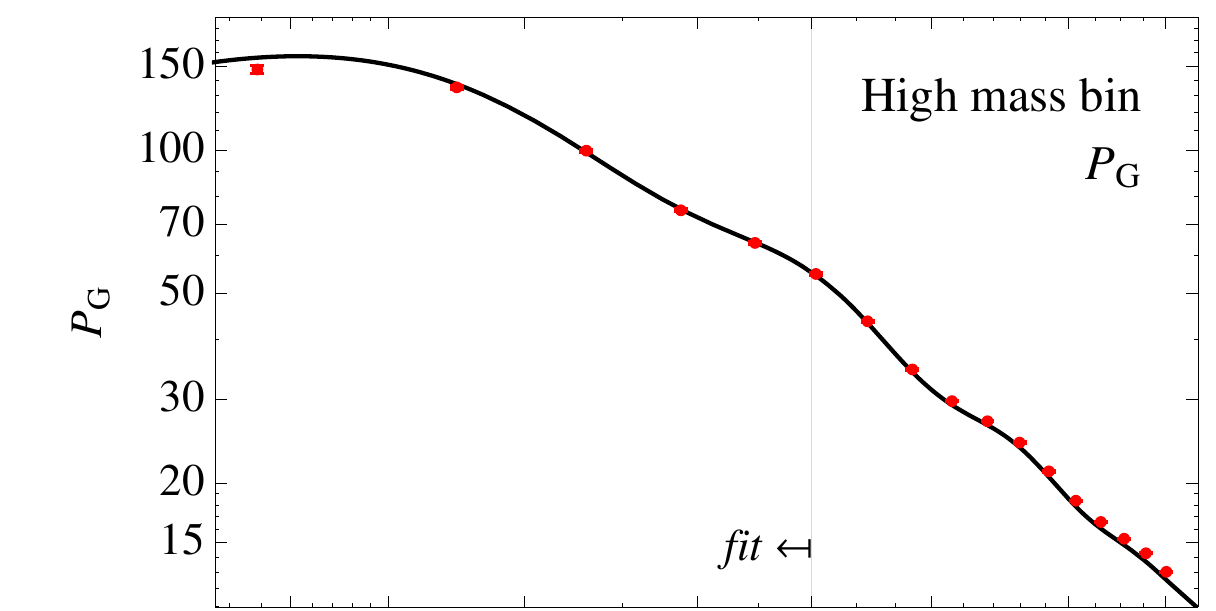}}
{\includegraphics[width=0.48\textwidth]{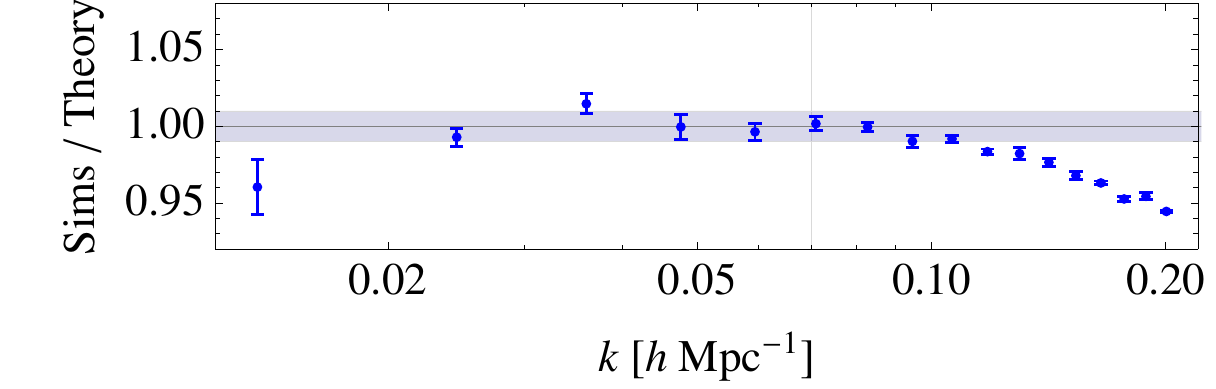}}
{\includegraphics[width=0.48\textwidth]{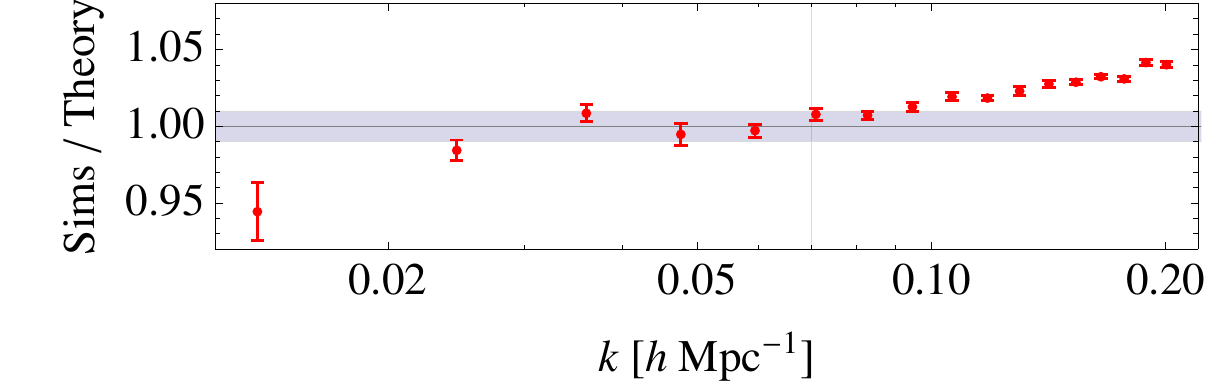}}
{\includegraphics[width=0.48\textwidth]{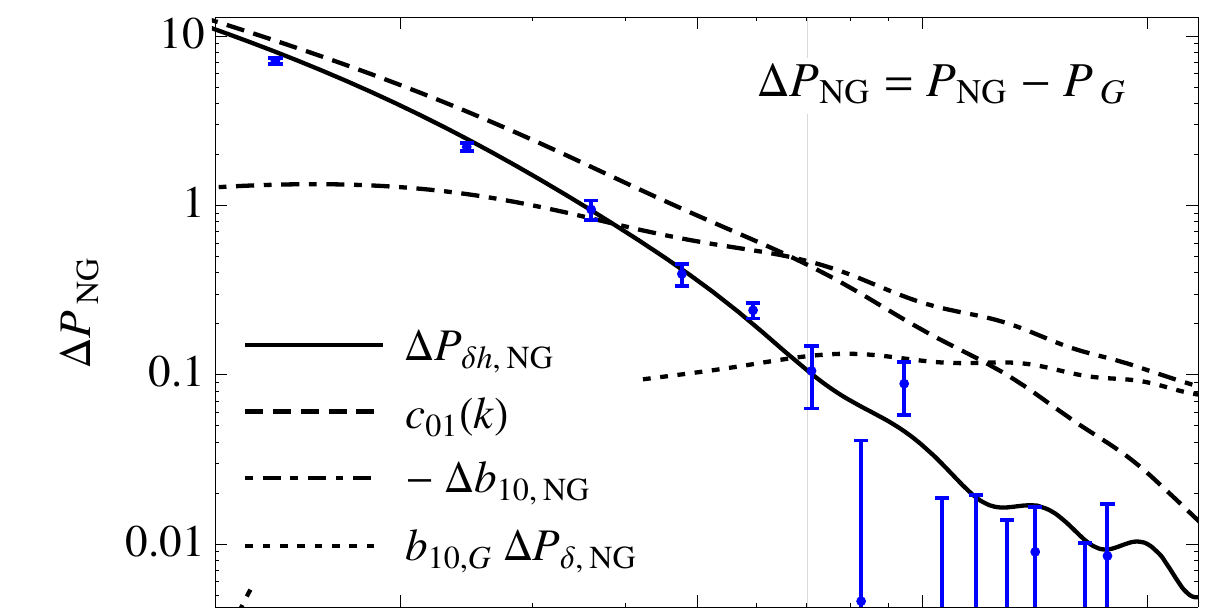}}
{\includegraphics[width=0.48\textwidth]{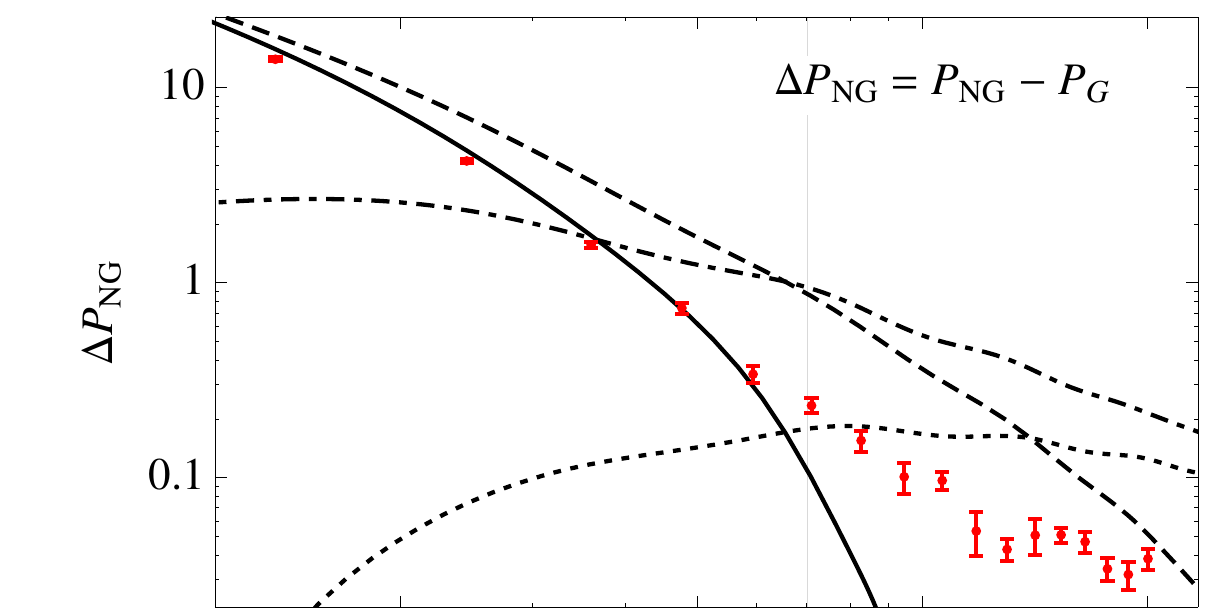}}
{\includegraphics[width=0.48\textwidth]{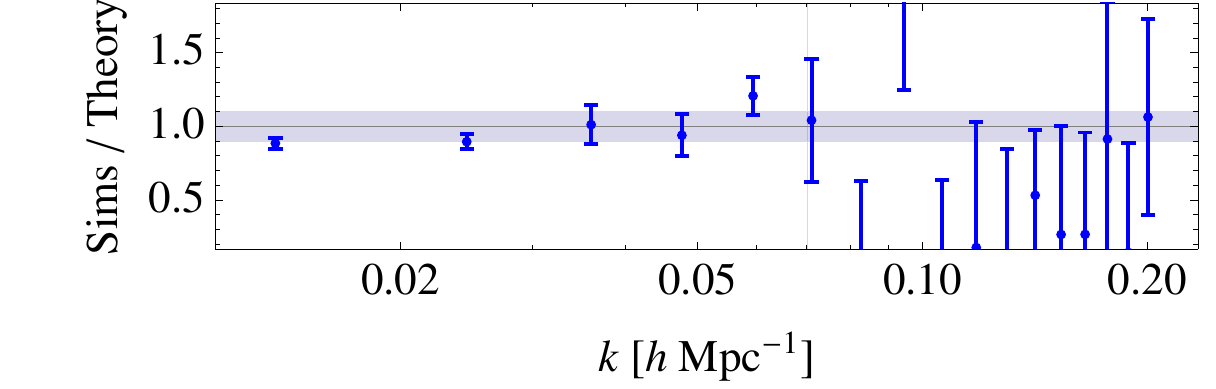}}
{\includegraphics[width=0.48\textwidth]{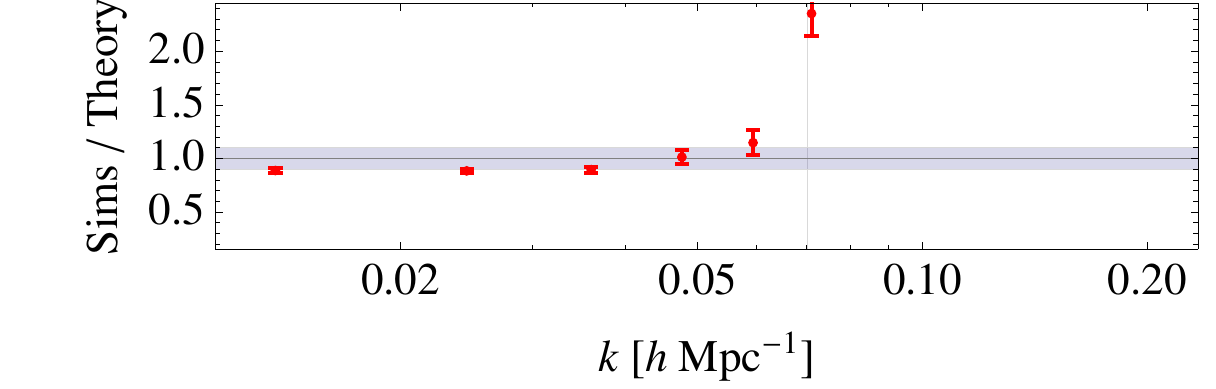}}
{\includegraphics[width=0.48\textwidth]{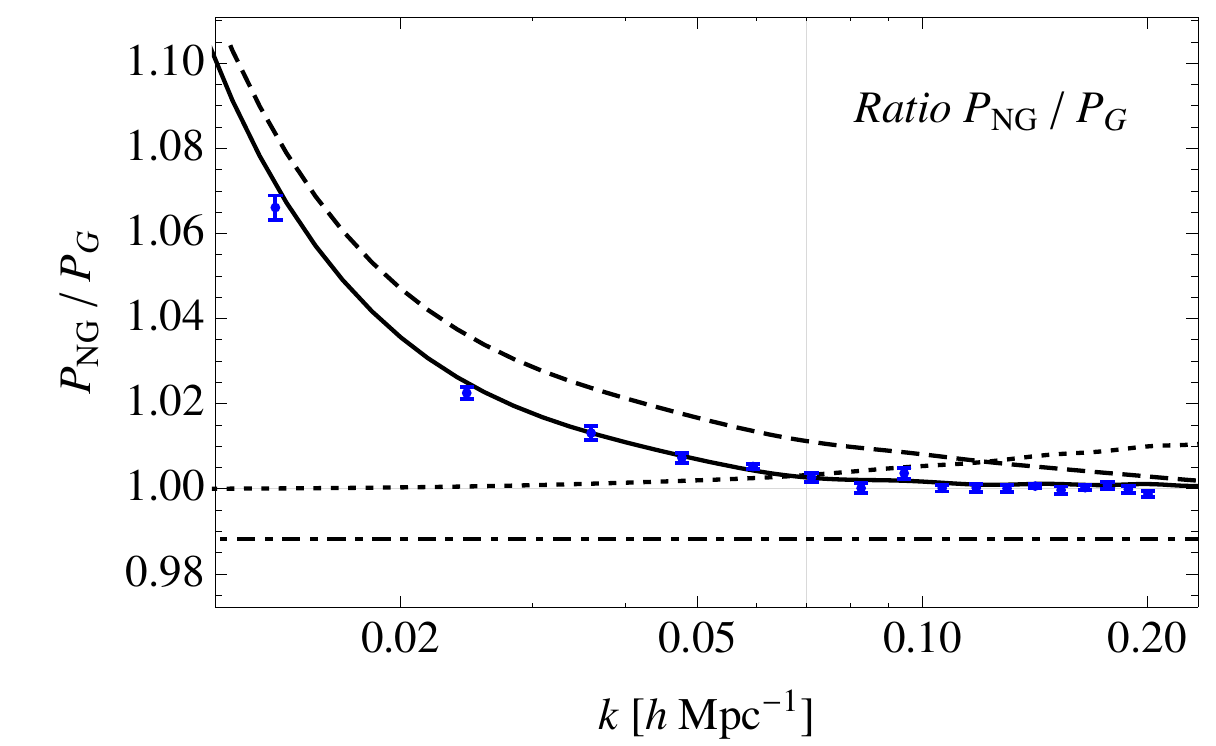}}
{\includegraphics[width=0.48\textwidth]{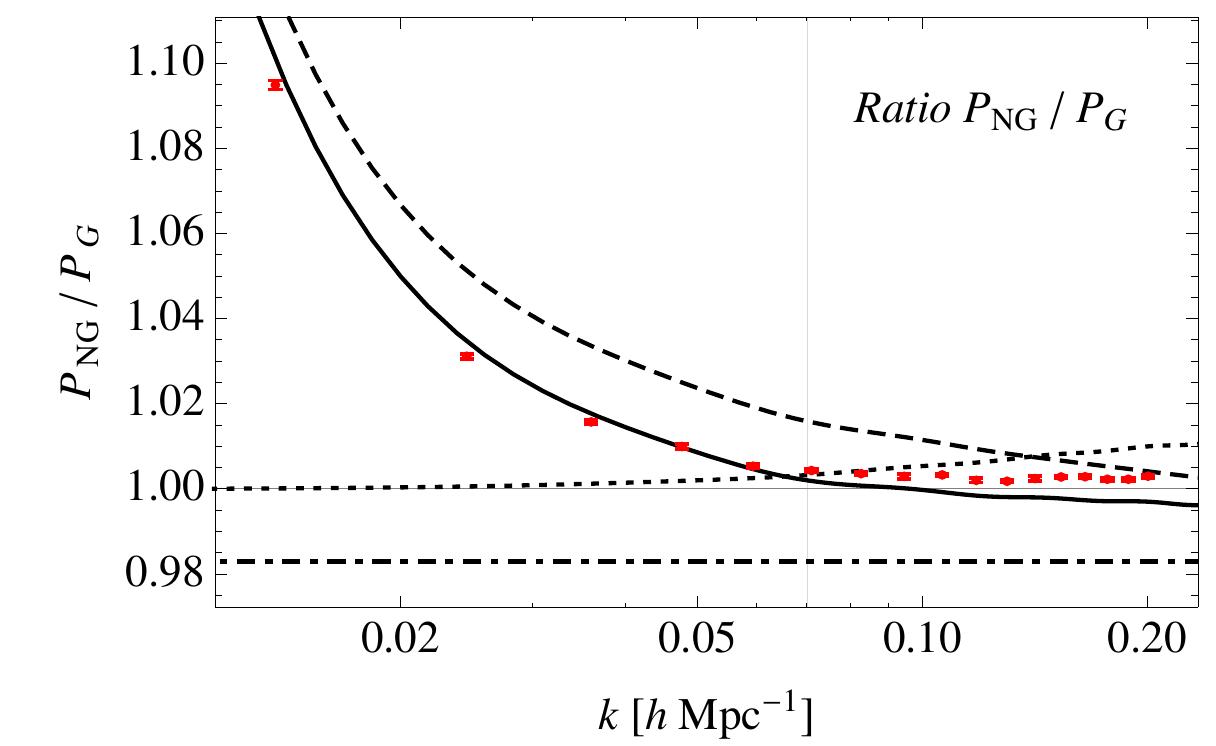}}
{\includegraphics[width=0.48\textwidth]{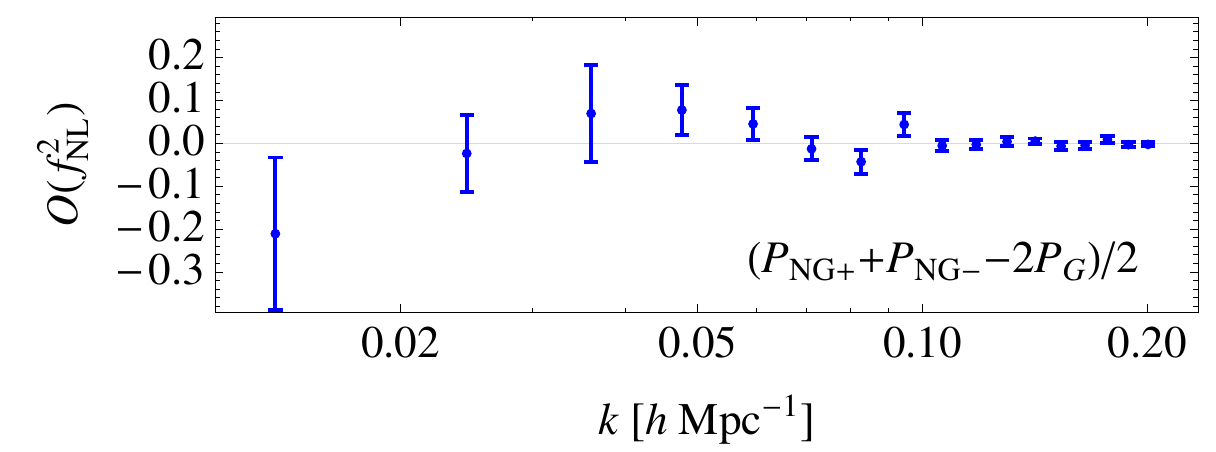}}
{\includegraphics[width=0.48\textwidth]{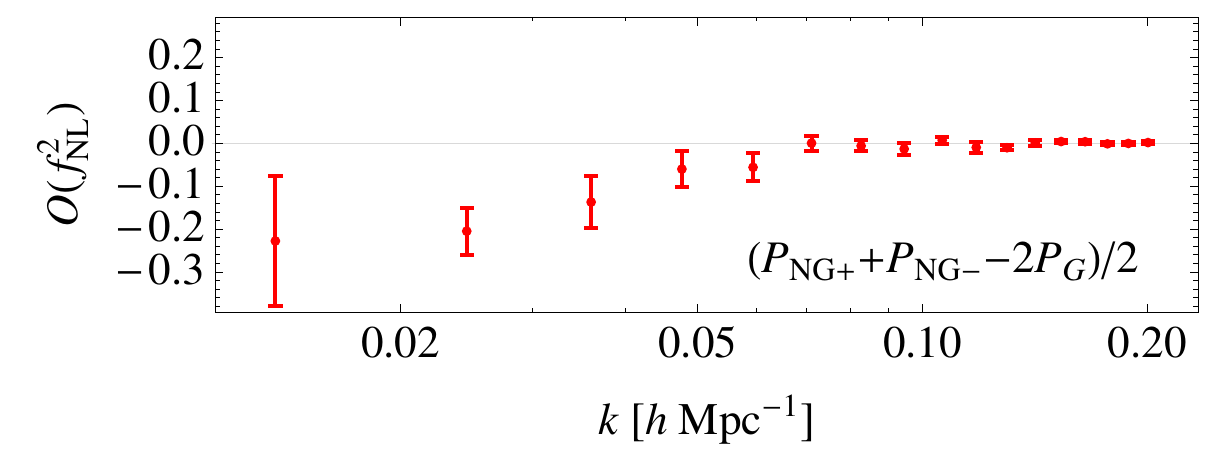}}
\caption{Cross matter-halo power spectrum, $P_{\d h}(k)$. See text for explanation.}
\label{fig:pmh}
\end{center}
\end{figure}

\begin{figure}[!p]
\begin{center}
\begin{center}{\bf Halo power spectrum, $P_{h}(k)$}\end{center}\vspace{0.2cm}
{\includegraphics[width=0.48\textwidth]{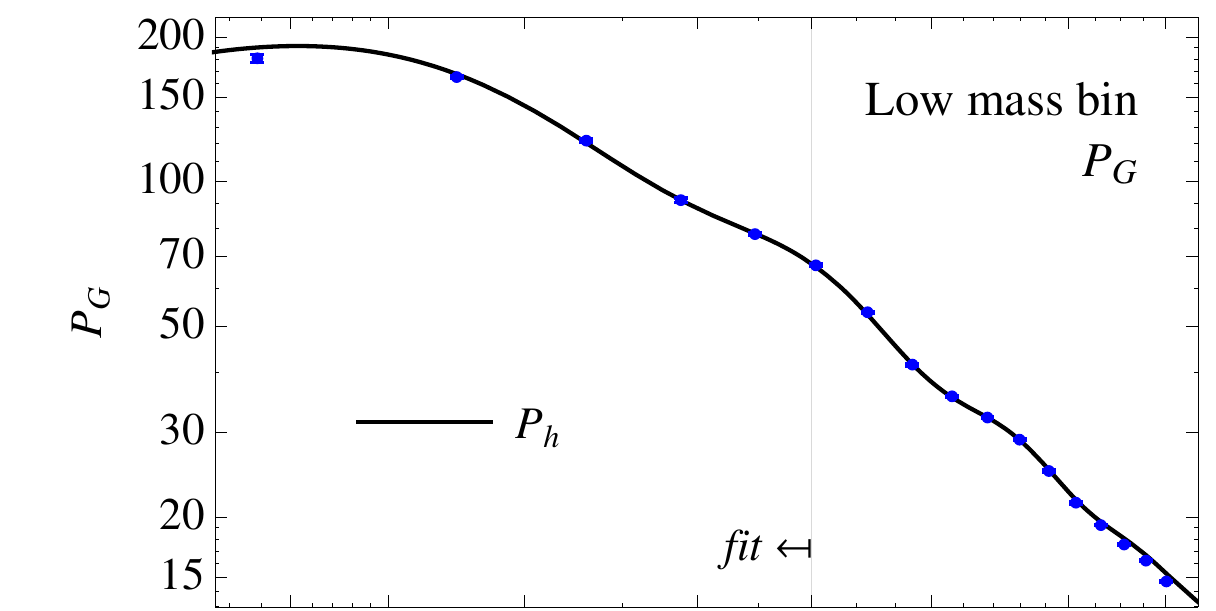}}
{\includegraphics[width=0.48\textwidth]{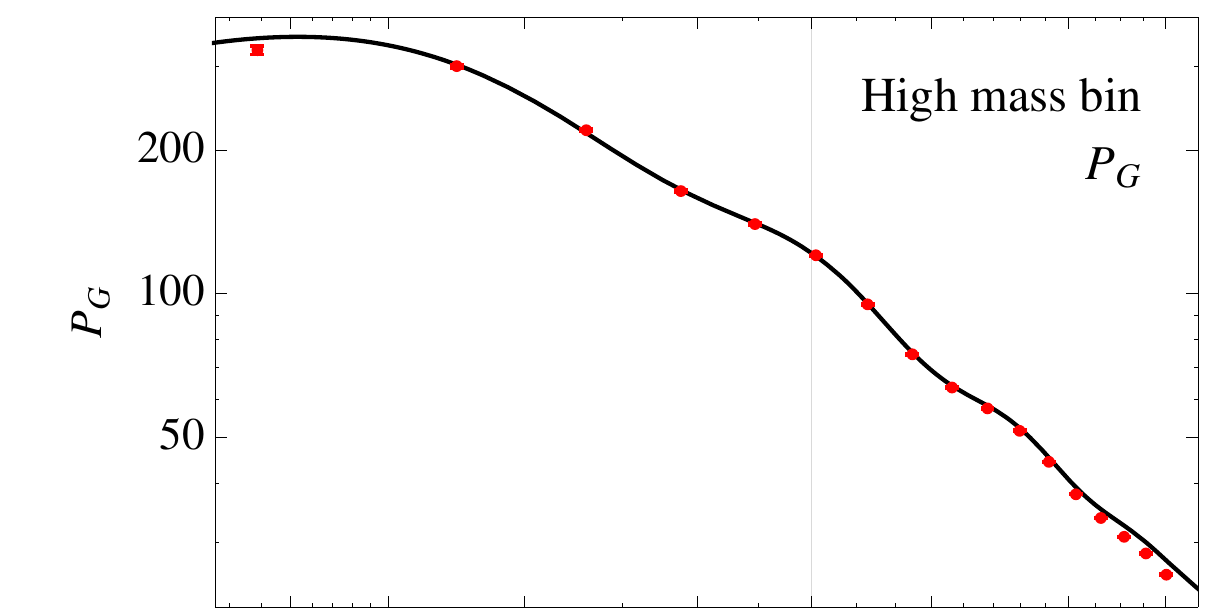}}
{\includegraphics[width=0.48\textwidth]{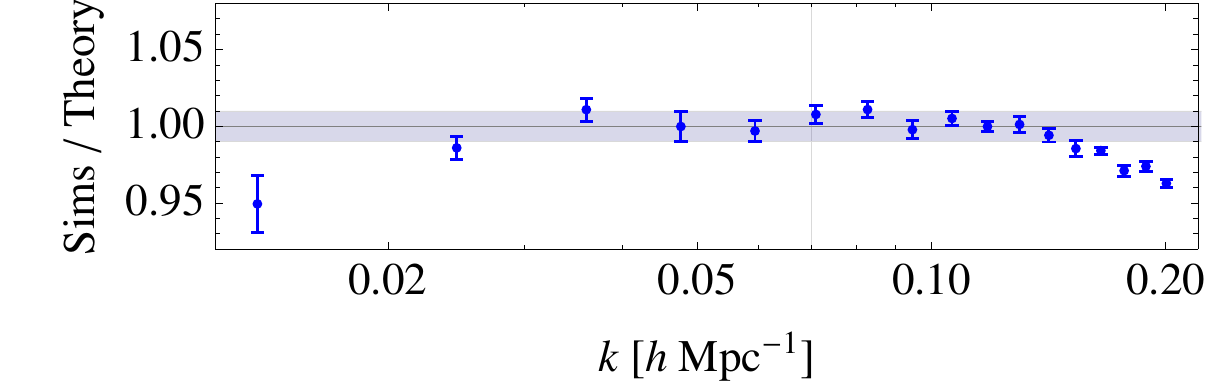}}
{\includegraphics[width=0.48\textwidth]{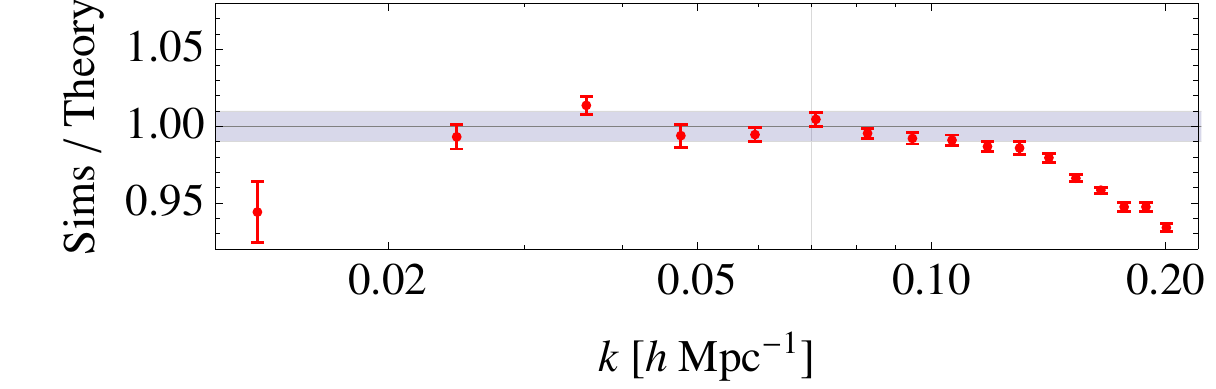}}
{\includegraphics[width=0.48\textwidth]{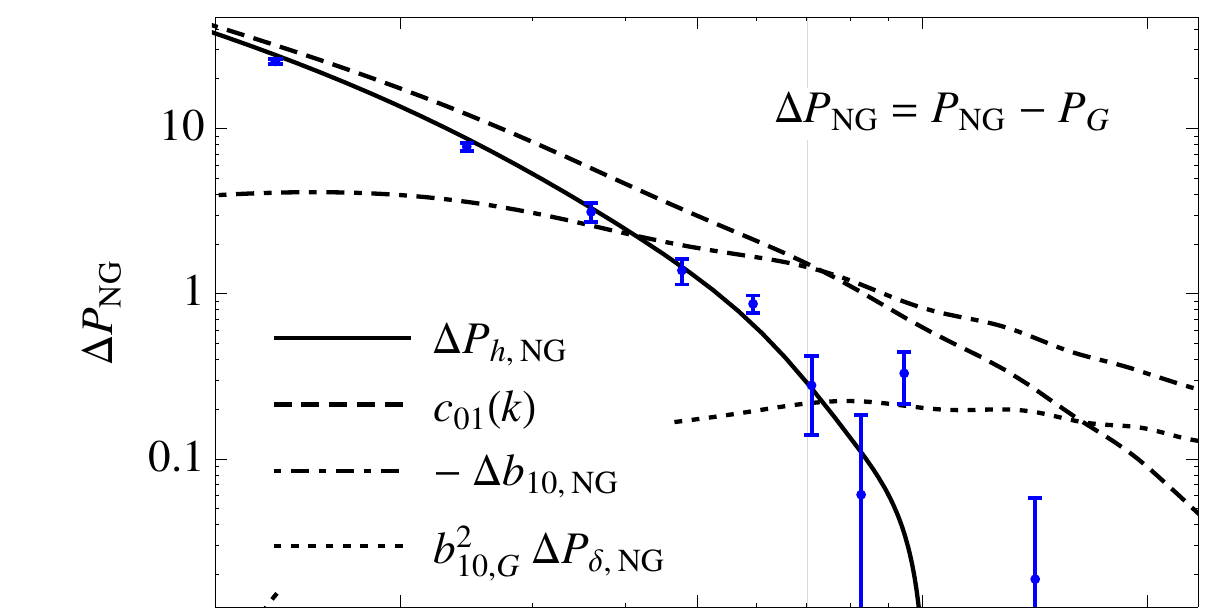}}
{\includegraphics[width=0.48\textwidth]{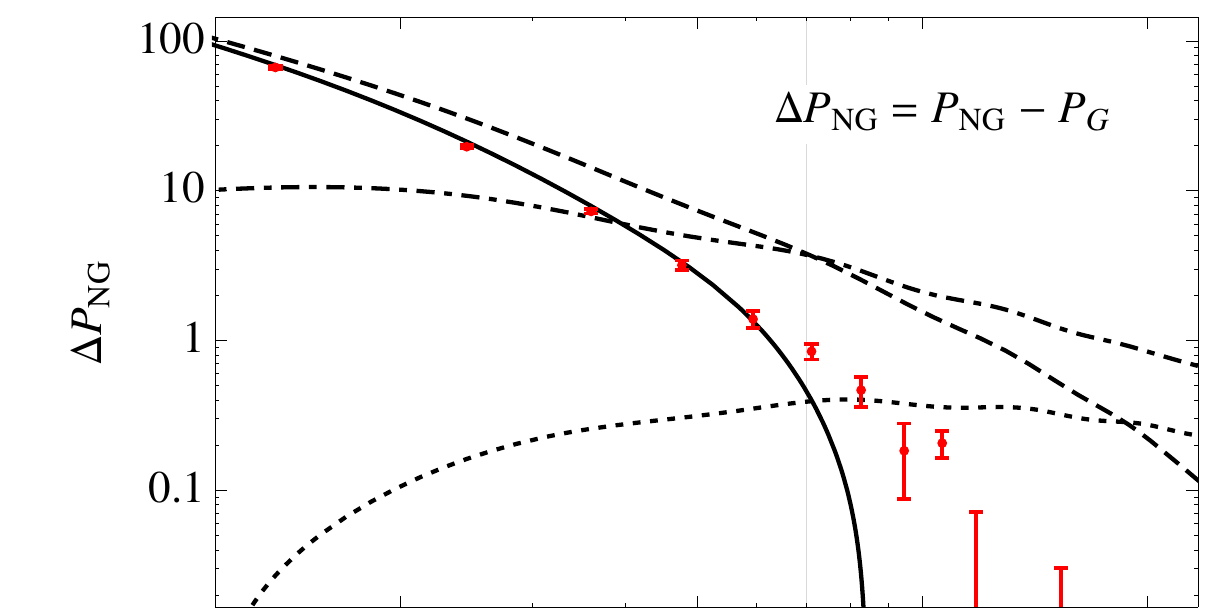}}
{\includegraphics[width=0.48\textwidth]{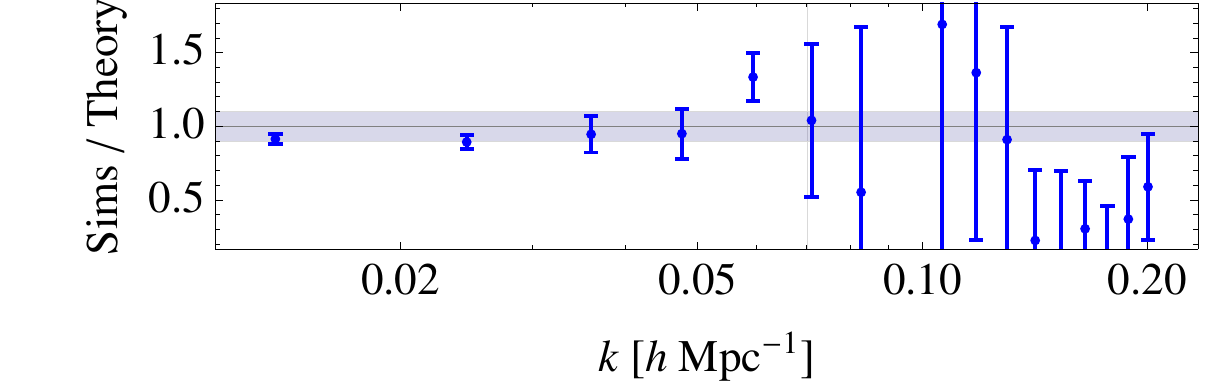}}
{\includegraphics[width=0.48\textwidth]{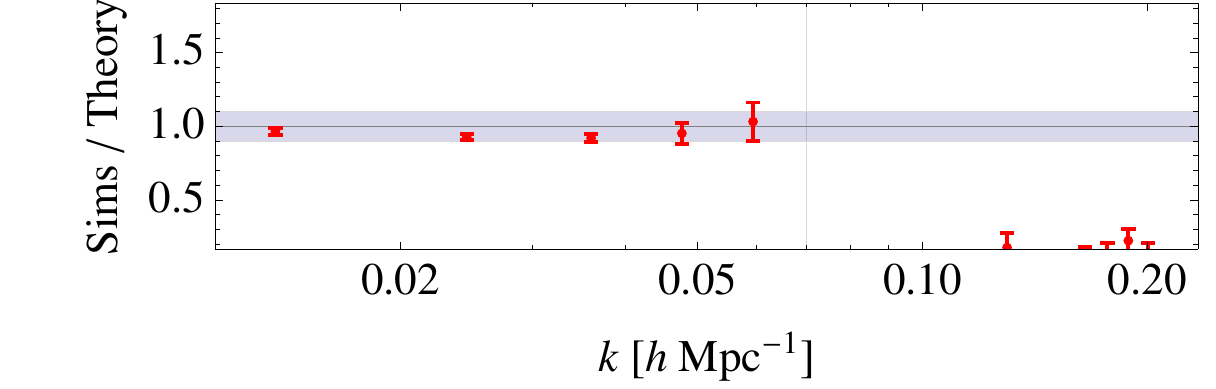}}
{\includegraphics[width=0.48\textwidth]{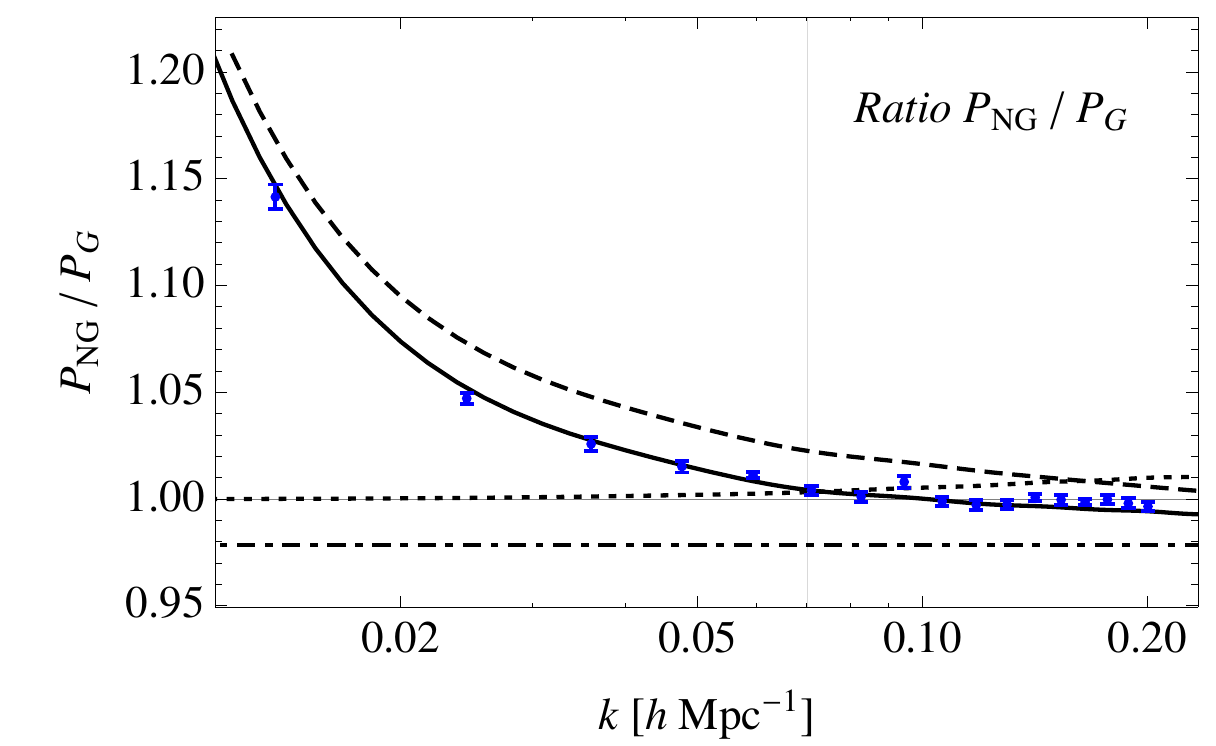}}
{\includegraphics[width=0.48\textwidth]{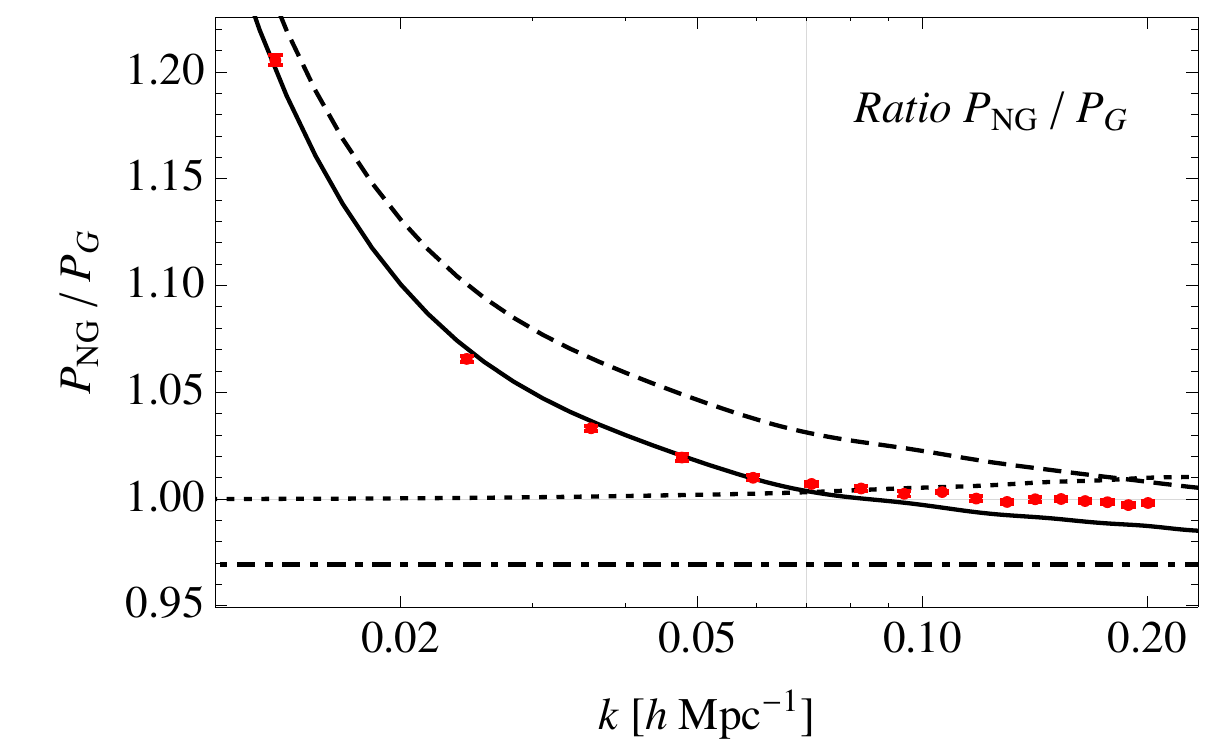}}
{\includegraphics[width=0.48\textwidth]{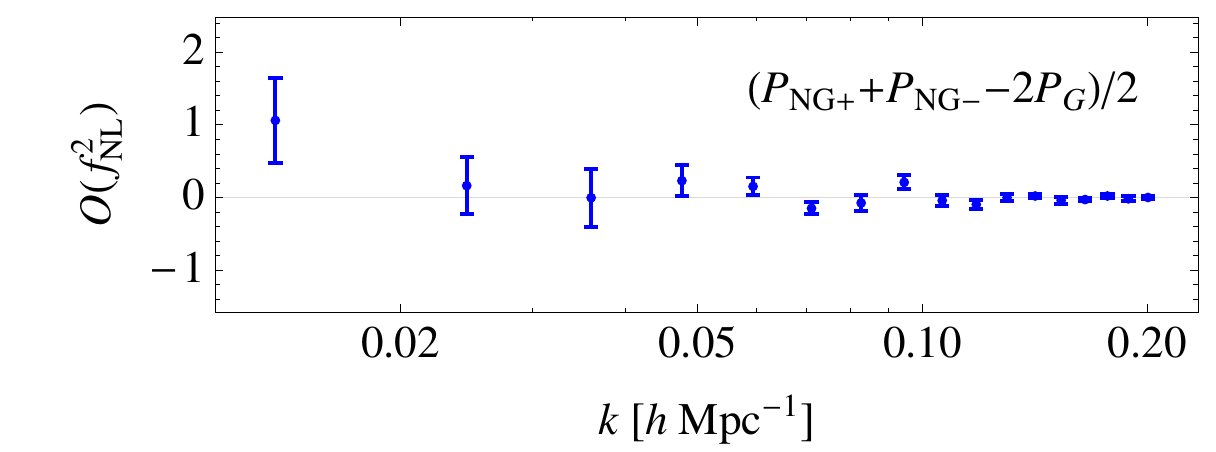}}
{\includegraphics[width=0.48\textwidth]{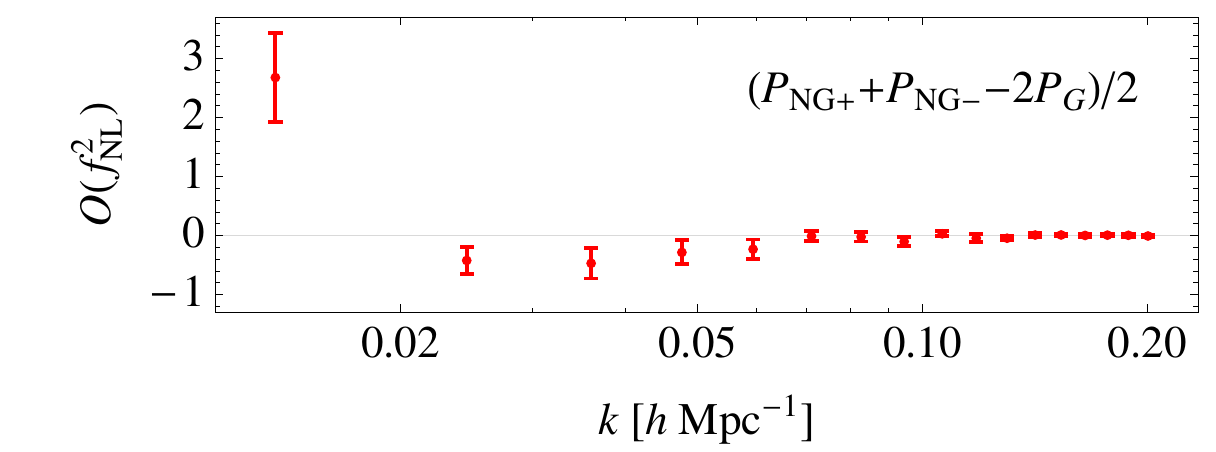}}
\caption{Halo power spectrum, $P_{h}(k)$. See text for explanation.}
\label{fig:ph}
\end{center}
\end{figure}

In Fig.~\ref{fig:pmh} and \ref{fig:ph}, we show results concerning the cross matter-halo power spectrum $P_{\d h}$ and the halo power spectrum $P_{h}$, respectively. Each figure displays the measurements of the Gaussian correlator $P_G$ and the residuals to the model fit ({\em first two rows}), the non-Gaussian correction $\D P_{NG}$ and the relative residuals ({\em third and fourth row}), the ratio non-Gaussian to Gaussian $P_{NG}/P_G$ and the measurement of the $\O(\fNL^2)$ component ({\em last row}). 
Left panels correspond to the low mass bin $8.8\times 10^{12}\Ms<M<1.6\times 10^{13}\Ms$ (data points in blue), while right panels correspond to the high mass bin $M>1.6\times 10^{13}\Ms$ (data points in red). The shaded gray area in the residual plots indicate a $1\%$ and $10\%$ error for the Gaussian power spectrum and the non-Gaussian correction, respectively. The thin vertical line indicate the maximum wavenumber, $k_{max}=0.07\kMpc$, used for the fits.

Let us consider first the matter-matter-halo power spectrum, Fig.~\ref{fig:pmh}. In the upper panels, the Gaussian measurements are compared to the simple theoretical prediction Eq.~(\ref{eq:PmhG}), where the linear bias parameter $b_{10,G}$ is fitted for and the nonlinear matter power spectrum is computed up to one-loop in PT. Since the fit is restricted to $k<0.07\kMpc$, such nonlinear corrections are not affecting the determination of the linear bias. At the same time, they do not improve much the agreement between the model and the simulations at smaller scales, where further corrections due to nonlinear bias should be significant. At large $k$, the model overestimates the data points for the low mass halos, and underestimates them for the large mass halos. As we will see shortly, this is consistent the Gaussian quadratic bias parameter being negative and positive, respectively.

In the second and third rows, the correction induced by PNG, $\D P_{\d h, NG}$, is compared to the theoretical prediction Eq.~(\ref{eq:dPmhNG}) ({\em continuous curves}). The various curves represent the contribution generated by the scale-dependent bias correction $c_{01}(k)$ ({\em dashed curve}), by the scale-independent bias correction $\D b_{10,NG}$ ({\em dot-dashed curves: the absolute value is shown as this term is negative}) and by the non-Gaussian corrections to the matter power spectrum $\D P_{\d,NG}$ ({\em dotted curves}). As already remarked in \citep{DesjacquesSeljakIliev2009} and \citep{GiannantonioPorciani2010}, adding the scale-independent correction $\D b_{10,NG}$ to the scale-dependent bias $c_{01}(k)$ significantly improves the agreement with the simulations. At smaller scales, further improvements can be achieved on including the additional non-Gaussian corrections derived in the peak-background split approach of \citep{DesjacquesJeongSchmidt2011B, ScoccimarroEtal2011}. Nevertheless, we shall ignore them here as we focus on the simplest model for the large-scale halo bispectrum. The fourth row shows the non-Gaussian to Gaussian ratio $P_{\d h,NG}/P_{\d h,G}$ with the same labeling for the various model components. Note that the contribution from the $\D b_{10,NG}$ term is negative. 

The lower panels show the combination $[P_{\d h,NG}(\fNL=+100)+P_{\d h,NG}(\fNL=-100)-2\,P_{\d h,G}(\fNL=0)]/2$, corresponding to $\O(\fNL^2)$ contributions the matter-halo power spectrum. These are consistent with zero for low mass halos whereas, for the high mass halos, there is some evidence for a signal at large scales. These second-order corrections $\fNL$ are negative, of the order of $5\%$ of the whole non-Gaussian correction $\D P_{\d h,NG}$. They might be due to our approximation for $c_{01}(k)\sim (b_{10}-1)\simeq (b_{10,G}-1)$. The inclusion of the scale-independent  correction on the linear bias parameter, $\D b_{10,NG}$ in the expression for $c_{01}(k)$, Eq.~(\ref{eq:b01Eb10}), leads to a second order effect in $\fNL$ which we neglect here and do not explore further.  

Fig.~\ref{fig:ph} shows the analogous quantities for the halo power spectrum. Considerations similar to those above apply to these figures. However, an interesting difference is the effect of nonlinear bias corrections to the Gaussian halo bias at small scales, that presents the same sign for both low and high halo masses. In addition, there is evidence for a $\fNL^2$ correction in the lowest $k$ bin for the high mass and, to a lesser extent, the low mass halos. Still, on intermediate scales $0.02\kMpc<k<0.05$, the effect appears to have a negative sign, like in the matter-halo power spectrum. Overall, it is at most $1\%$ of the Gaussian halo power spectrum $P_{h,G}$, and at most $10\%$ of the total non-Gaussian correction $\Delta P_{h,NG}$. Again, its inclusion might improve marginally the agreement of model with the measured $\Delta P_{h,NG}$. 
 
Our power spectrum measurements agree broadly with those of previous studies \citep{DalalEtal2008, DesjacquesSeljakIliev2009, GrossiEtal2009, PillepichPorcianiHahn2010, WagnerVerde2011, DesjacquesJeongSchmidt2011A, ScoccimarroEtal2011}. Our choice to {\em fit} for the non-Gaussian correction to the linear bias $\D b_{10,NG}$ can account, among others, for the additional large-scale corrections predicted by the approach of \citep{DesjacquesJeongSchmidt2011A, ScoccimarroEtal2011} on top of the standard contribution, $c_{01}(k)$. Notice that, unlike \citep{WagnerVerde2011} for instance, we do not add an extra fudge factor to the amplitude of the scale-dependent correction. Fitting the non-Gaussian corrections $\D P_{\d h,NG}$ and $\D P_{h,NG}$ for both $\D b_{10,NG}$ and this additional parameter would lead to results consistent with those of \citep{WagnerVerde2011}, yet significantly complicate the bispectrum analysis. We postpone a more detailed analysis to future work.

Overall the results of this section show that the linear bias models for $P_{\d h,NG}$ and $P_{h,NG}$ agree at the percent level with the simulations results up to $k \sim 0.07 h/Mpc$. The non-Gaussian correction, in fact, represent, over the range of scales probed, a 10\% correction at most, described by the simple model for $\D P_{\d h,NG}$ and $\D P_{h,NG}$ with an accuracy of a few percent.

\subsection{Bispectra}

We now test the models for the cross matter-matter-halo and halo bispectra. As explained above, for the analysis of the Gaussian $B_{\d\d h,G}$ we assume the best-fit value for the Gaussian linear bias $b_{10,G}$ obtained, in turn, from the analysis of the matter-halo power spectrum. We therefore determine from $B_{\d\d h,G}$ only the quadratic bias parameter $b_{20,G}$, which we then use, together with $b_{10,G}$ and $\D b_{10,NG}$ from $P_{\d h}$, in the fit of the non-Gaussian correction $\D B_{\d\d h,NG}$. Again, the latter is a one-parameter fit in terms of $\D b_{20,NG}$. The same applies, independently, to the halo bispectrum components $B_{h,G}$ and $\D B_{h,NG}$.

The fits to bispectrum measurements are performed for all the triangular configurations with sides smaller than or equal to a given $k_{max}$. Therefore, in order to provide a complete picture of the goodness of the fit, we show in the first place the $\chi^2$ value {\em as a function} of $k_{max}$. This will allow us, in particular, to justify our choice for $k_{max}=0.07\kMpc$ as the wavenumber below which we expect our theoretical description to be accurate. 

The $\chi^2$ as a function of $k_{max}$ is defined as
\beq
\chi^2=\sum_{k_1\ge k_2\ge k_3}^{k_{max}}\frac{[B_{data}(k_1,k_2;k_3)-B_{model}(k_1,k_2;k_3)]^2}{\Delta B_{data}^2}\,,
\eeq
where the variance is defined as the square of the error on the mean measured from the twelve available realizations, and where the sum runs over all triangles defined by wavenumbers below or equal to $k_{max}$. We consider values of $k_{max}$ ranging from $\Delta k=3k_f=0.012\kMpc$ up to $0.2\kMpc$. The  best-fit bias parameters are determined from configurations with sides less than $k_{max}$ solely. For the sake of consistency, the values for the linear bias parameters $b_{10,G}$ and $\D b_{10,NG}$ are determined from power spectra measurements at wavenumbers up to $k_{max}$, and similarly for the value of $b_{20,G}$ in the analysis of $\D B_{\d\d h,NG}$ and $\D B_{h,NG}$. Notice that the number of degrees of freedom grows considerably as a function of $k_{max}$, as does the number of triangles included. 

\begin{figure}[!t]
\begin{center}
\begin{center}{\bf Matter-matter-halo bispectrum}, $B_{\d\d h}$\end{center}\vspace{0.2cm}
{\includegraphics[width=0.48\textwidth]{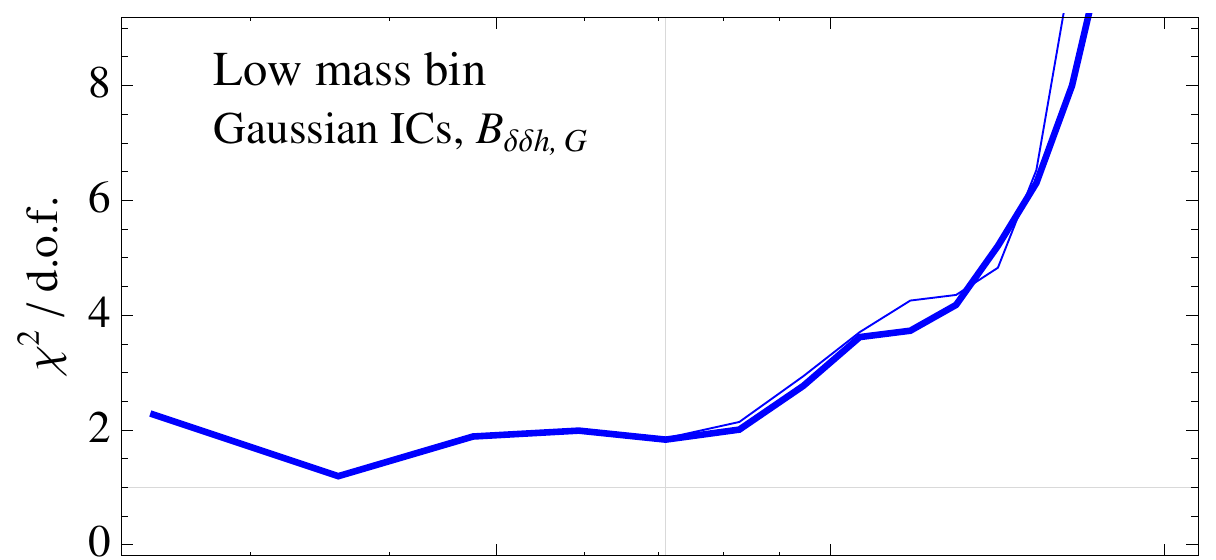}}
{\includegraphics[width=0.48\textwidth]{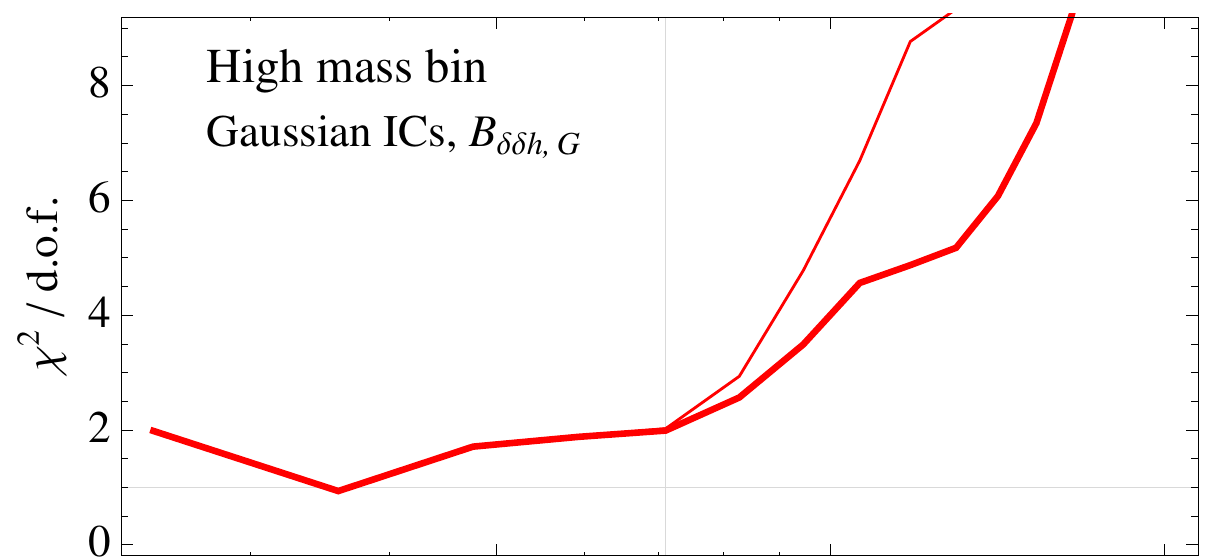}}
{\includegraphics[width=0.48\textwidth]{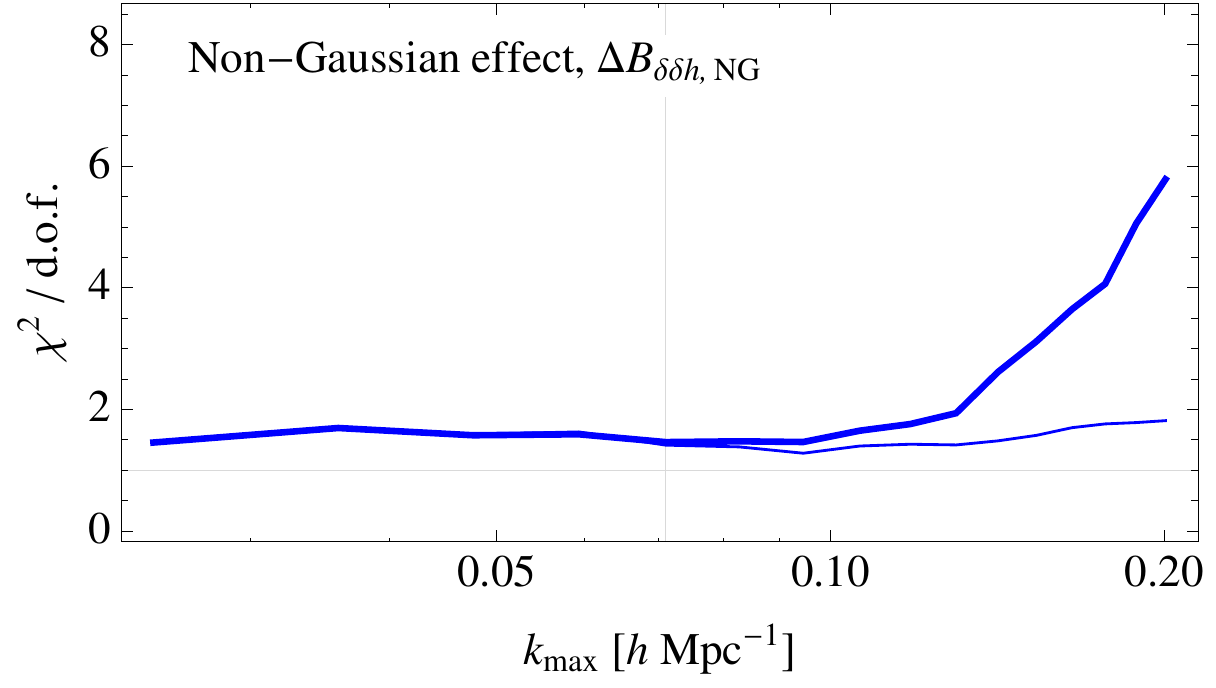}}
{\includegraphics[width=0.48\textwidth]{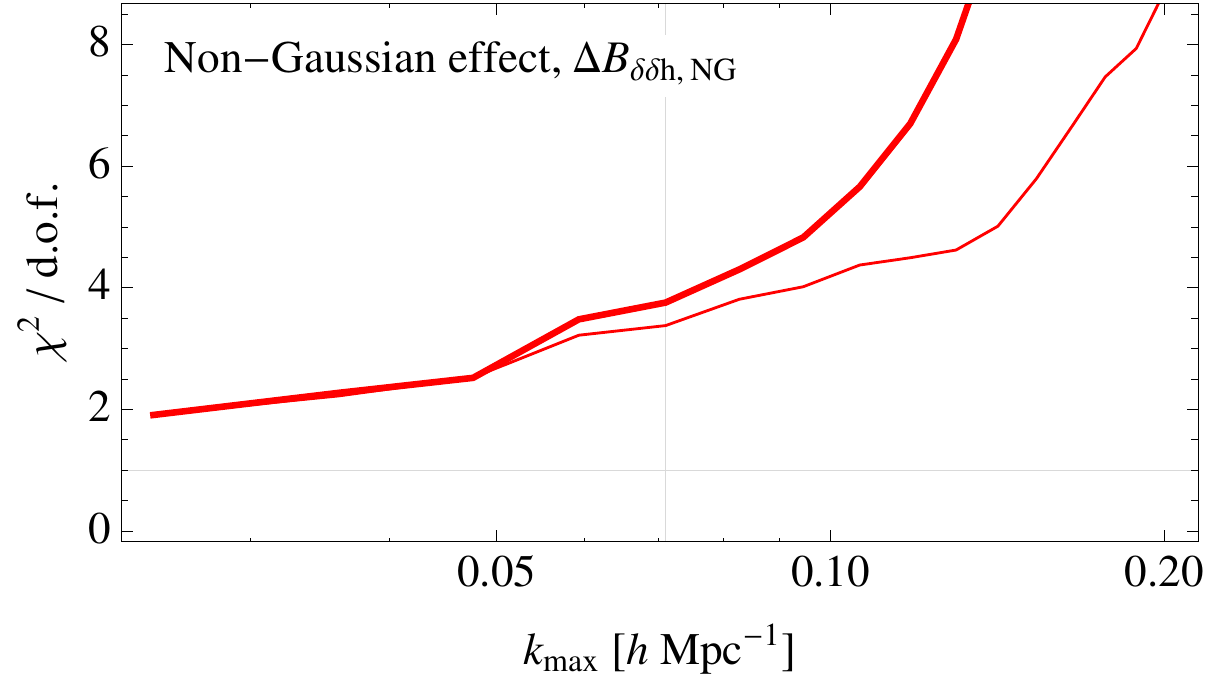}}
\begin{center}{\bf Halo bispectrum}, $B_{h}$\end{center}\vspace{0.2cm}
{\includegraphics[width=0.48\textwidth]{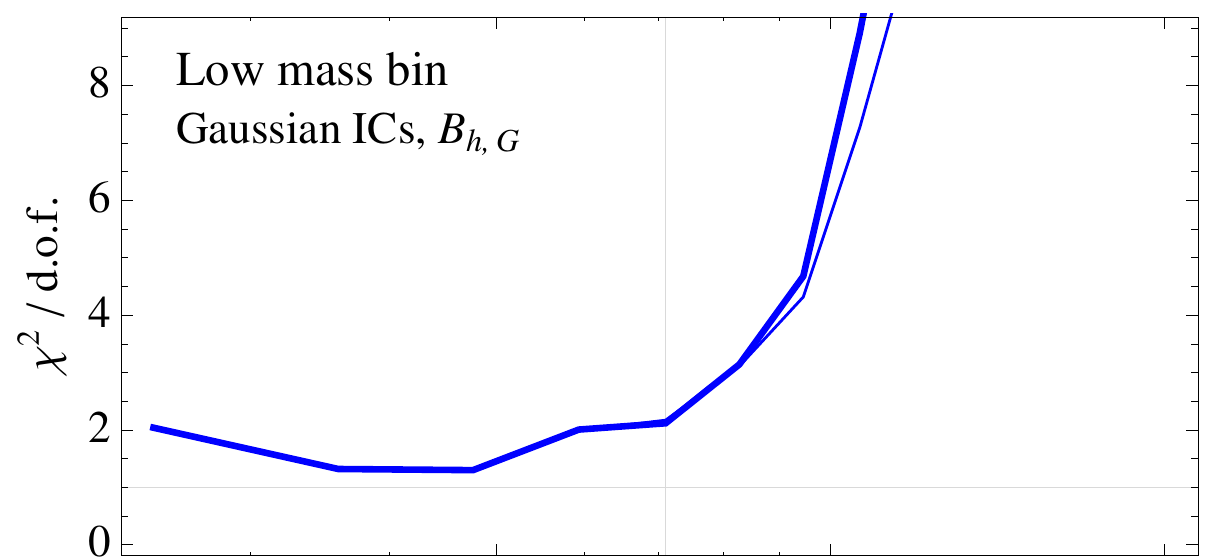}}
{\includegraphics[width=0.48\textwidth]{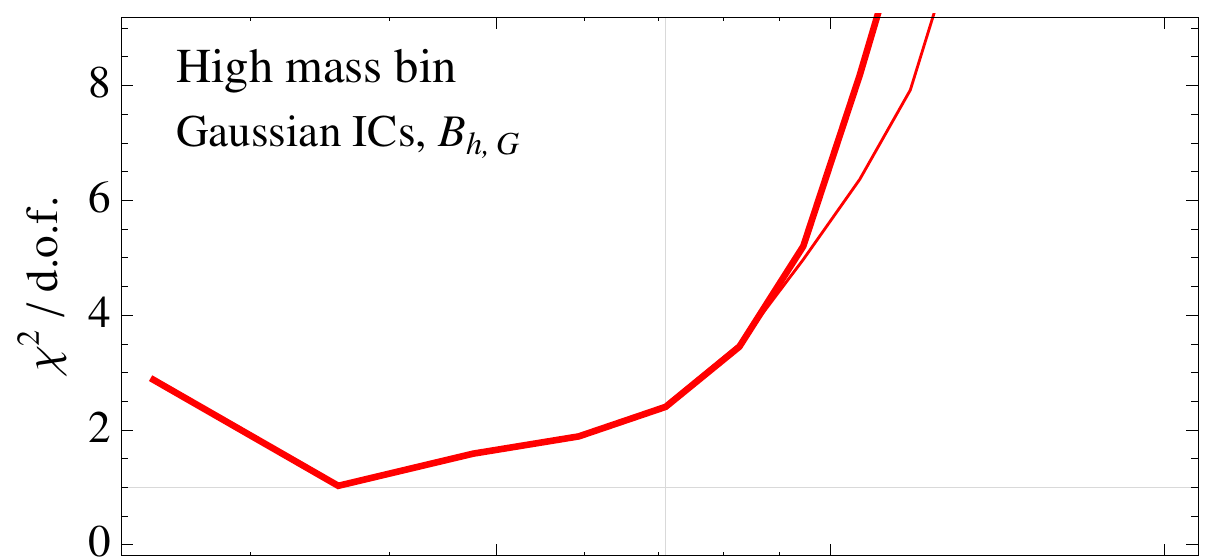}}
{\includegraphics[width=0.48\textwidth]{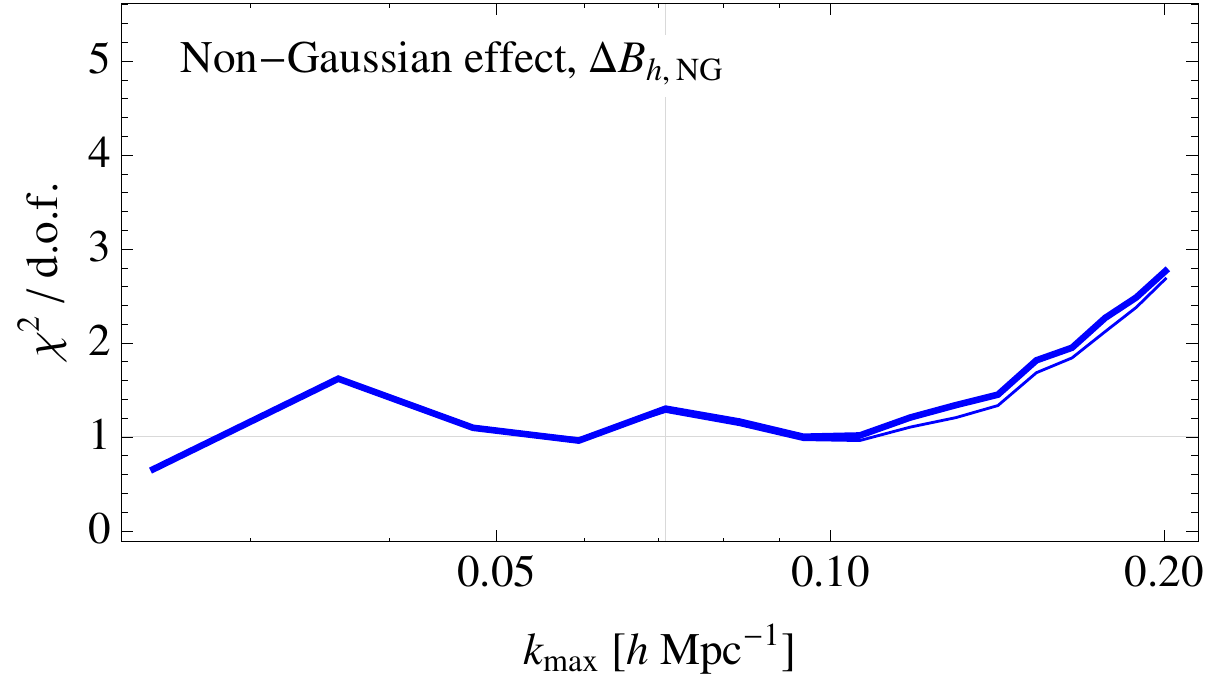}}
{\includegraphics[width=0.48\textwidth]{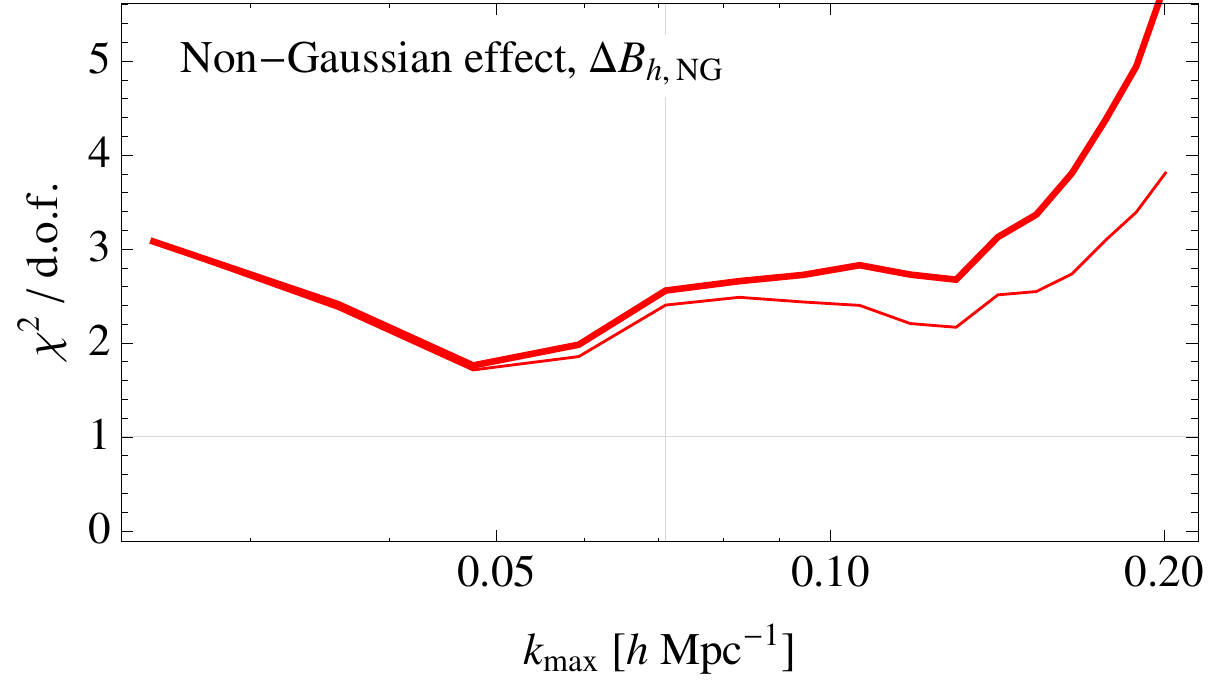}}
\caption{Reduced $\chi^2$ for the matter-matter-halo bispectrum fits. Thin lines correspond to the same quantity obtained assuming a tree-level approximation for matter correlators.}
\label{fig:Chi2}
\end{center}
\end{figure}

In Fig.~\ref{fig:Chi2} we show the $\chi^2$ {\em per degree of freedom} as a function of $k_{max}$ for both $B_{\d\d h}$ and $B_h$ ({\em thick continuous curves}). More precisely, the first two rows show the $\chi^2$ corresponding to the Gaussian component of the cross-bispectrum $B_{\d\d h,G}$ and to the non-Gaussian correction $\Delta B_{\d\d h,NG}$. The same quantities for the halo bispectrum components $B_{h,G}$ and $\Delta B_{h,NG}$ are shown in the third and fourth row. Panels on the left correspond to the low mass halos, while panels on the right to the high mass halos. The thin curves represent the same quantities obtained assuming the tree-level predictions for the matter bispectra and linear matter power spectra. 

Firstly, it should be noted that the $\chi^2$ per degree of freedom for the Gaussian components $B_{\d\d h,G}$ and $B_{h,G}$ is roughly constant, between values of 1 to 2 up to scales $\lesssim 0.1\kMpc$. This indicates that, on large scales, the tree-level approximation from the bias expansion captures well the scale and shape dependence of the halo bispectrum measurements for both low and high mass halos, thereby justifying our choice of fitting the bias parameters using all triangles formed by wavenumbers equal or smaller than $0.07\kMpc$. If smaller scales or larger wavenumbers are included, then this simple approximation breaks down, but it may be improved by adding further nonlinear corrections in the bias expansion or non-local terms. However, this problem appears to be more severe for the Gaussian components than for the non-Gaussian corrections $\D B_{\d\d h,NG}$ and $\D B_{h,NG}$. 

Secondly, a comparison of the thick and thin lines suggests that the inclusion of one-loop corrections to the matter correlators does not improves the fits significantly at large scales ($k<0.07\kMpc$). The effect at smaller scales cannot be properly assessed as nonlinearities in the bias expansion, which clearly play a major role, are not accounted for. 

In the following figures, we compare directly the measured bispectra with the model for a choice of specific triangular configurations. The chosen subsets of triangles are defined as follows
\begin{itemize}
\item {\bf squeezed configurations}: $k_3$ is kept constant at $k_3=\Delta k=0.012\kMpc$ (the smallest measured wavenumber) while $k_1=k_2=k$ vary from $\Delta k$ to $0.02\kMpc$; the first data point corresponds therefore to a large-scale triangular configuration while as $k$ grows we obtain increasingly squeezed configurations;
\item {\bf generic configurations (I)}: $k_1$ and $k_2$ are kept constant at $k_1=0.071\kMpc$ and $k_2=0.082\kMpc$ while the angle $\theta$ between $\kv_1$ and $\kv_2$ varies from $0$ to $\pi$; this corresponds to $k_3$ taking values from $k_1+k_2=0.15\kMpc$ to $k_1-k_2=\Delta k=0.012\kMpc$ ({\em left to right in the plot, as $\theta$ increases}); in this case, the first data point corresponds to a folded triangle ($k_3= k_1+k_2$) at relatively small scales while $\theta\rightarrow \pi$ corresponds to a squeezed limit\footnote{{\em Nota bene}: the angle $\theta$ is defined in terms of the vectors $\kv_1$, $\kv_2$ and $\kv_3$ satisfying the property $\kv_1+\kv_2+\kv_3=0$, {\em not} as an internal angle of the triangle with sides $k_1$, $k_2$ and $k_3$.};
\item {\bf generic configurations (II)}: $k_1$ and $k_2$ are kept constant at $k_1=0.047\kMpc$ and $k_2=0.071\kMpc$ while the angle $\theta$ between $\kv_1$ and $\kv_2$ varies from $0$ to $\pi$; this corresponds to $k_3$ taking values from $k_1+k_2=0.12\kMpc$ to $k_1-k_2=\Delta k=0.024\kMpc$; these triangles are almost all scalene and approximately equally distant from equilateral or squeezed configurations;
\item {\bf equilateral configurations}: $k_1=k_2=k_3=k$ with $k$ varying from $\Delta k=0.012\kMpc$ to $0.02\kMpc$.  
\end{itemize}
These configurations are chosen to given a fair assessment of the effects of non-Gaussianity on the halo bispectrum which, as we will see, is {\em not} limited to squeezed configurations.

\begin{figure}[!p]
\begin{center}
\begin{center} {\bf Squeezed configurations}, $B_{\d\d h}(\D k,k,k)$\end{center}\vspace{0.2cm}
{\includegraphics[width=0.48\textwidth]{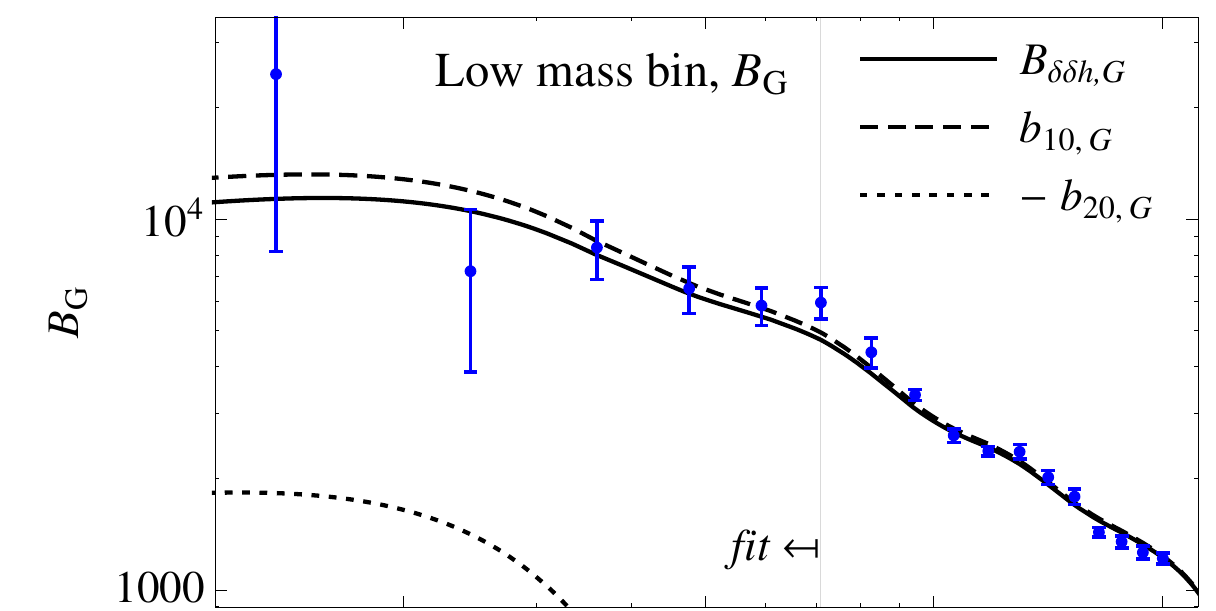}}
{\includegraphics[width=0.48\textwidth]{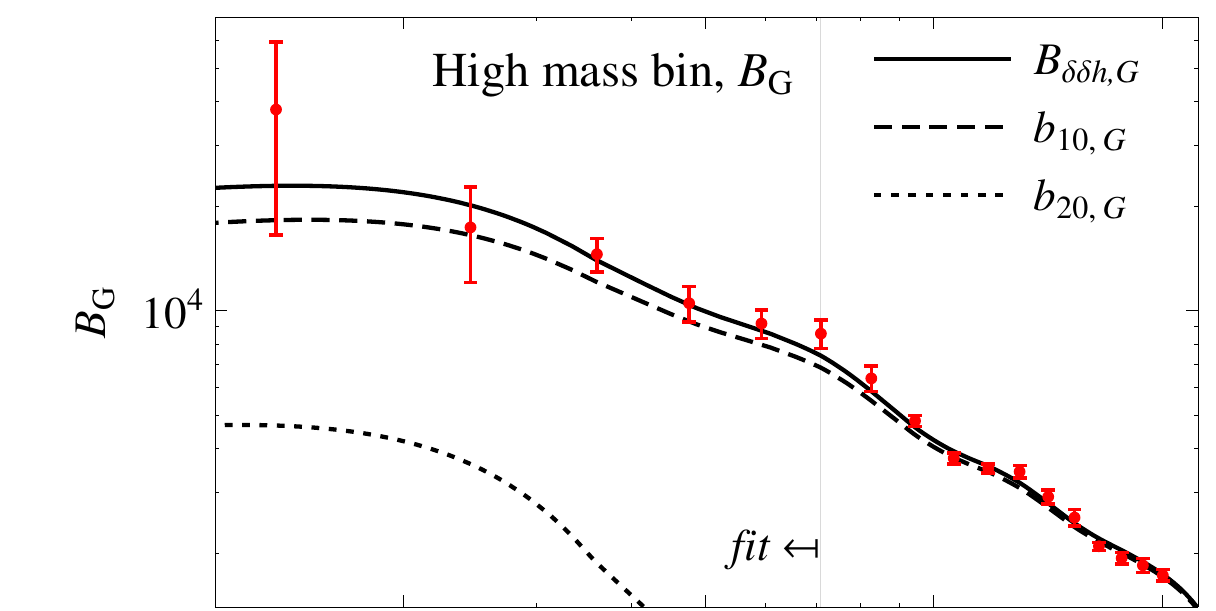}}
{\includegraphics[width=0.48\textwidth]{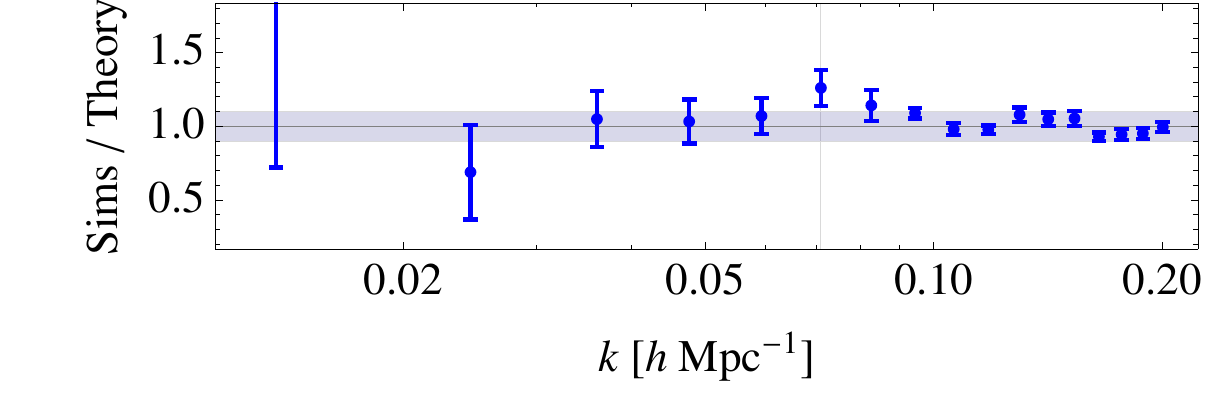}}
{\includegraphics[width=0.48\textwidth]{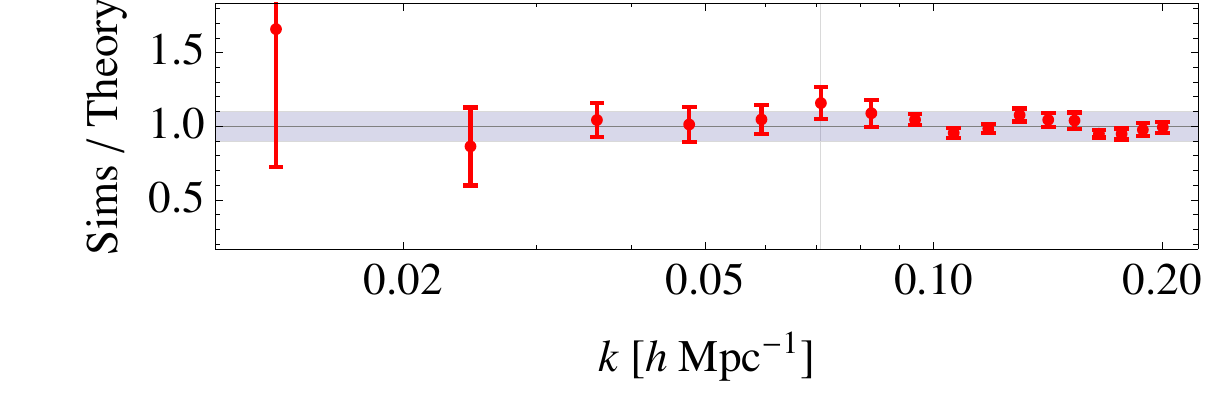}}
{\includegraphics[width=0.48\textwidth]{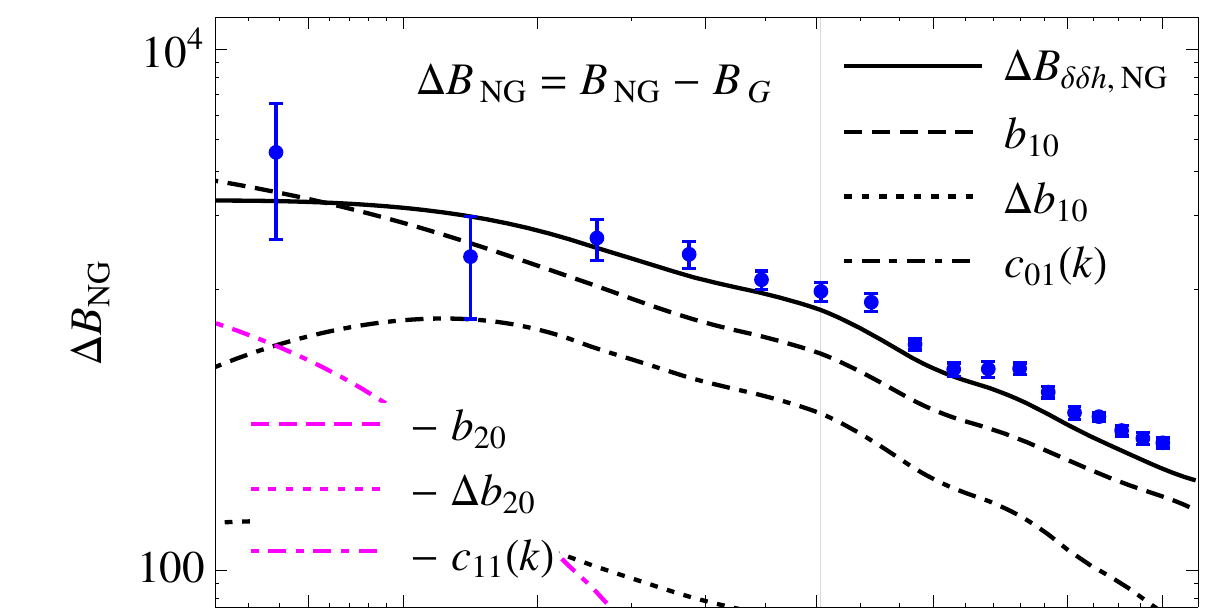}}
{\includegraphics[width=0.48\textwidth]{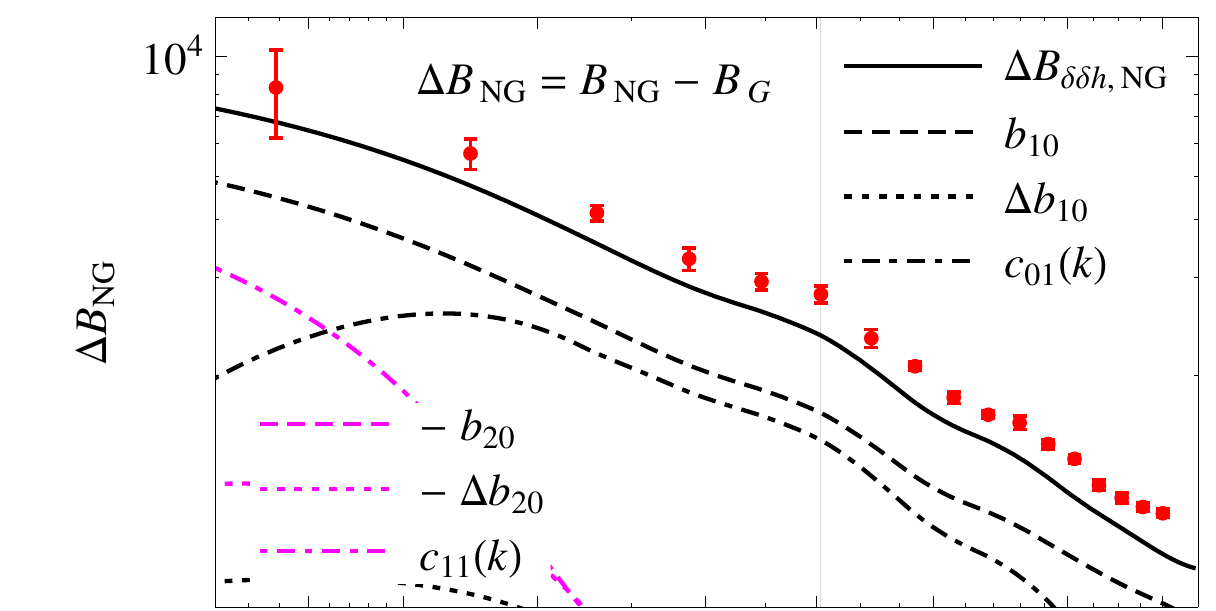}}
{\includegraphics[width=0.48\textwidth]{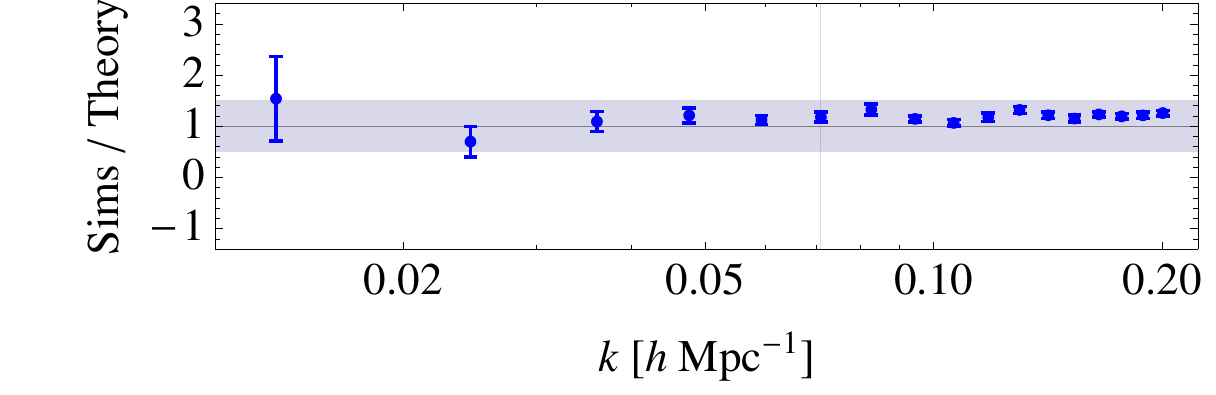}}
{\includegraphics[width=0.48\textwidth]{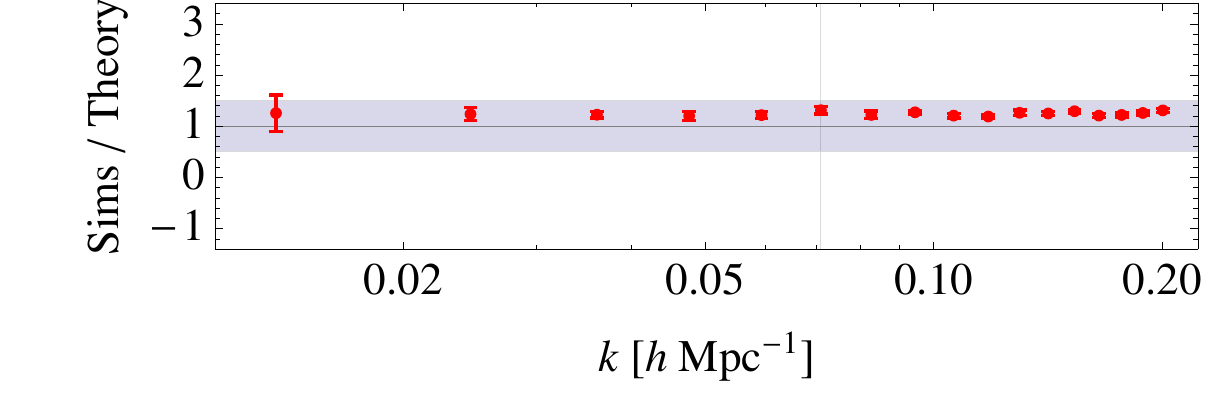}}
{\includegraphics[width=0.48\textwidth]{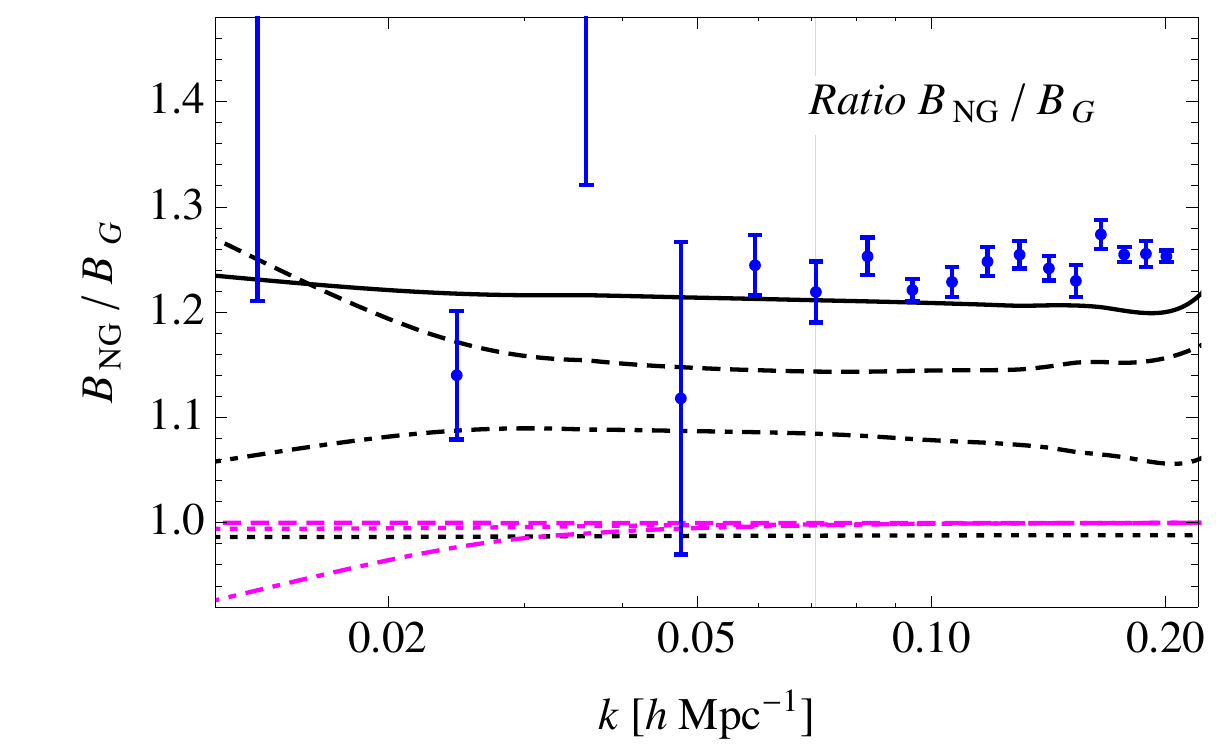}}
{\includegraphics[width=0.48\textwidth]{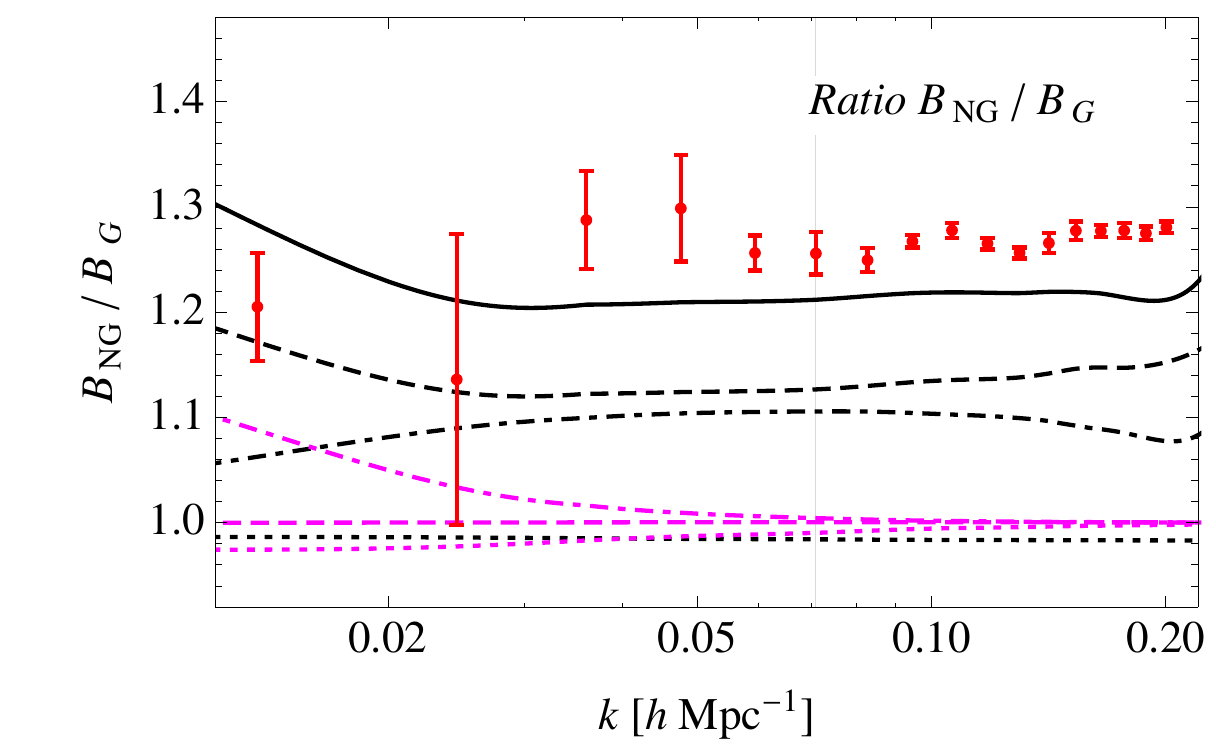}}
{\includegraphics[width=0.48\textwidth]{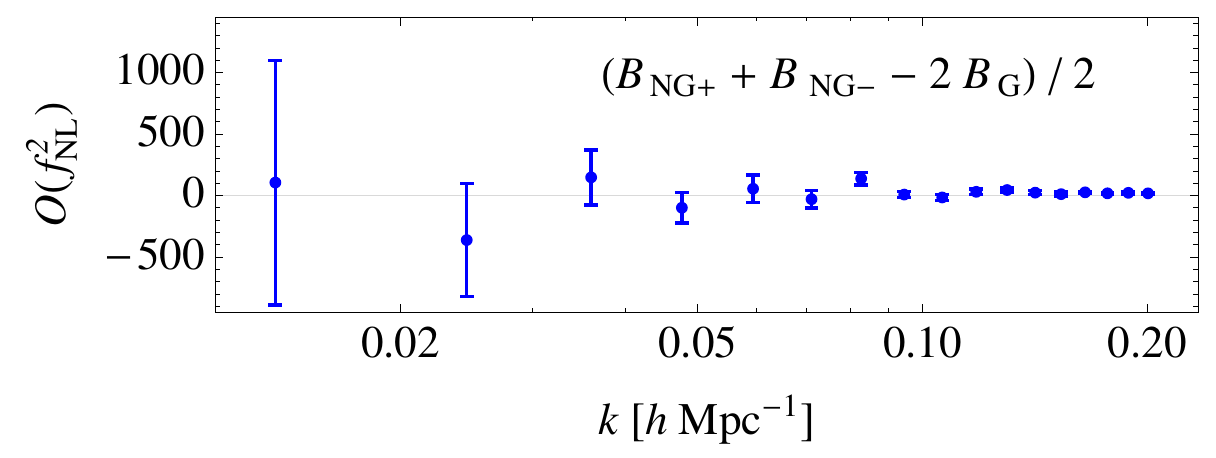}}
{\includegraphics[width=0.48\textwidth]{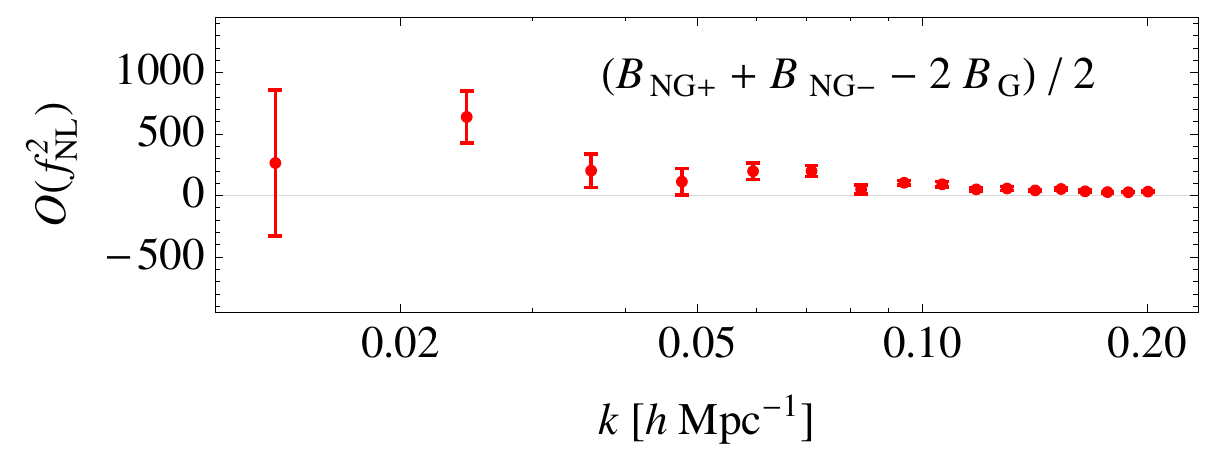}}
\caption{Squeezed configurations of the cross matter-matter-halo bispectrum, $B_{\d\d h}(\D k,k,k)$. See text for explanation.}
\label{fig:bmmhGsq}
\end{center}
\end{figure}

\begin{figure}[!p]
\begin{center}
\begin{center}{\bf Generic configurations (I)}, $B_{\d\d h}(k_1,k_2,\theta)$, $k_1=0.07\kMpc$, $k_2=0.08\kMpc$\end{center}\vspace{0.2cm}
{\includegraphics[width=0.48\textwidth]{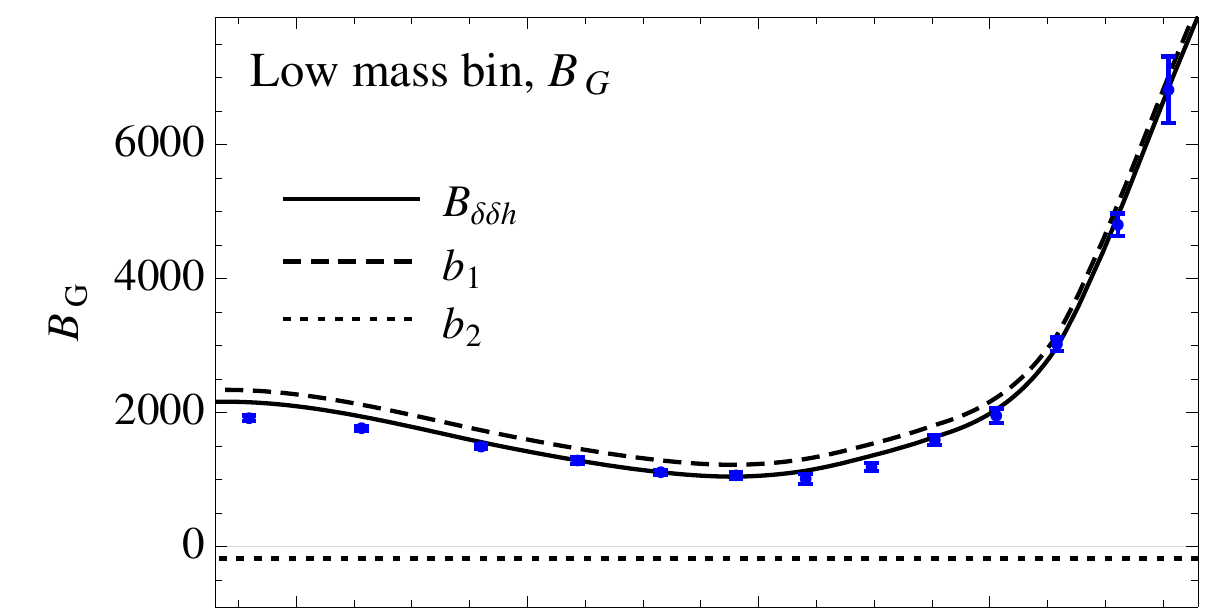}}
{\includegraphics[width=0.48\textwidth]{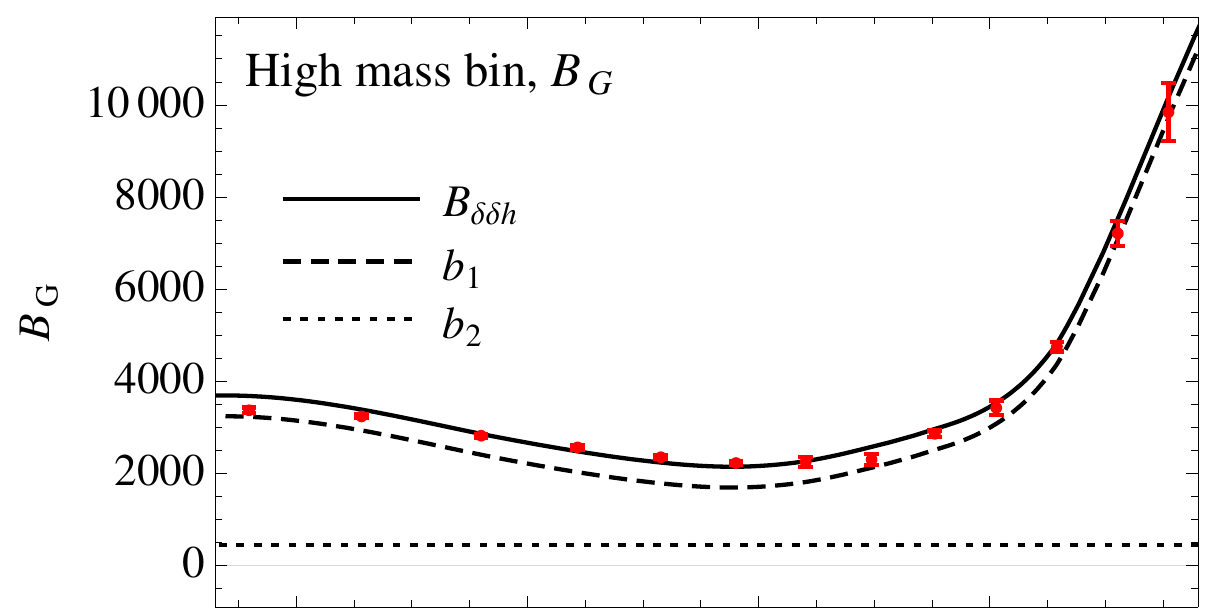}}
{\includegraphics[width=0.48\textwidth]{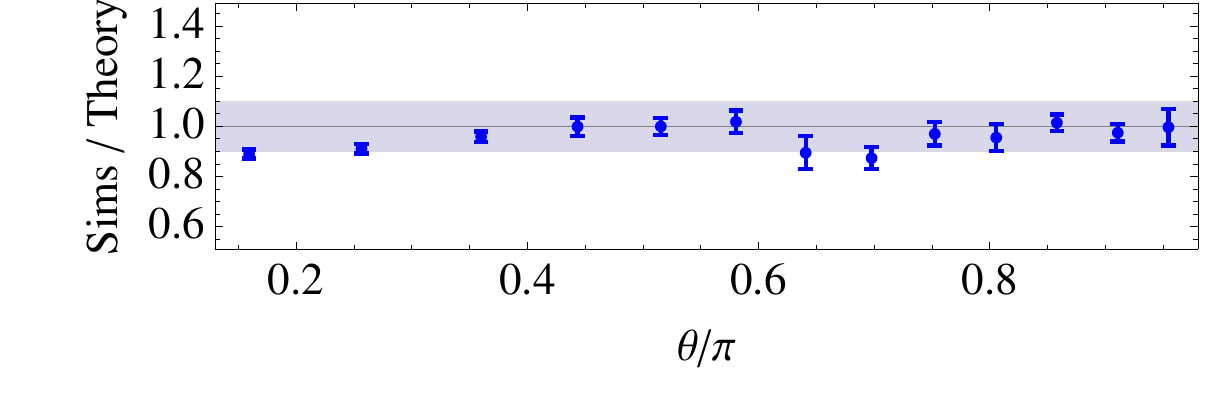}}
{\includegraphics[width=0.48\textwidth]{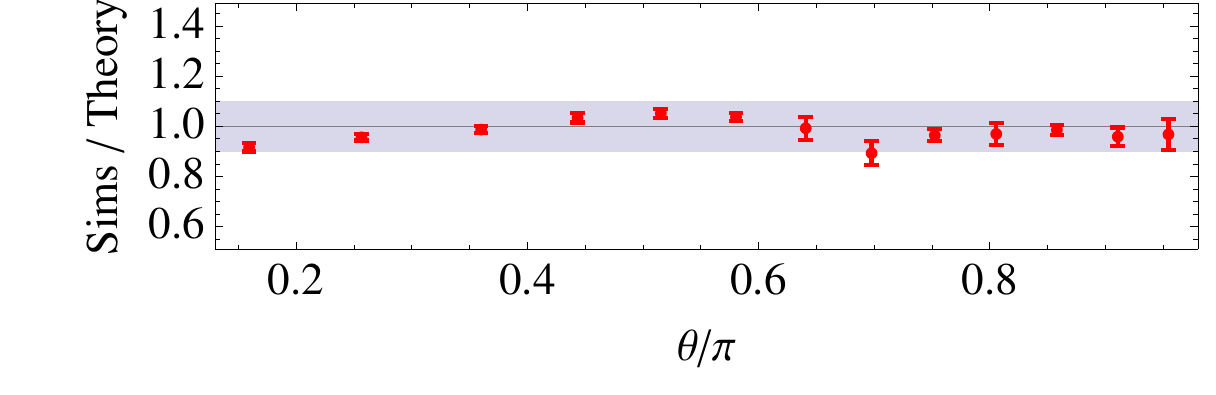}}
{\includegraphics[width=0.48\textwidth]{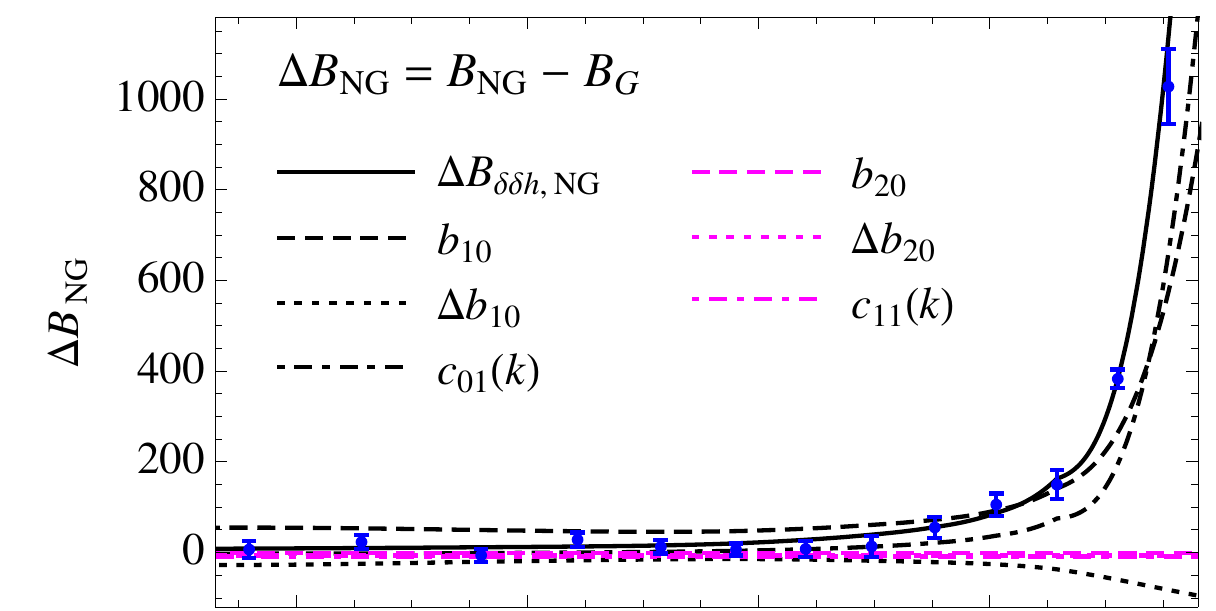}}
{\includegraphics[width=0.48\textwidth]{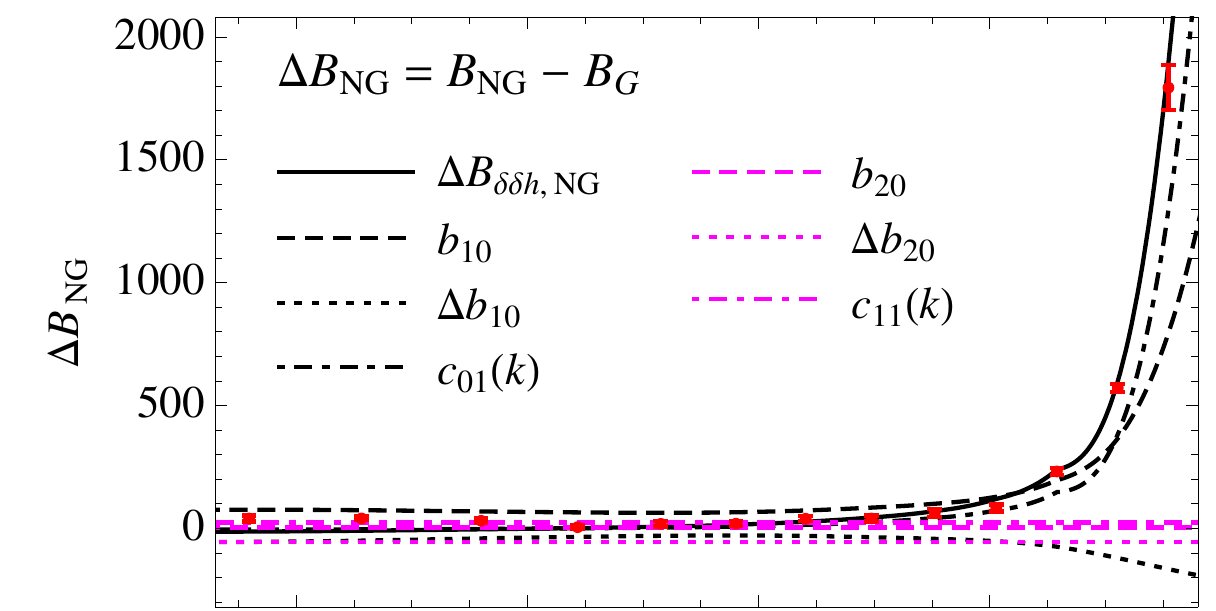}}
{\includegraphics[width=0.48\textwidth]{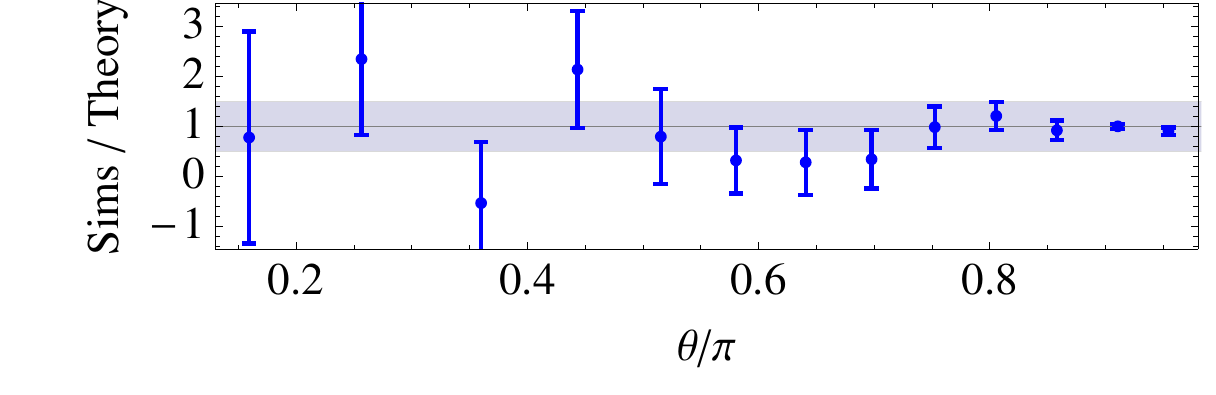}}
{\includegraphics[width=0.48\textwidth]{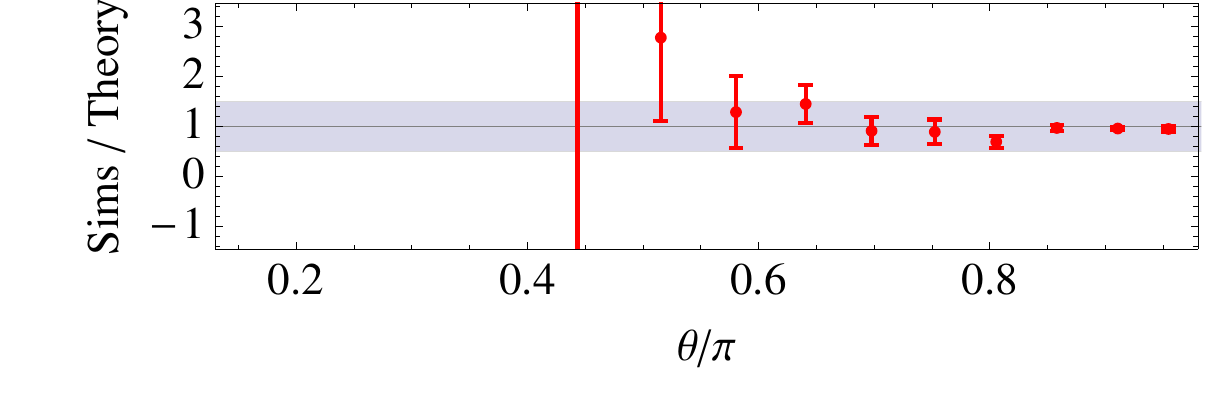}}
{\includegraphics[width=0.48\textwidth]{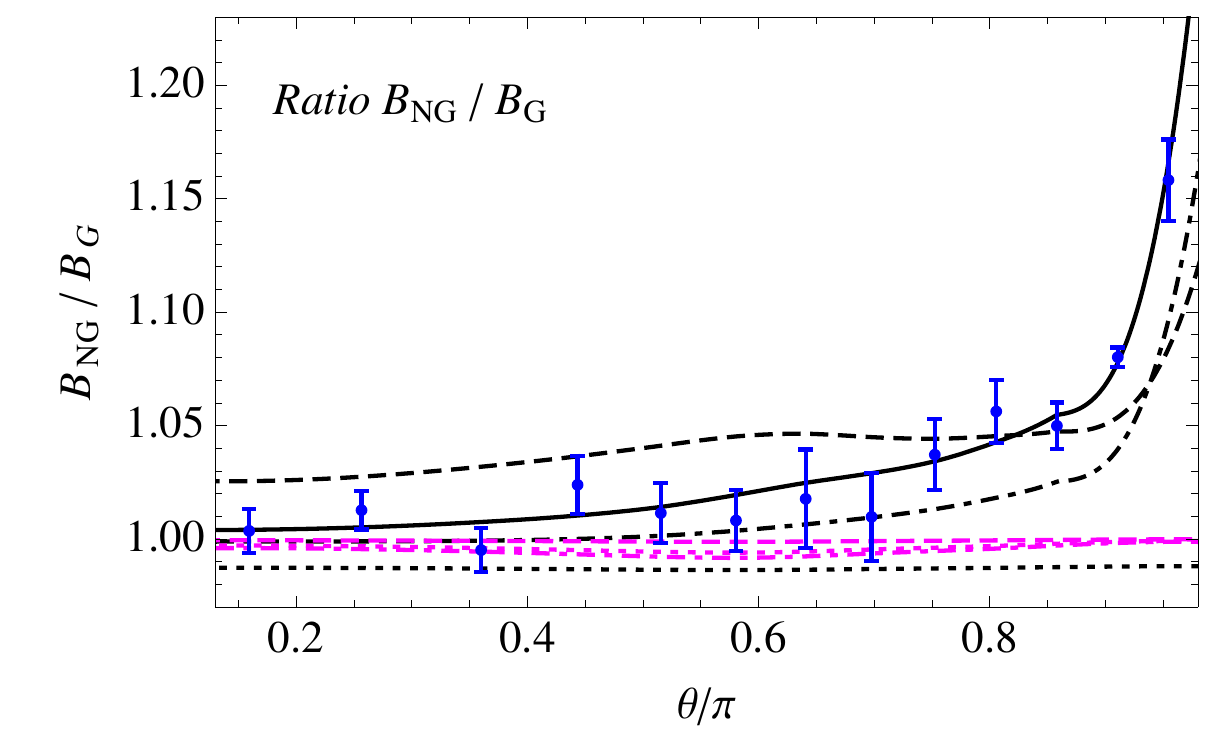}}
{\includegraphics[width=0.48\textwidth]{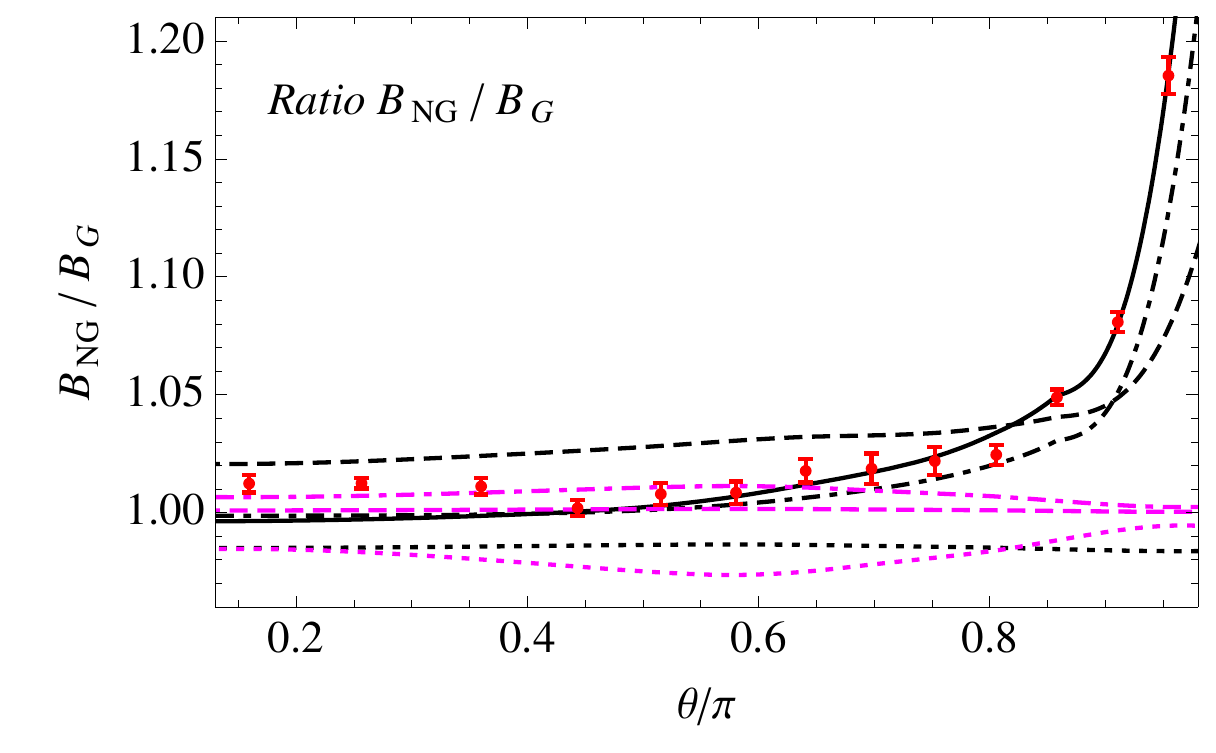}}
{\includegraphics[width=0.48\textwidth]{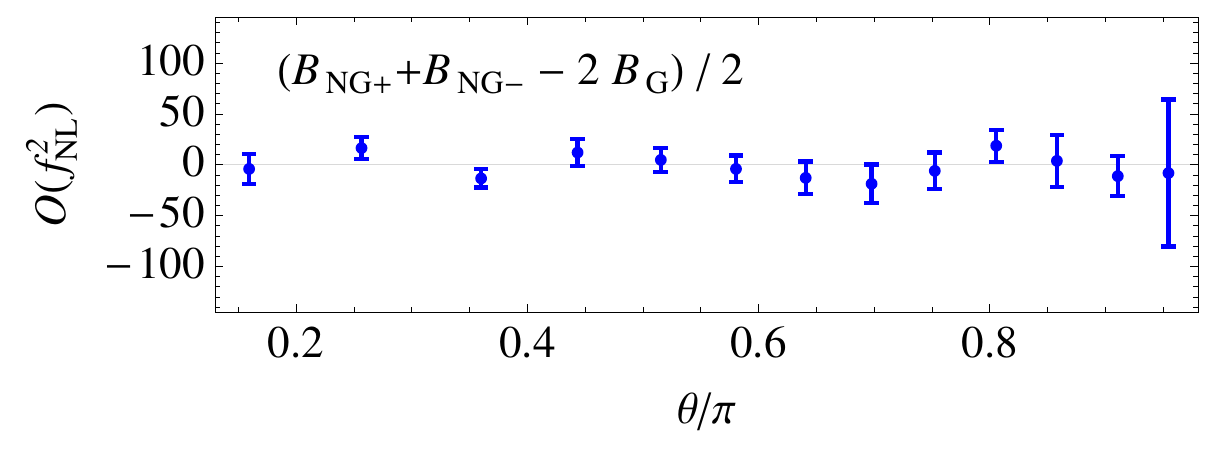}}
{\includegraphics[width=0.48\textwidth]{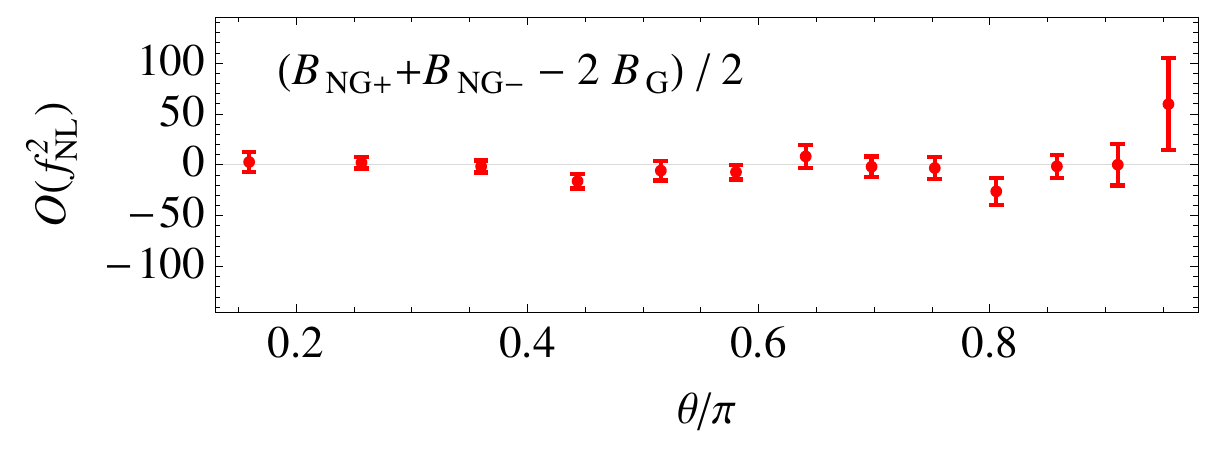}}
\caption{Generic configurations (I) of the cross-bispectrum, $B_{\d\d h}(k_1,k_2,\theta)$, with $k_1=0.07\kMpc$, $k_2=0.08\kMpc$. See text for explanation.}
\label{fig:bmmhGg1}
\end{center}
\end{figure}

\begin{figure}[!p]
\begin{center}
\begin{center}{\bf Generic configurations (II)}, $B_{\d\d h}(k_1,k_2,\theta)$, $k_1=0.05\kMpc$, $k_2=0.07\kMpc$\end{center}\vspace{0.2cm}
{\includegraphics[width=0.48\textwidth]{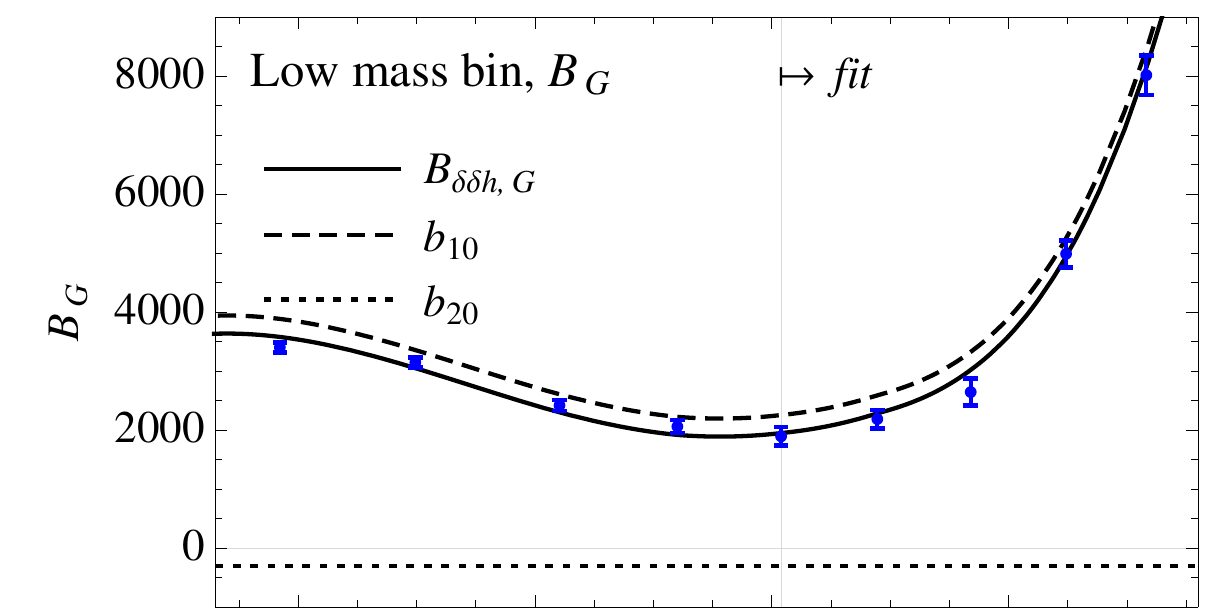}}
{\includegraphics[width=0.48\textwidth]{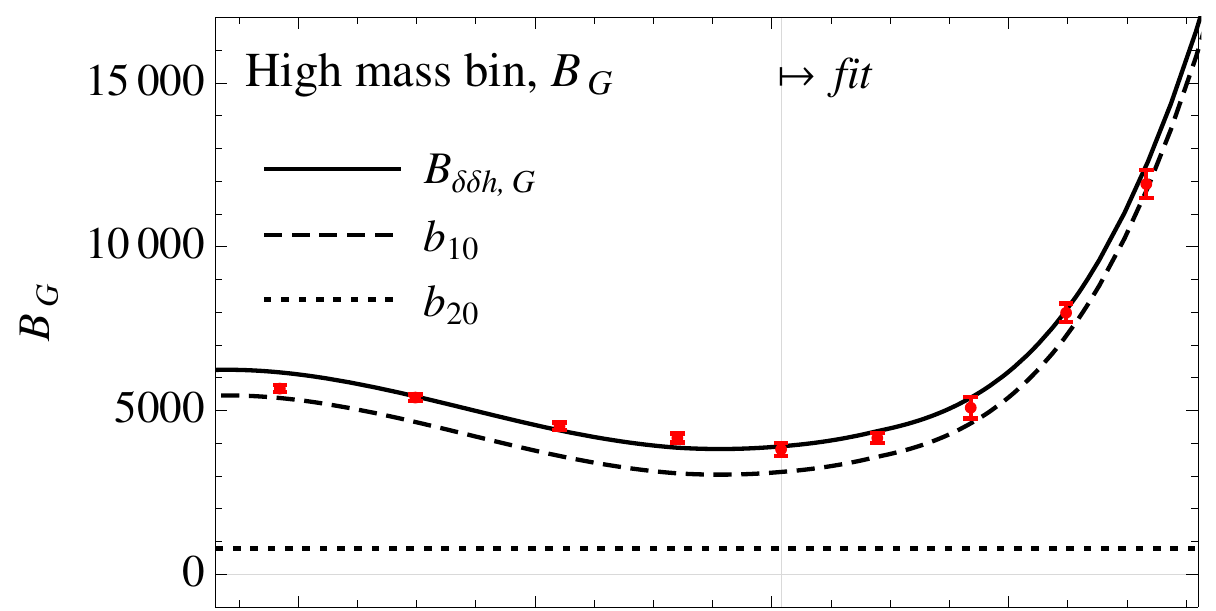}}
{\includegraphics[width=0.48\textwidth]{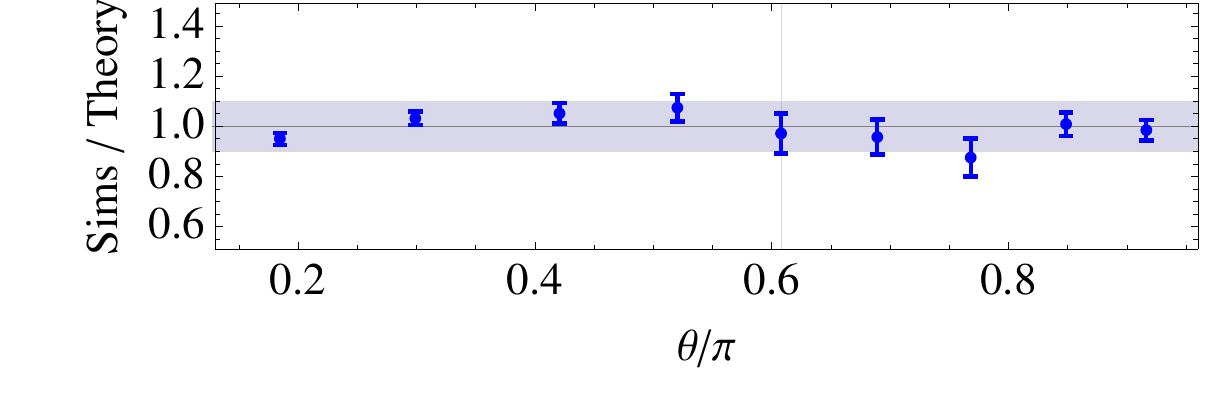}}
{\includegraphics[width=0.48\textwidth]{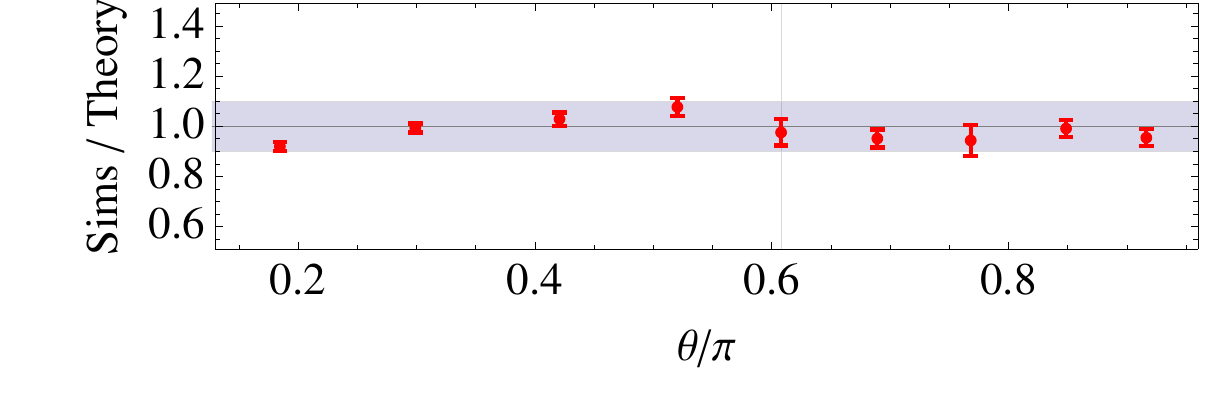}}
{\includegraphics[width=0.48\textwidth]{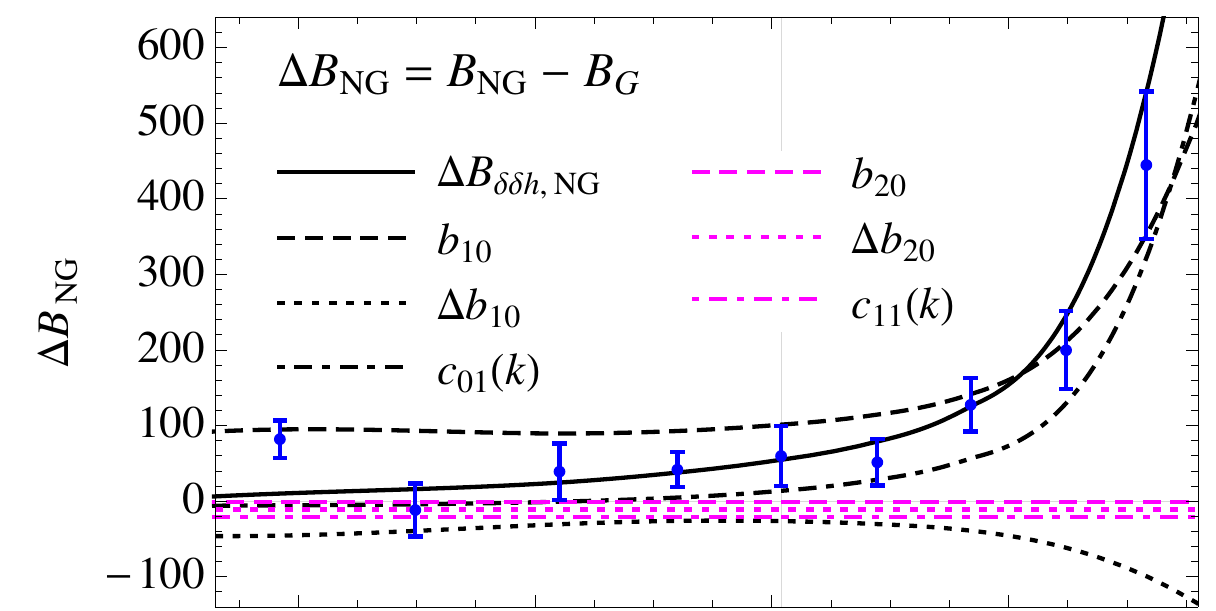}}
{\includegraphics[width=0.48\textwidth]{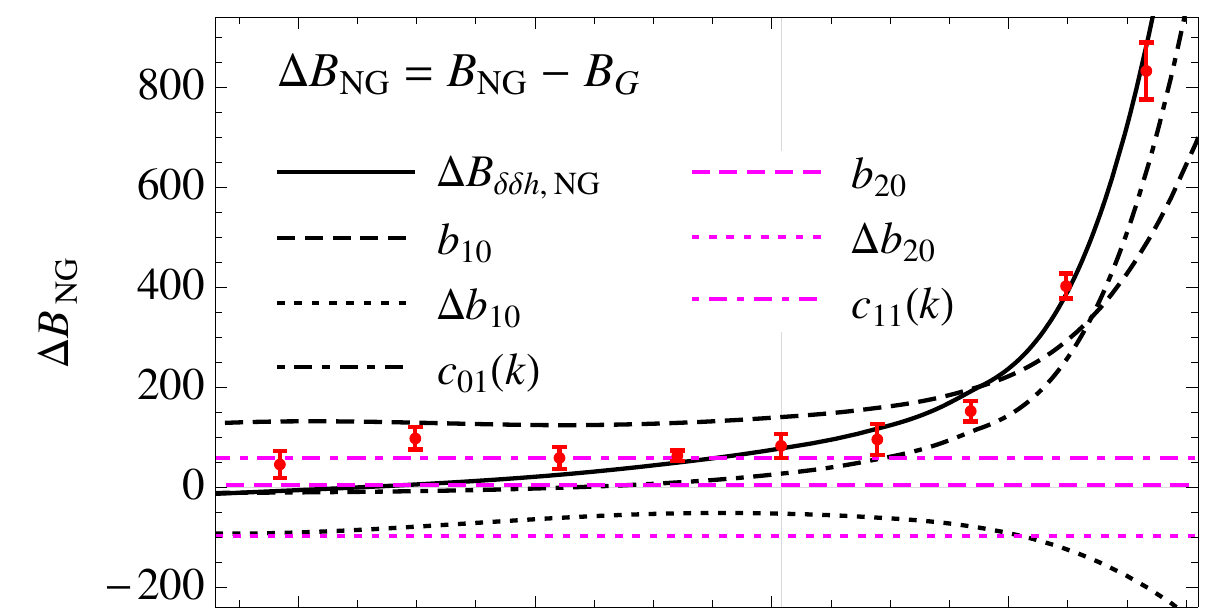}}
{\includegraphics[width=0.48\textwidth]{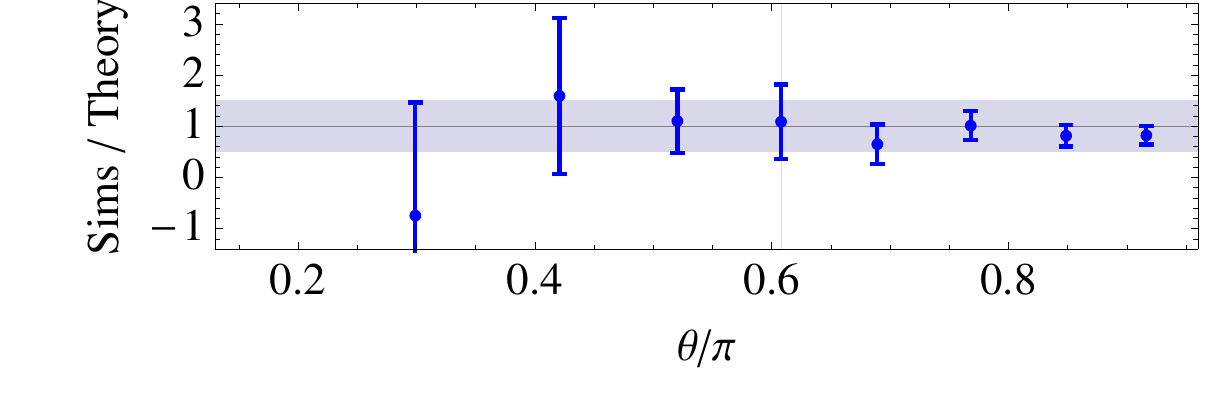}}
{\includegraphics[width=0.48\textwidth]{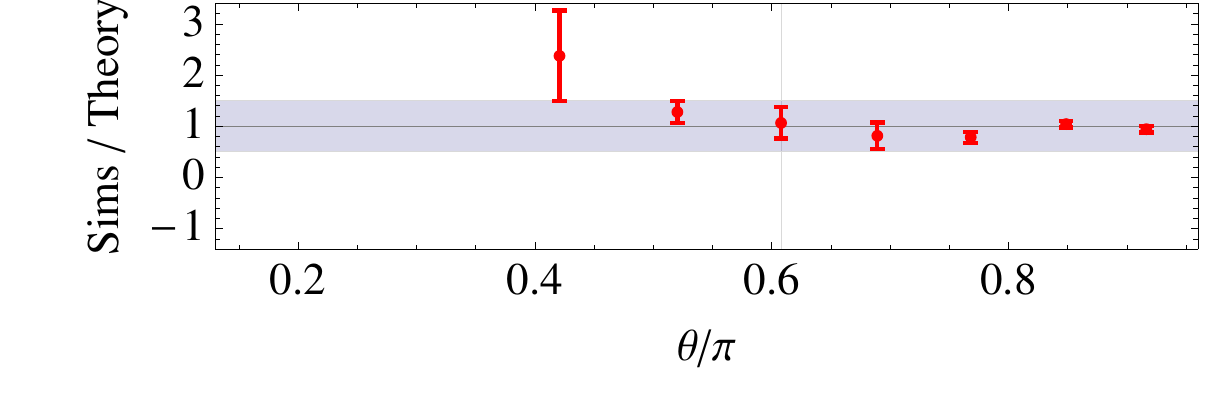}}
{\includegraphics[width=0.48\textwidth]{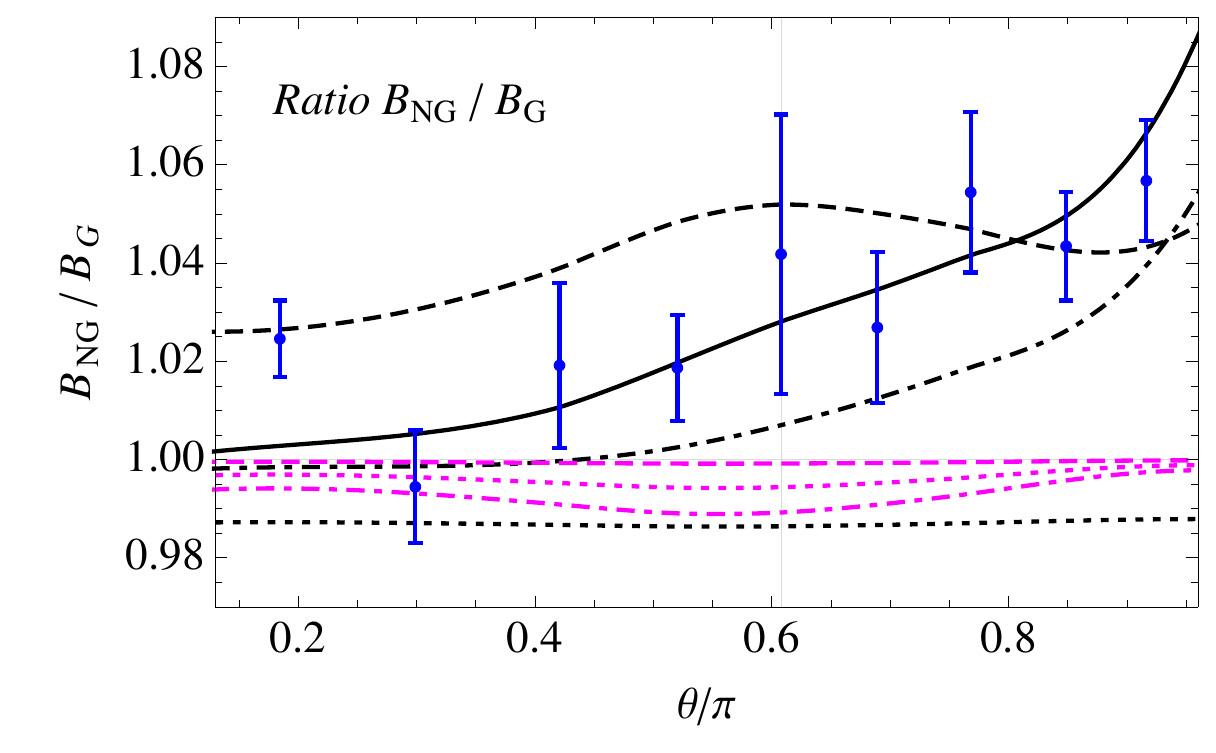}}
{\includegraphics[width=0.48\textwidth]{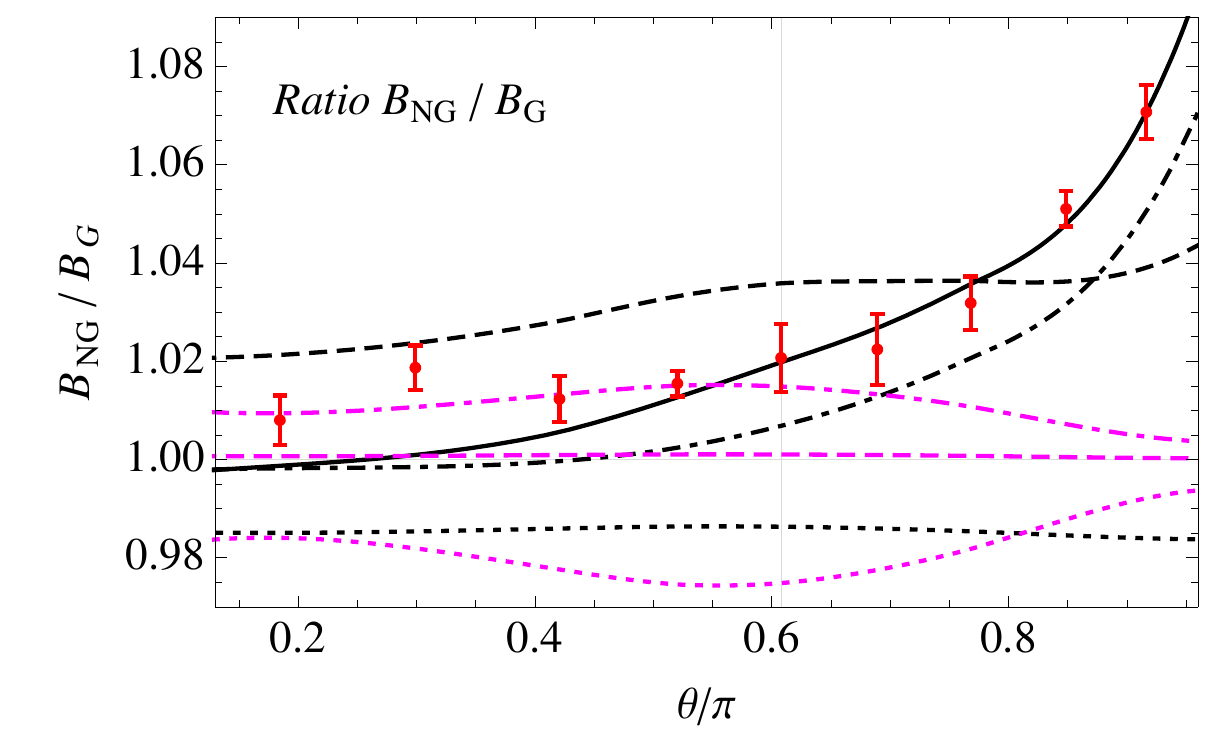}}
{\includegraphics[width=0.48\textwidth]{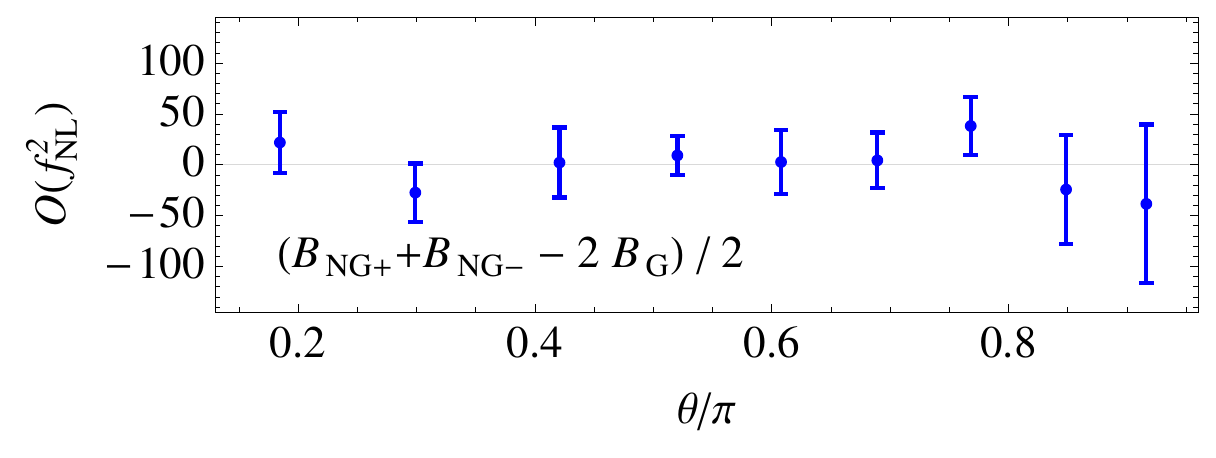}}
{\includegraphics[width=0.48\textwidth]{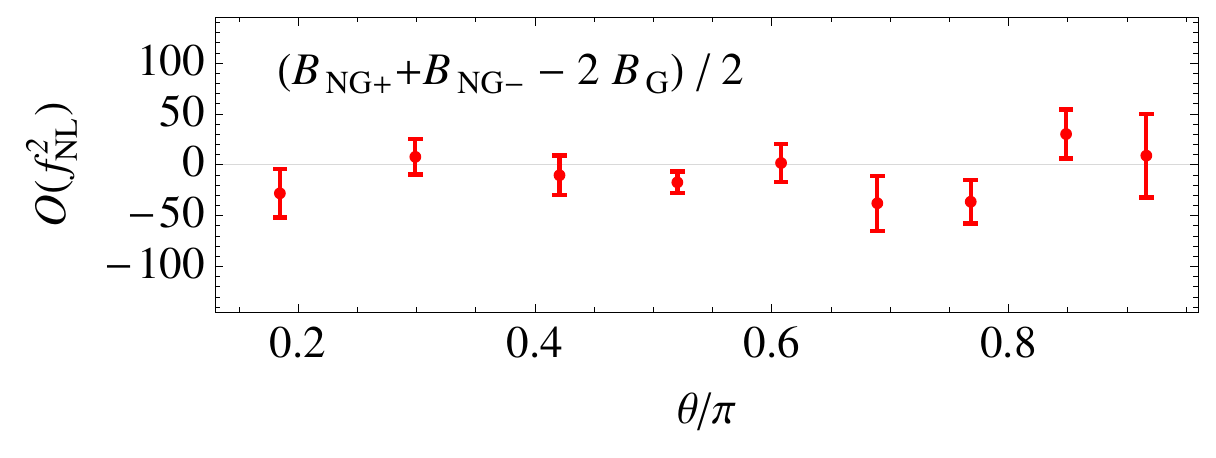}}
\caption{Generic configurations (II) of the cross-bispectrum, $B_{\d\d h}(k_1,k_2,\theta)$, with $k_1=0.05\kMpc$, $k_2=0.07\kMpc$. See text for explanation.}
\label{fig:bmmhGg2}
\end{center}
\end{figure}

\begin{figure}[!p]
\begin{center}
\begin{center}{\bf Equilateral configurations}, $B_{\d\d h}(k,k,k)$\end{center}\vspace{0.2cm}
{\includegraphics[width=0.48\textwidth]{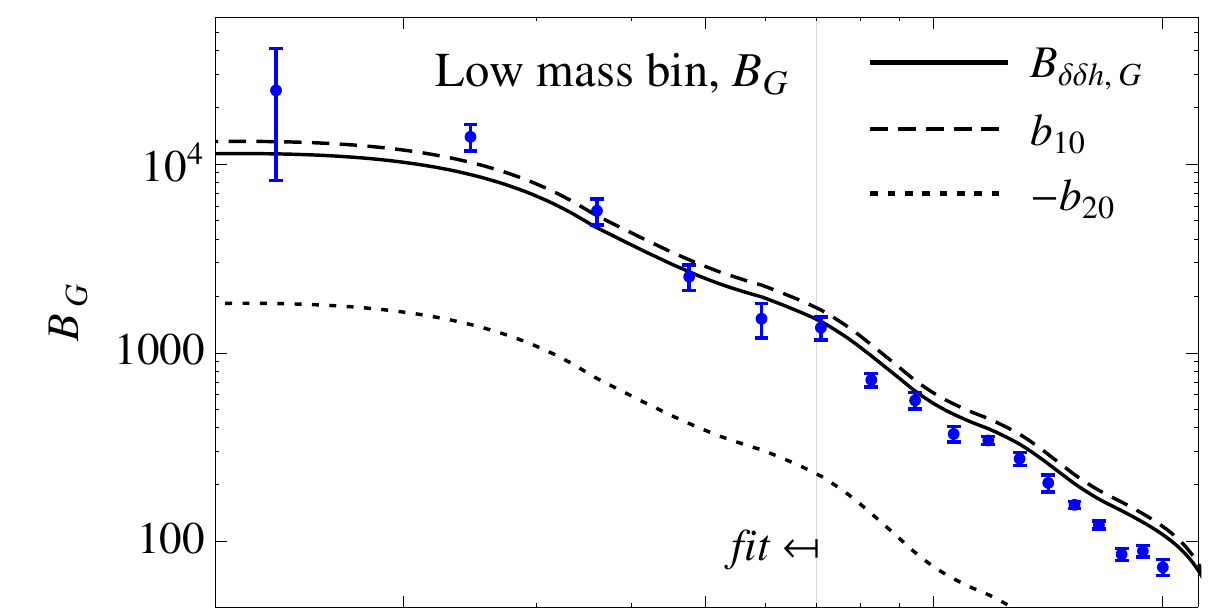}}
{\includegraphics[width=0.48\textwidth]{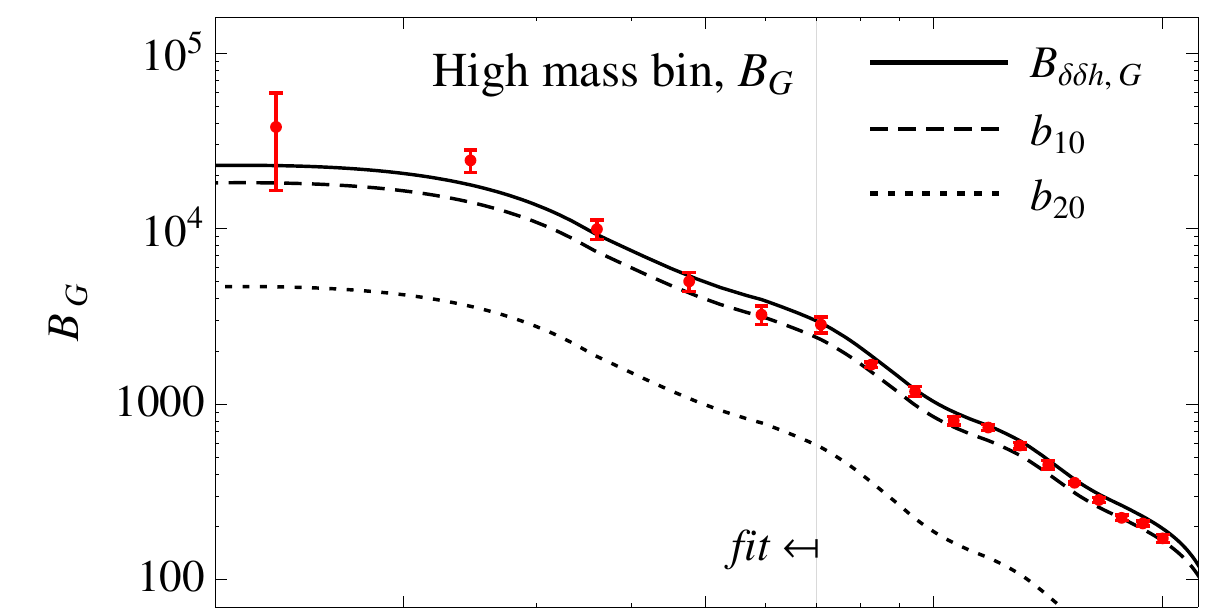}}
{\includegraphics[width=0.48\textwidth]{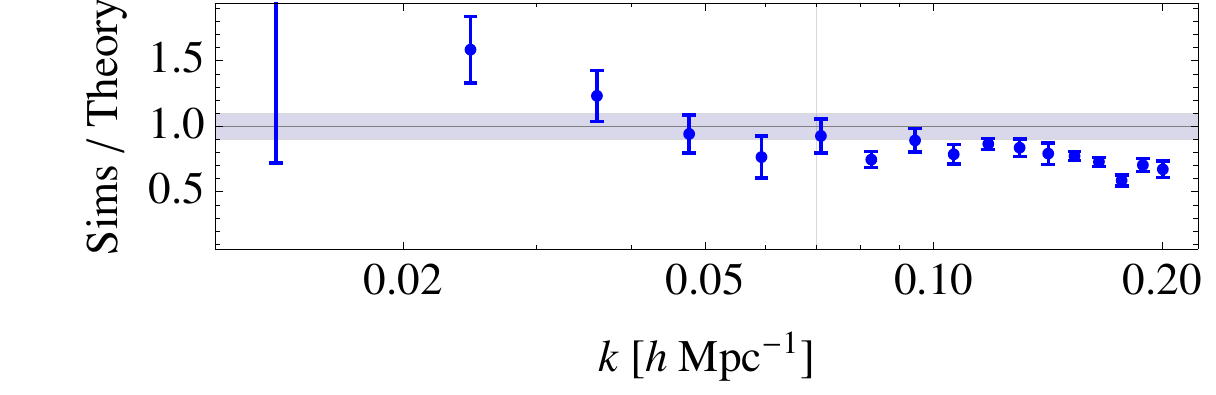}}
{\includegraphics[width=0.48\textwidth]{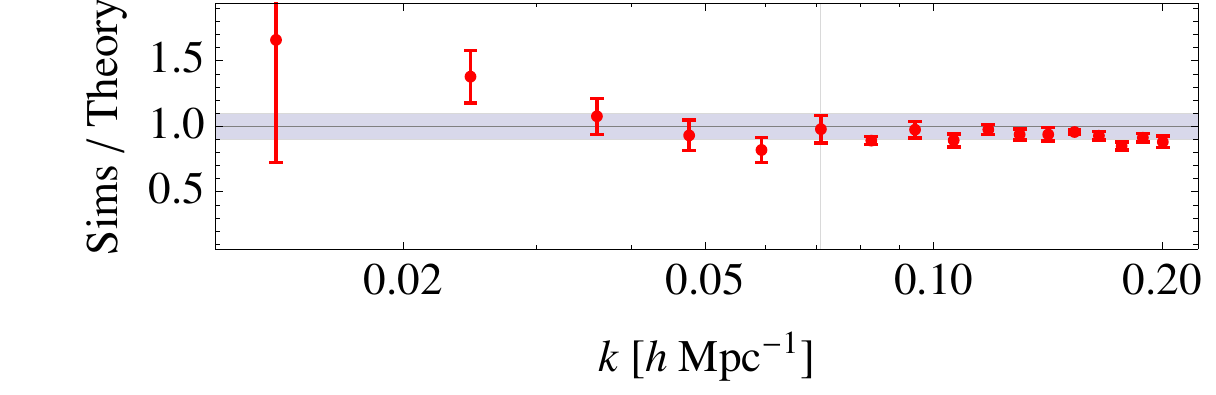}}
{\includegraphics[width=0.48\textwidth]{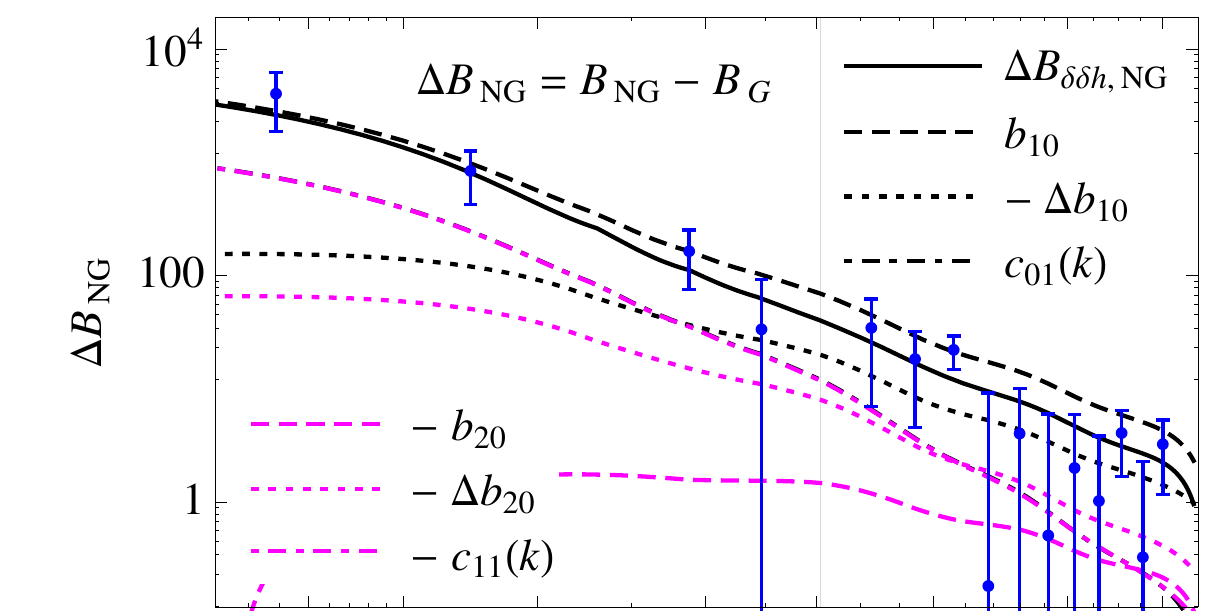}}
{\includegraphics[width=0.48\textwidth]{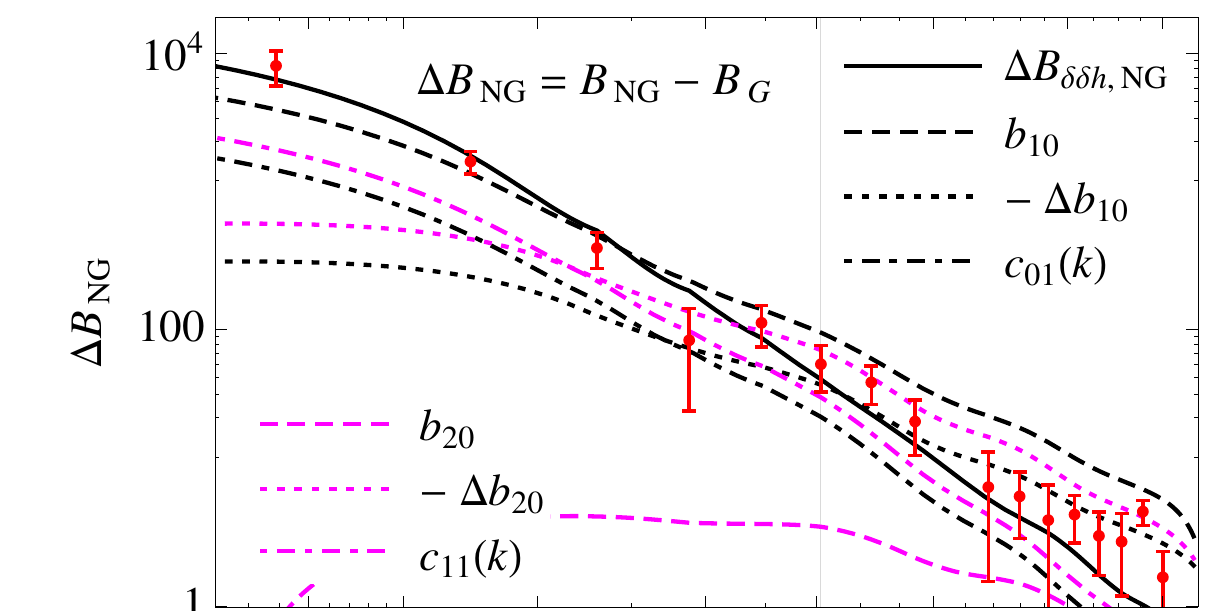}}
{\includegraphics[width=0.48\textwidth]{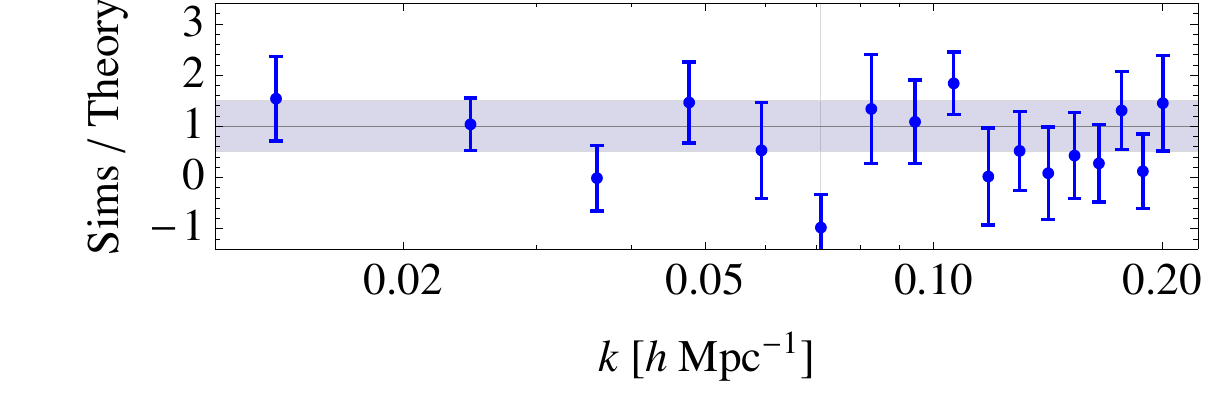}}
{\includegraphics[width=0.48\textwidth]{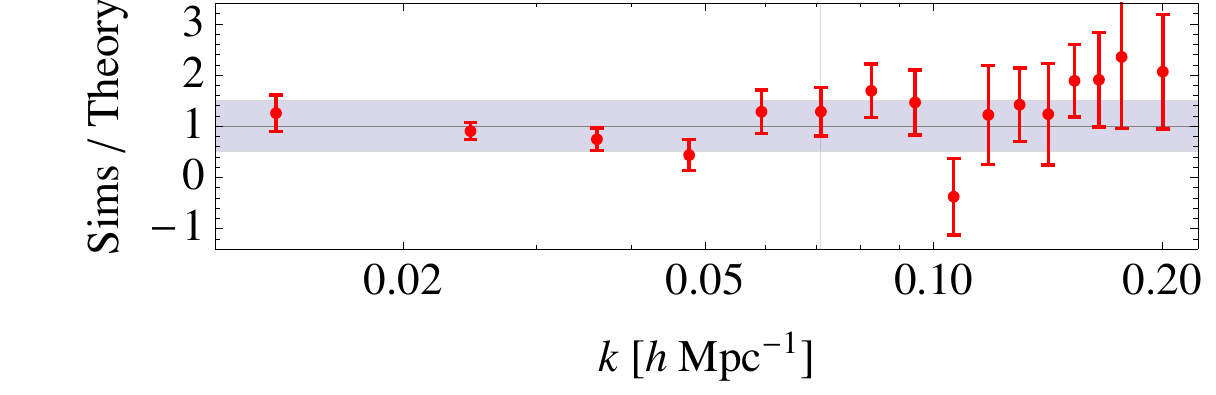}}
{\includegraphics[width=0.48\textwidth]{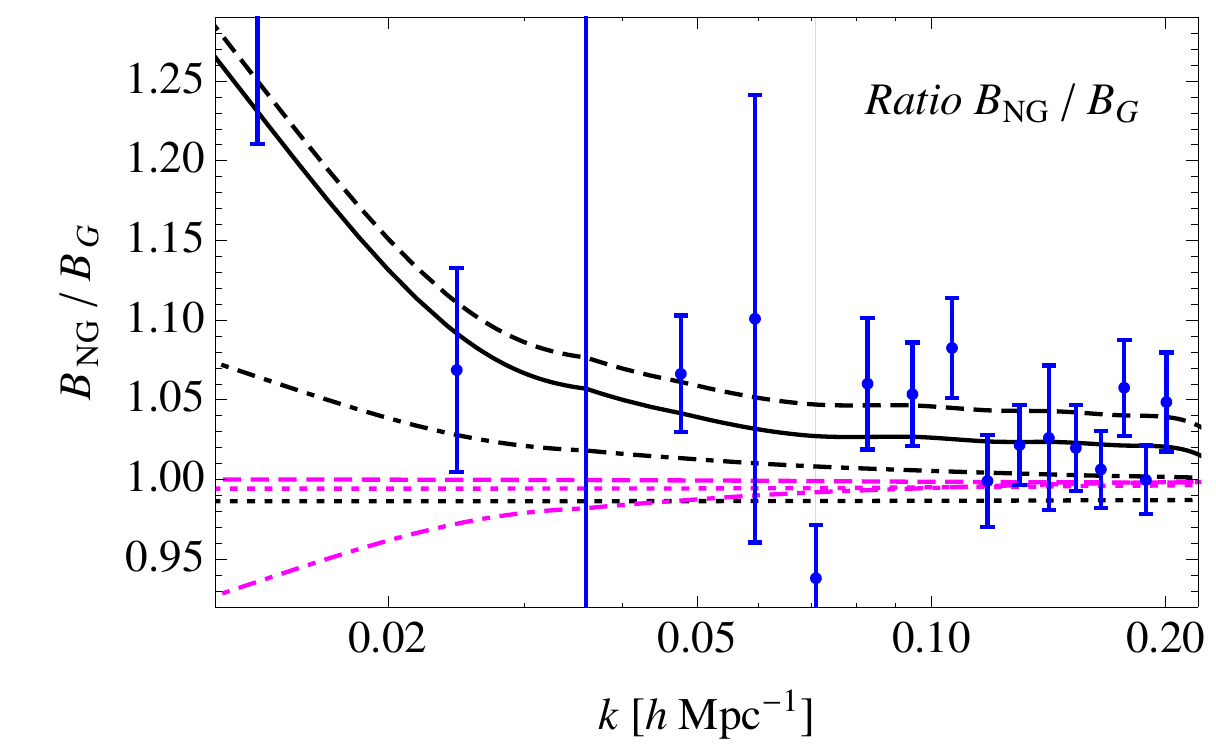}}
{\includegraphics[width=0.48\textwidth]{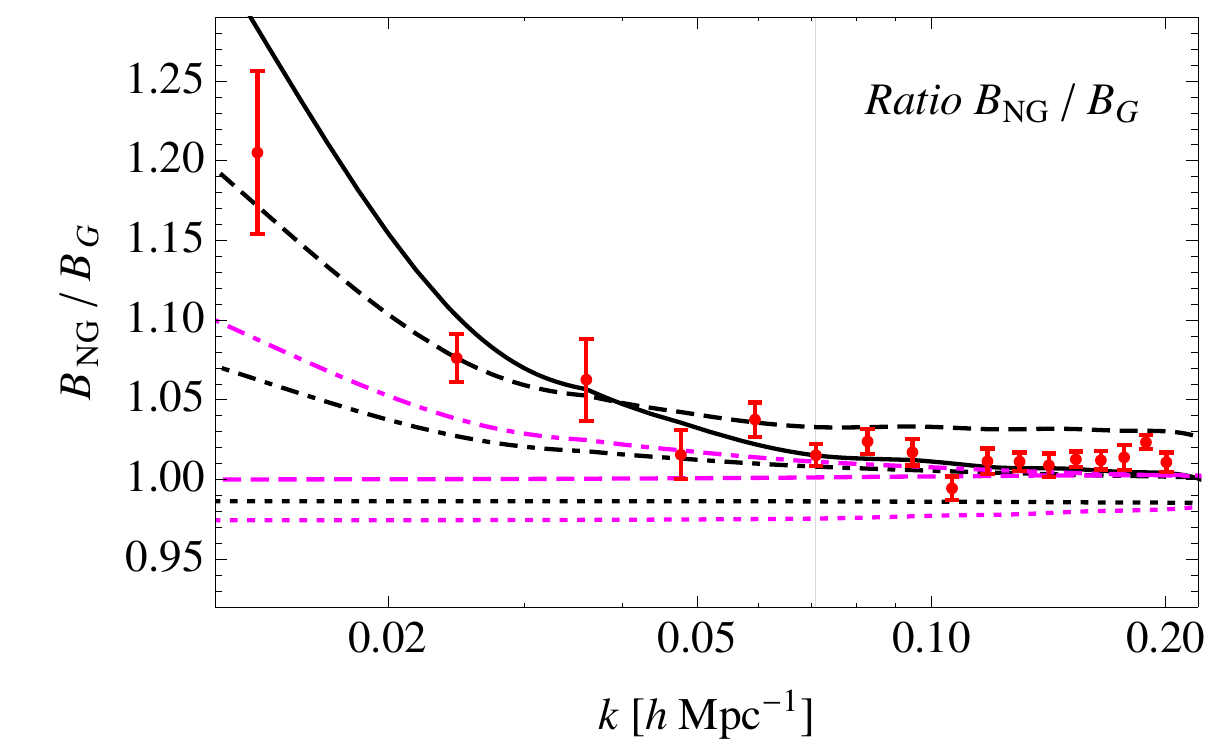}}
{\includegraphics[width=0.48\textwidth]{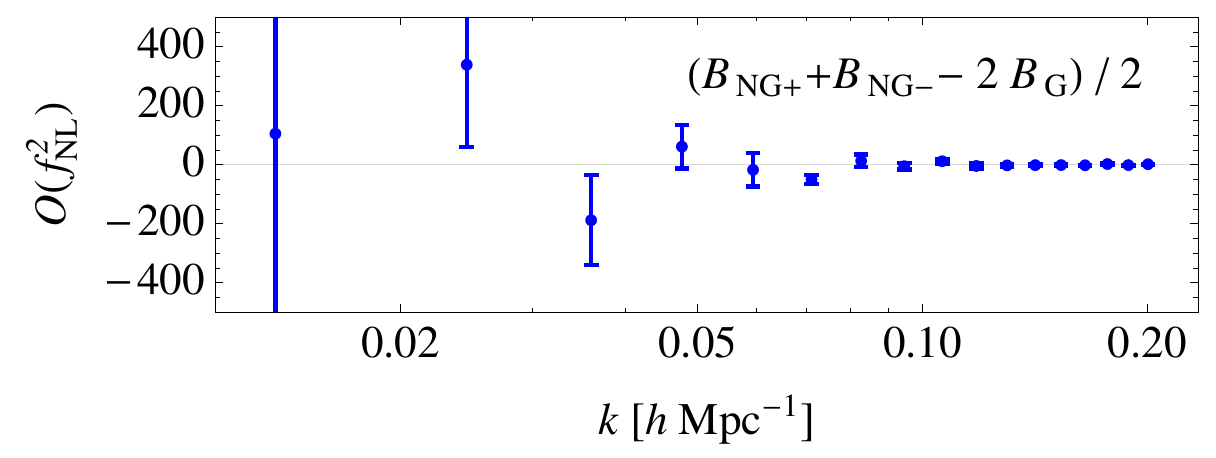}}
{\includegraphics[width=0.48\textwidth]{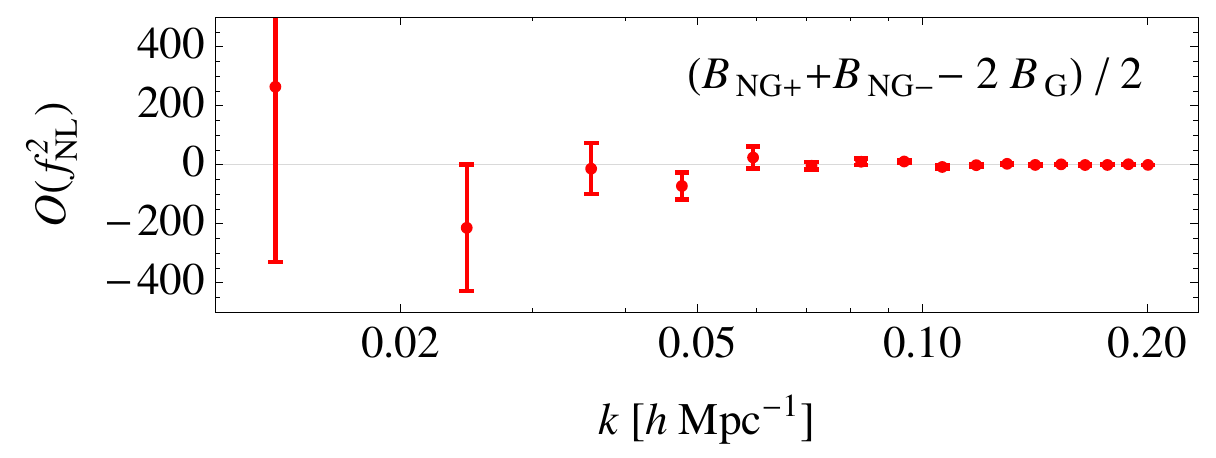}}
\caption{Equilateral configurations of the cross matter-matter-halo bispectrum, $B_{\d\d h}(k,k,k)$ as a function of $k$. See text for explanation.}
\label{fig:bmmhGeq}
\end{center}
\end{figure}

Figures \ref{fig:bmmhGsq}, \ref{fig:bmmhGg1}, \ref{fig:bmmhGg2} and \ref{fig:bmmhGeq} show, respectively, the four different subsets of triangles described above for the matter-matter-halo bispectrum $B_{\d\d h}(k_1,k_2;k_3)$. Notice that the third variable $k_3$ corresponds to the halo density contrast $\d_{h}(\kv_3)$.    

As for the power spectrum plots, for each set of configurations we show the Gaussian component $B_{\d\d h,G}$ and the model residuals ({\em upper two rows}), the non-Gaussian correction $\D B_{\d\d h,NG}$ and the corresponding residuals ({\em third and fourth row}), the non-Gaussian to Gaussian ratio  $B_{\d\d h,NG}/ B_{\d\d h,G}$ ({\em fifth row}) and the $\O(\fNL^2)$ component determined from the average of $[B_{\d\d h,NG}(\fNL=+100)+B_{\d\d h,NG}(\fNL=-100)]-\,B_{\d\d h,G}(\fNL=0)$ measured in each realization ({\em last row}). Left and right columns show results for the low and high-mass halos, respectively. 

For the Gaussian piece of the cross bispectrum, the model is given in terms of two components, the first and second terms on the r.h.s.~of Eq.~(\ref{eq:BmmhG}), shown in the plots as a dashed and dotted curve, respectively. These terms are simply labelled by $b_{10}$ and $b_{20}$, since the latter are the bias parameters controlling their amplitude. We implicitly assign to $b_{10}$ and $b_{20}$ the value corresponding to Gaussian initial conditions, \ie $b_{10,G}$ and $b_{20,G}$. For the low mass halos, $b_{20,G}$ is negative as is the corresponding contribution to the bispectrum. For this reason, we show its absolute value in the log-log plots of squeezed and equilateral configurations. When a given contribution is negative, we denote it in the plot legend by a negative sign in front of the related symbol. In the plots of the models residuals for the Gaussian term $B_{\d\d h,G}$, the shaded area indicate a 10\% deviation. The thin, gray vertical line at $k=0.07$ shows the subset of data points being part of the larger subset used for the model fit. Notice that, in the case of the generic configurations (I) for which $k_2=0.082\kMpc$, {\em none} of the triangles shown is used for the fit. 
 
For nearly all triangles, with the exception of the equilateral ones, the model and the data generally agree within 10\% at large scales. For squeezed configurations in particular, such an agreement persists well beyond $k_{max}=0.07\kMpc$. 

For the non-Gaussian correction $\D B_{\d\d h, NG}$ we have several distinct components. The first three terms on the r.h.s.~of Eq.~(\ref{eq:dBmmhNG}) are shown in the plots respectively as dashed, dotted and dot-dashed black curves, and labelled as $b_{10}$, $\D b_{10}$ and $c_{01}(k)$ since they loosely correspond to the {\em linear} halo bias and its corrections. The remaining three terms on the r.h.s.~of Eq.~(\ref{eq:dBmmhNG}), which are corrections related to the quadratic bias, are shown as dashed, dotted and dot-dashed magenta (or light gray) curves and labelled as $b_{20}$, $\D b_{20}$ and $c_{11}(k)$. The continuous black curve shows the sum of all the terms. In the residual plots, the shaded area for $\D B_{\d\d h, NG}$ indicate a less than $50\%$ deviation between the model and the data. The same notation is assumed for the various curves in the ratio plots on the fifth row.

Even for the correction $\D B_{\d\d h, NG}$ in the case of the squeezed configurations shown in Fig.~\ref{fig:bmmhGsq}, the agreement between the model and simulations extends to relatively small scales. More interestingly, the non-Gaussian signal in the squeezed limit results from a comparable effects of PNG on the matter bispectrum and on the linear bias, $c_{01}(k)$. For these specific configurations, the effect on the quadratic bias is very small and even vanishes in the squeezed limit, because the small, constant side of the triangle is in this case $k_3$, the wavenumber corresponding to the halo overdensity, while the scale-dependent corrections $c_{11}(k_1)$ and $c_{11}(k_2)$ are suppressed for large values of $k_1$ and $k_2$. 
 
Figures \ref{fig:bmmhGg1} and \ref{fig:bmmhGg2} confirm that non-Gaussian corrections, now smaller for generic configurations, are the results of two distinct contributions which are roughly of the same order. Since for these configurations $k_1$ and $k_2$ are fixed, the non-Gaussian correction to the quadratic bias is constant and, in this case, negative. As a consequence, only the primordial component to the matter bispectrum and the correction to the linear bias contribute to the non-Gaussian effect in the squeezed limit, which is attained for the triangles with $\theta\rightarrow \pi$. This is particularly obvious in figure \ref{fig:bmmhGg1} where $k_1$ and $k_2$ take similar values. The constant correction to $b_2$ is most evident for nearly equilateral configurations, especially in figure \ref{fig:bmmhGg2} where it amounts to a reduction of the overall non-Gaussian correction to the cross-bispectrum. What is more interesting is the fact that, since we are only fitting for the value of $\D b_{20,NG}$, the model correctly predicts the shape-dependence of $\D B_{\d\d h, NG}$. This is not trivial, since such dependency is given by the peculiar combination of the two terms mentioned above. 

No significant correction beyond linear order in $\fNL$ to the cross bispectrum is detected for generic triangles and for all halos, with the exception of high mass halos in the squeezed configurations. However, it is small and does not affect the analysis of the overall non-Gaussian correction in terms of a model linear in $\fNL$.  

\begin{figure}[!p]
\begin{center}
\begin{center} {\bf Squeezed configurations}, $B_{h}(\D k,k,k)$\end{center}\vspace{0.2cm}
{\includegraphics[width=0.48\textwidth]{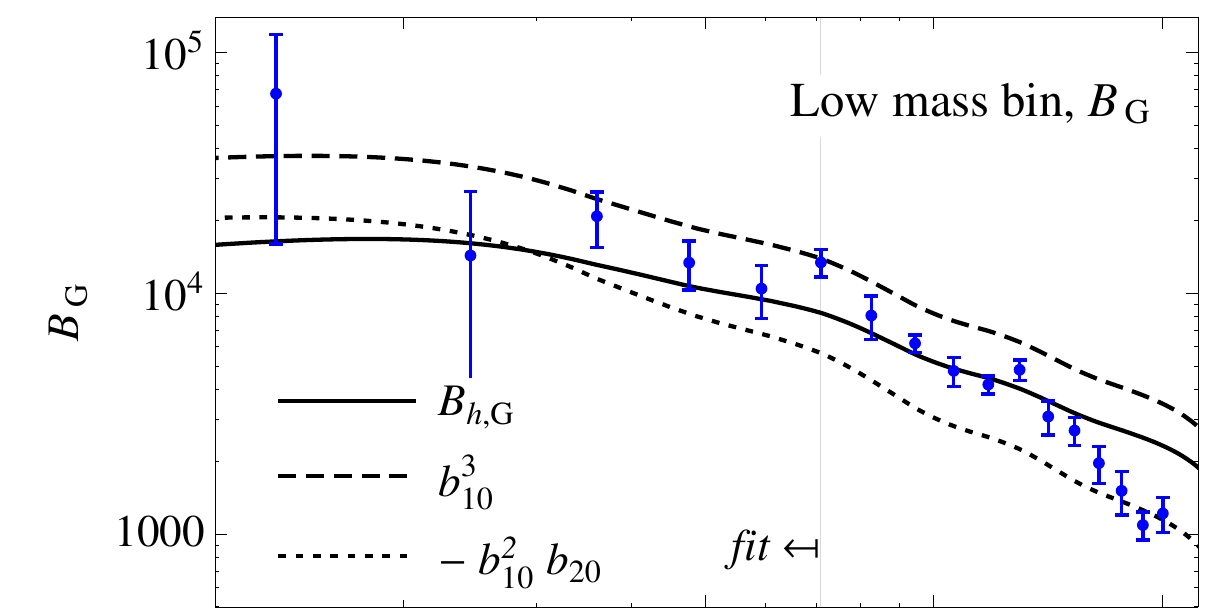}}
{\includegraphics[width=0.48\textwidth]{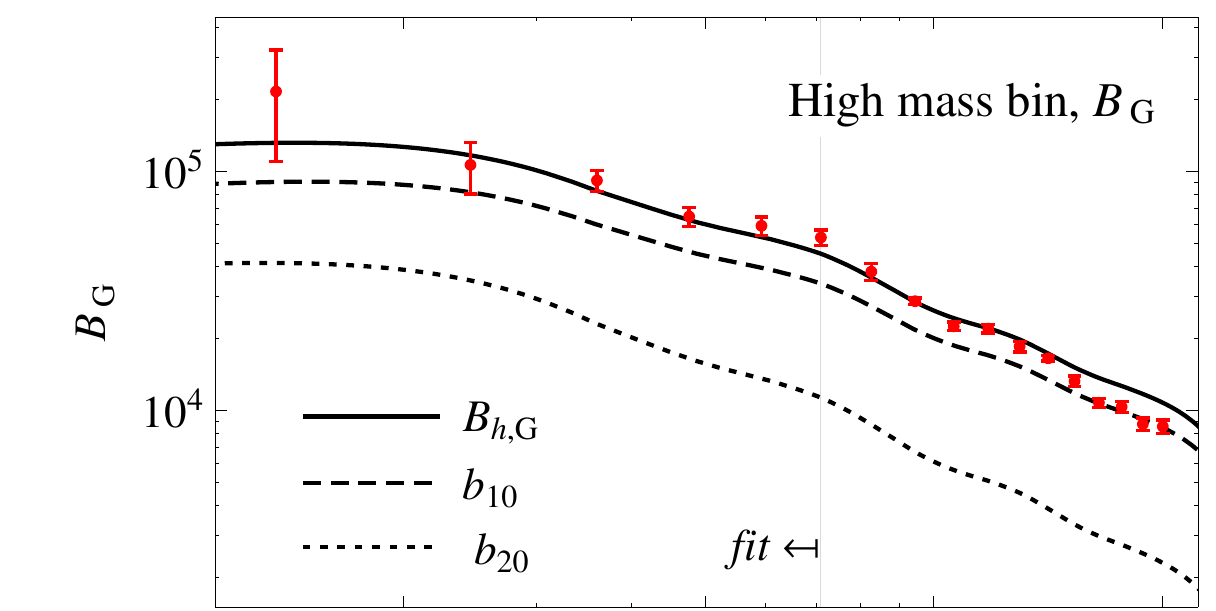}}
{\includegraphics[width=0.48\textwidth]{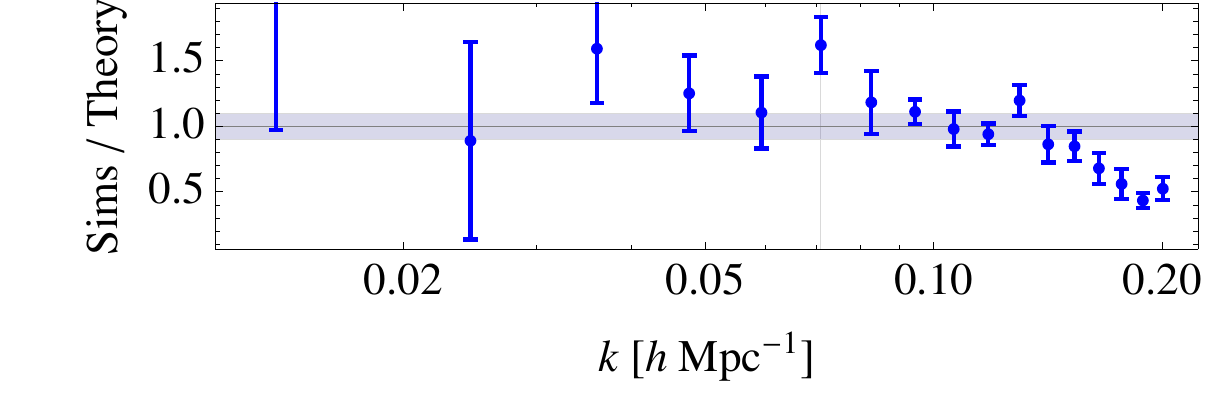}}
{\includegraphics[width=0.48\textwidth]{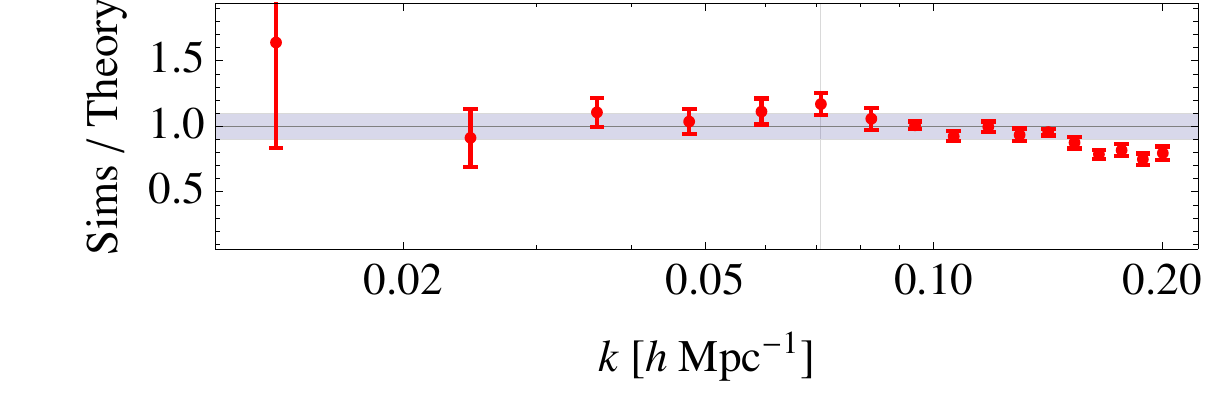}}
{\includegraphics[width=0.48\textwidth]{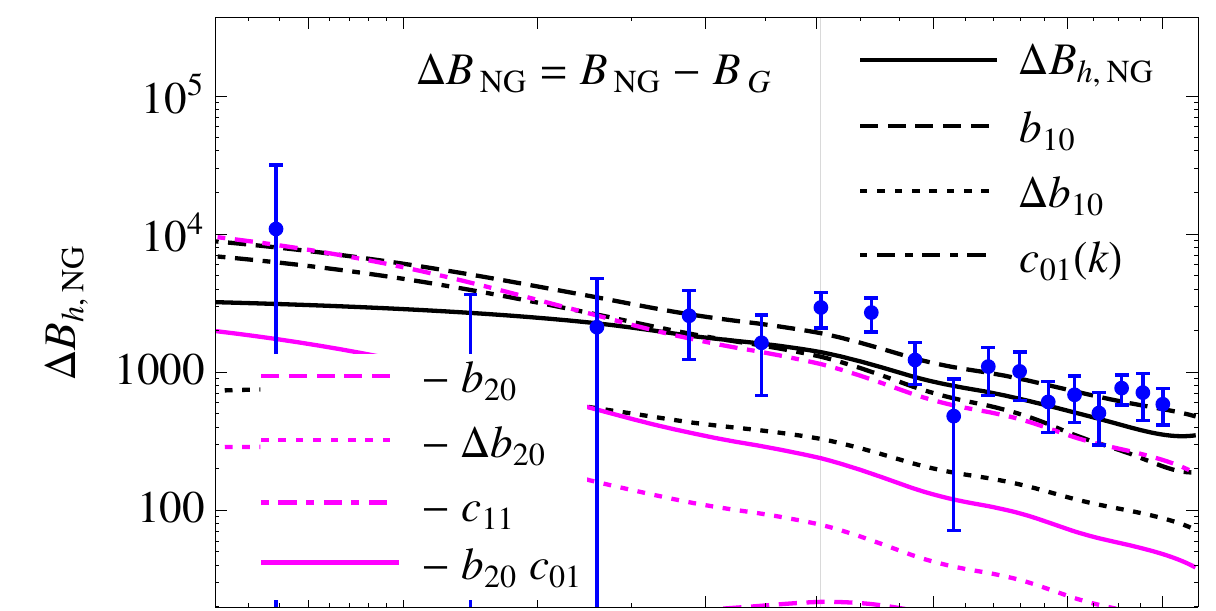}}
{\includegraphics[width=0.48\textwidth]{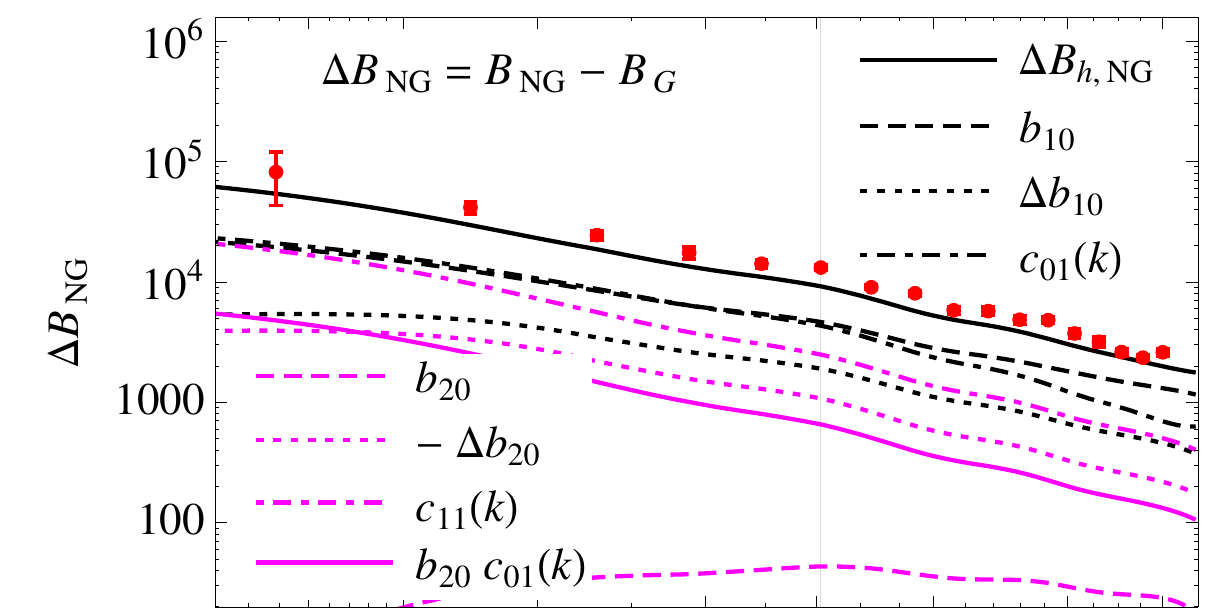}}
{\includegraphics[width=0.48\textwidth]{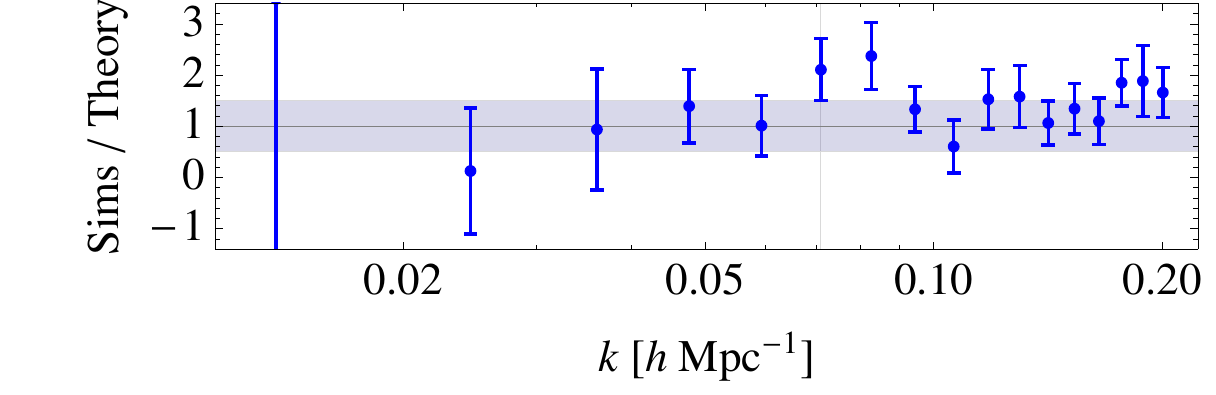}}
{\includegraphics[width=0.48\textwidth]{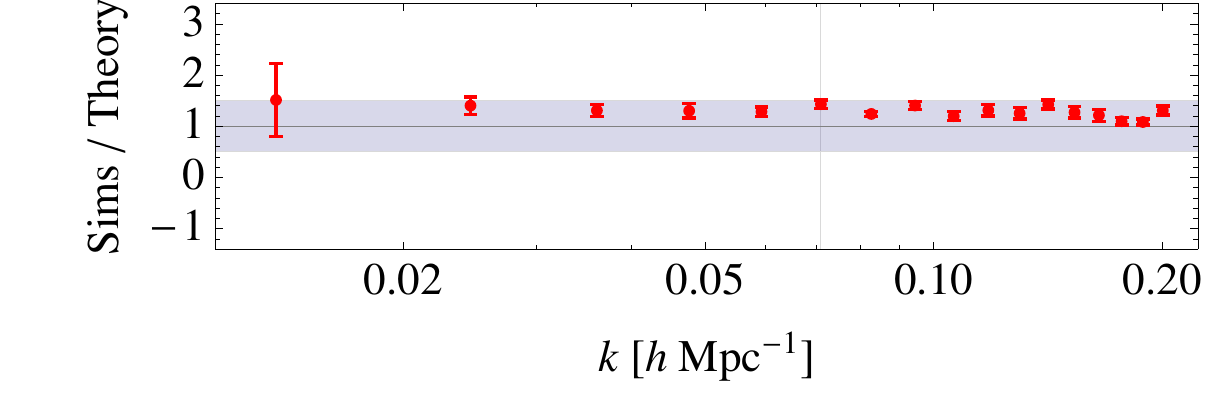}}
{\includegraphics[width=0.48\textwidth]{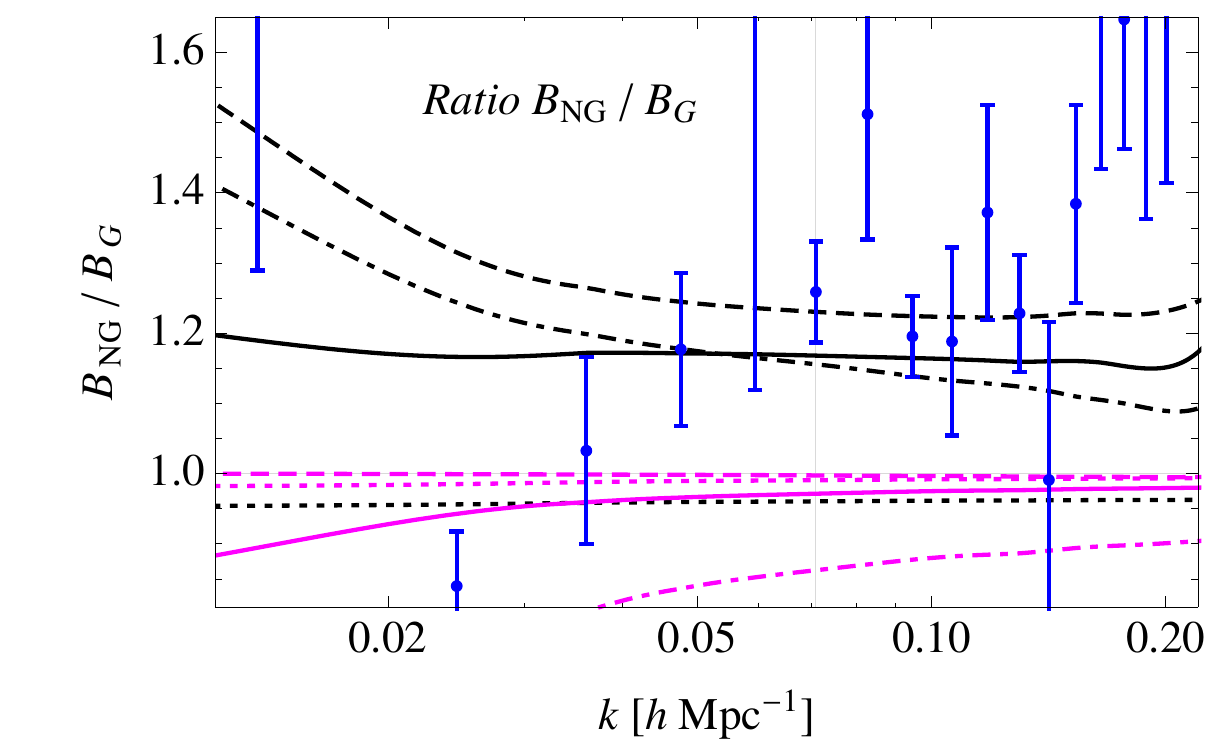}}
{\includegraphics[width=0.48\textwidth]{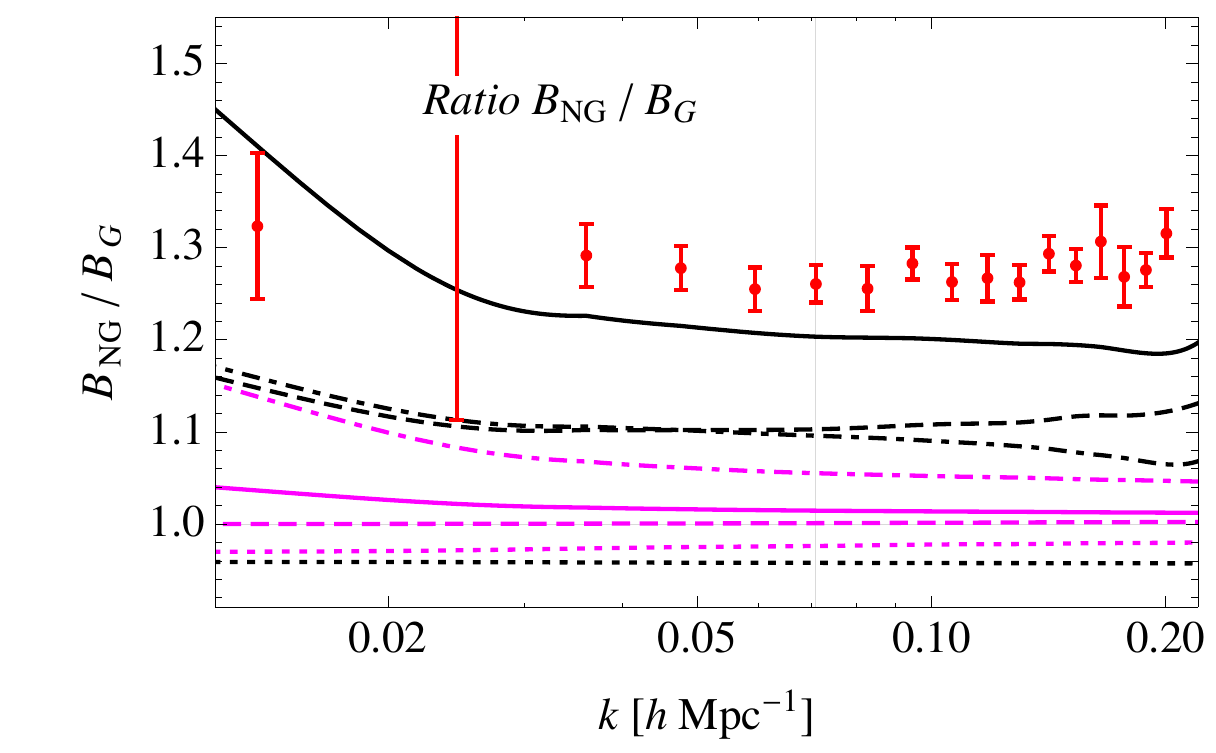}}
{\includegraphics[width=0.48\textwidth]{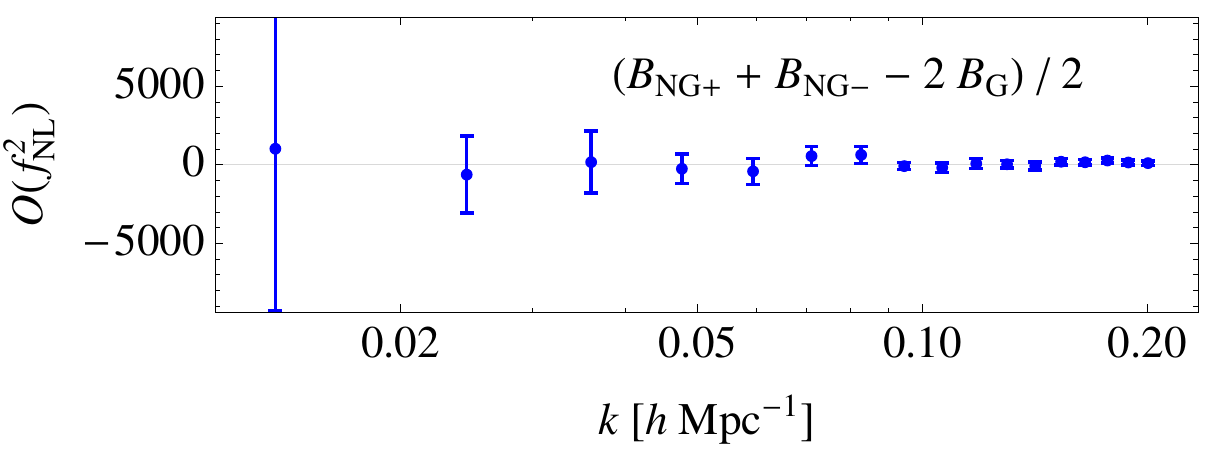}}
{\includegraphics[width=0.48\textwidth]{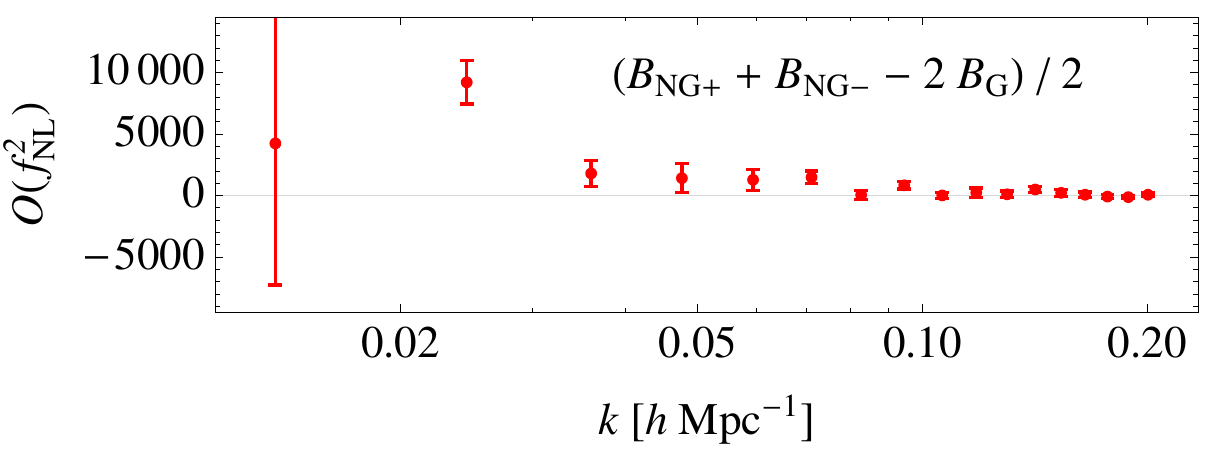}}
\caption{Equilateral configurations of the halo bispectrum, $B_{h}(\d k,k,k)$, compared with the theoretical prediction assuming the best fit values for the bias parameters $b_1$ and $b_2$. High mass bin ({\em right}).}
\label{fig:bhGsq}
\end{center}
\end{figure}

\begin{figure}[!p]
\begin{center}
\begin{center}{\bf Generic configurations (I)}, $B_{h}(k_1,k_2,\theta)$, $k_1=0.07\kMpc$, $k_2=0.08\kMpc$\end{center}\vspace{0.2cm}
{\includegraphics[width=0.48\textwidth]{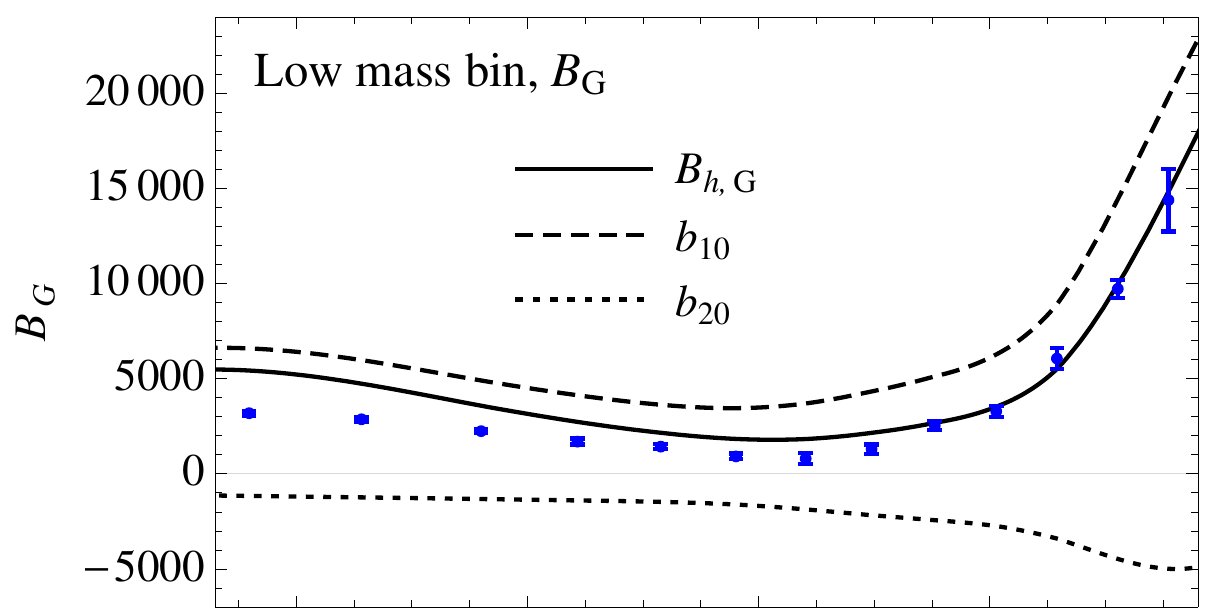}}
{\includegraphics[width=0.48\textwidth]{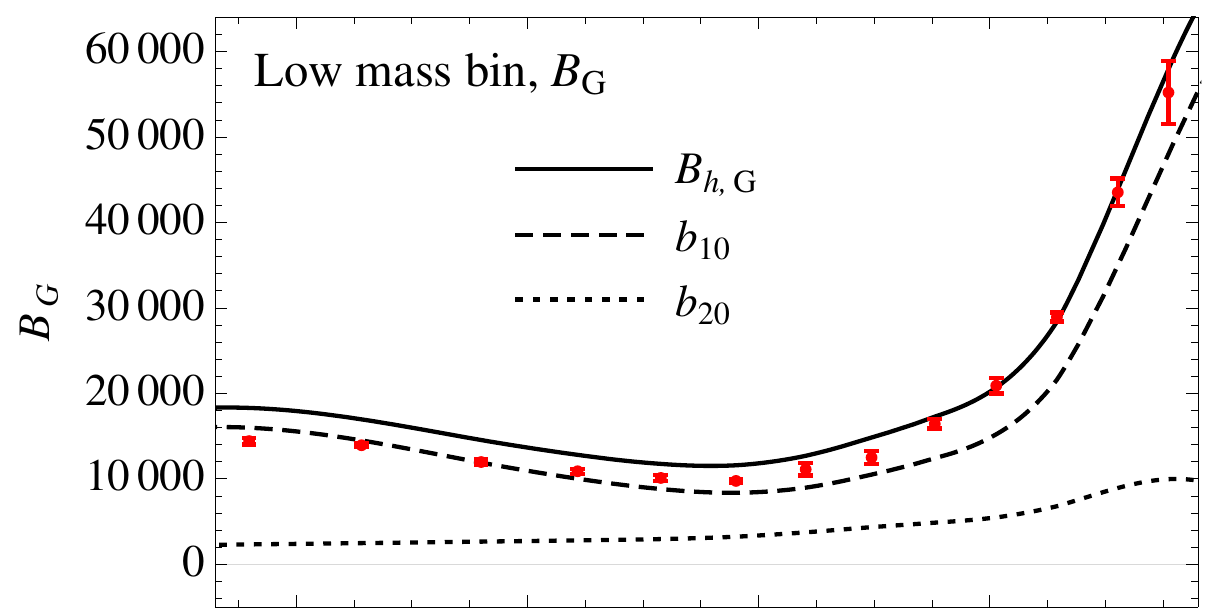}}
{\includegraphics[width=0.48\textwidth]{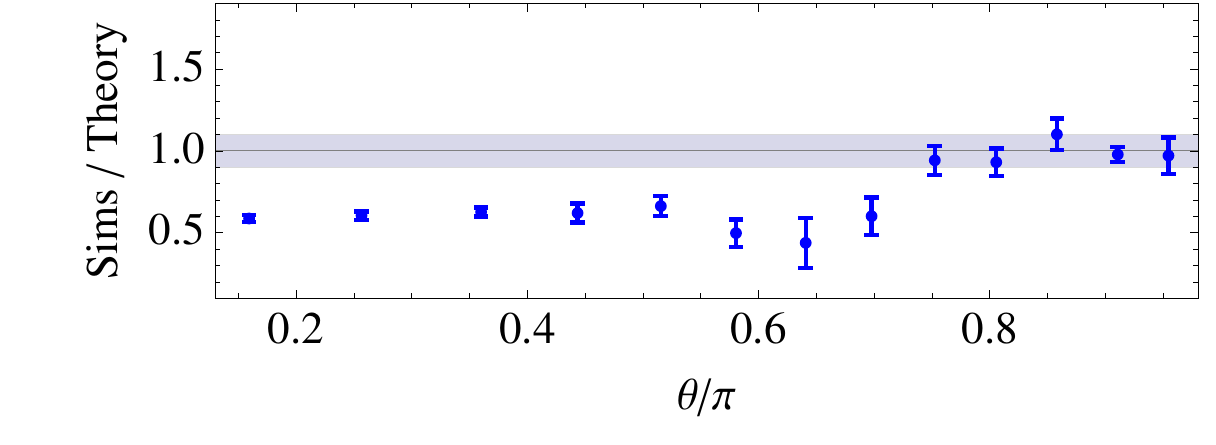}}
{\includegraphics[width=0.48\textwidth]{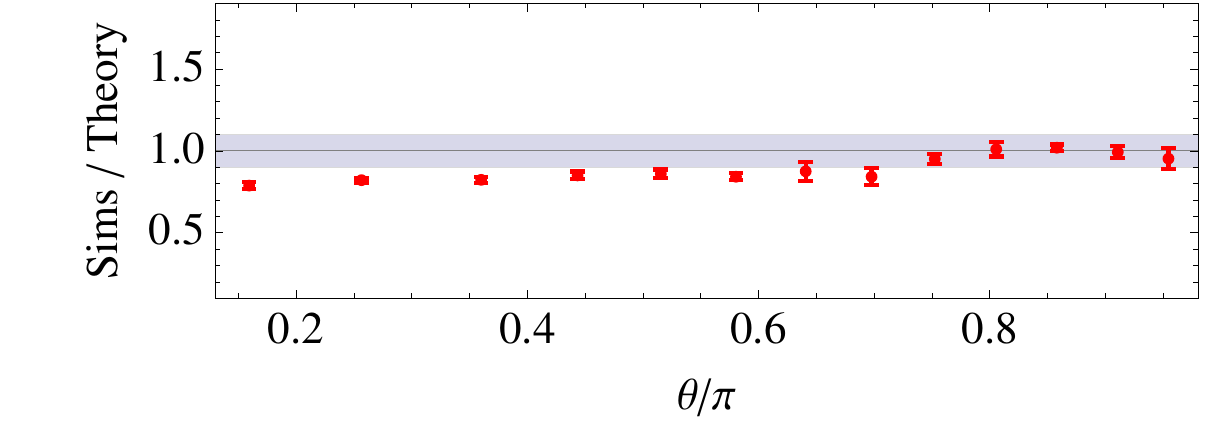}}
{\includegraphics[width=0.48\textwidth]{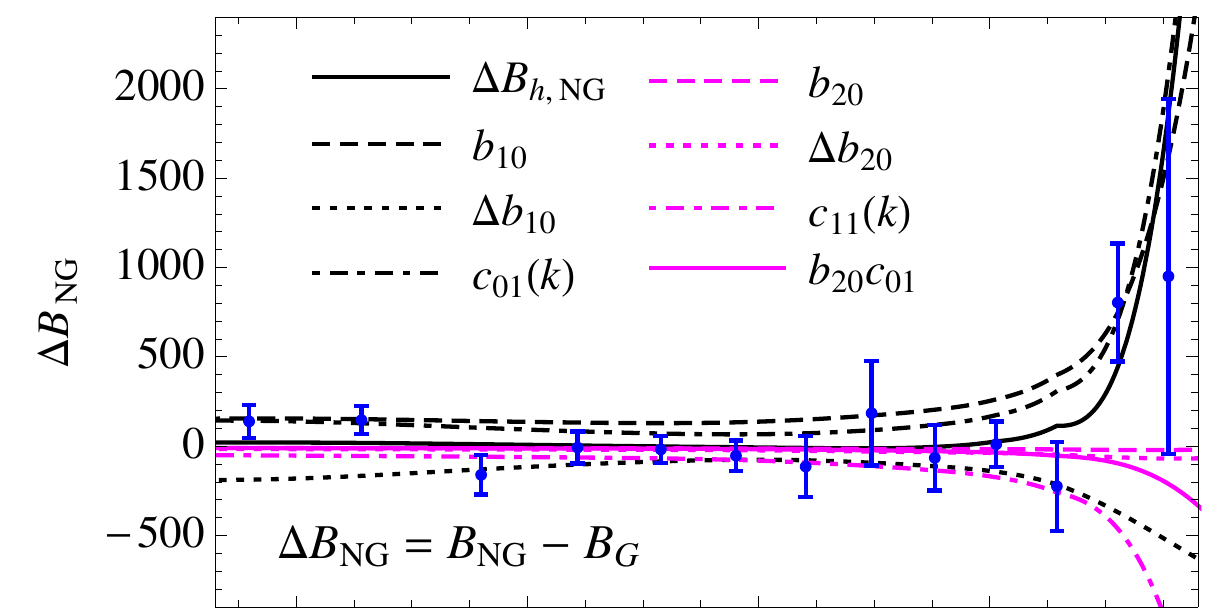}}
{\includegraphics[width=0.48\textwidth]{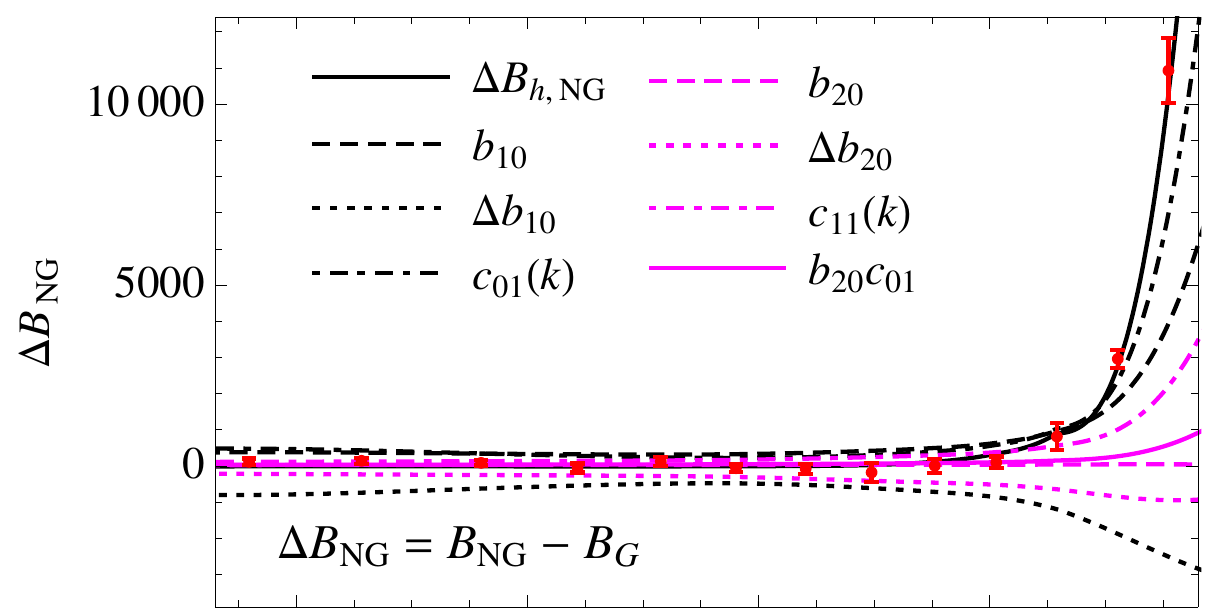}}
{\includegraphics[width=0.48\textwidth]{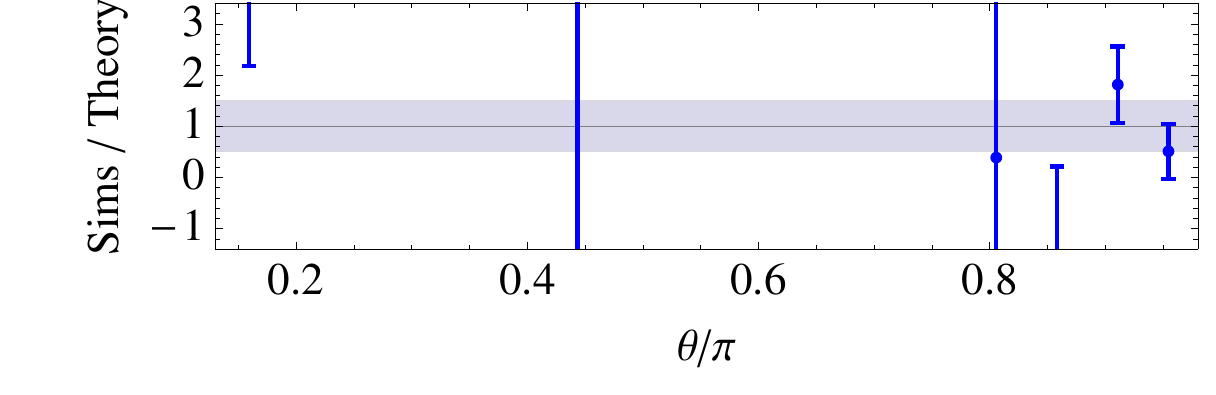}}
{\includegraphics[width=0.48\textwidth]{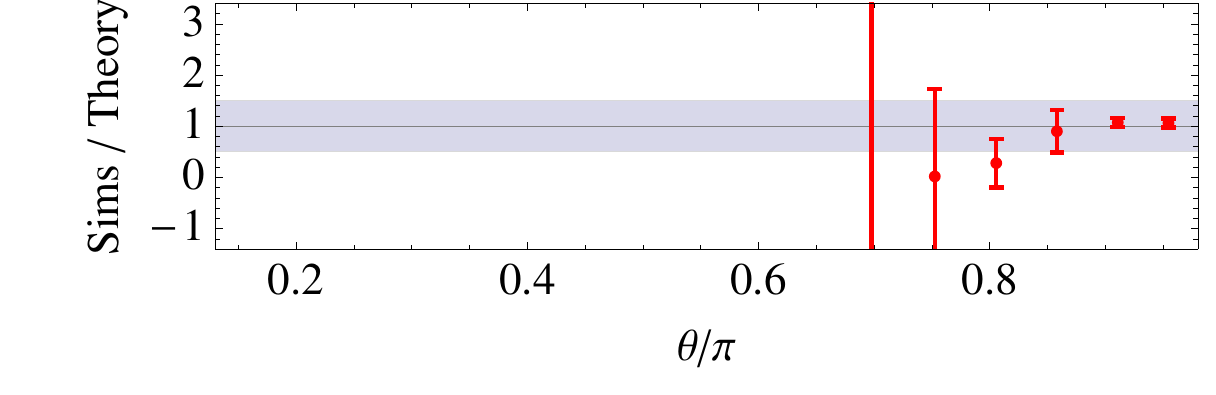}}
{\includegraphics[width=0.48\textwidth]{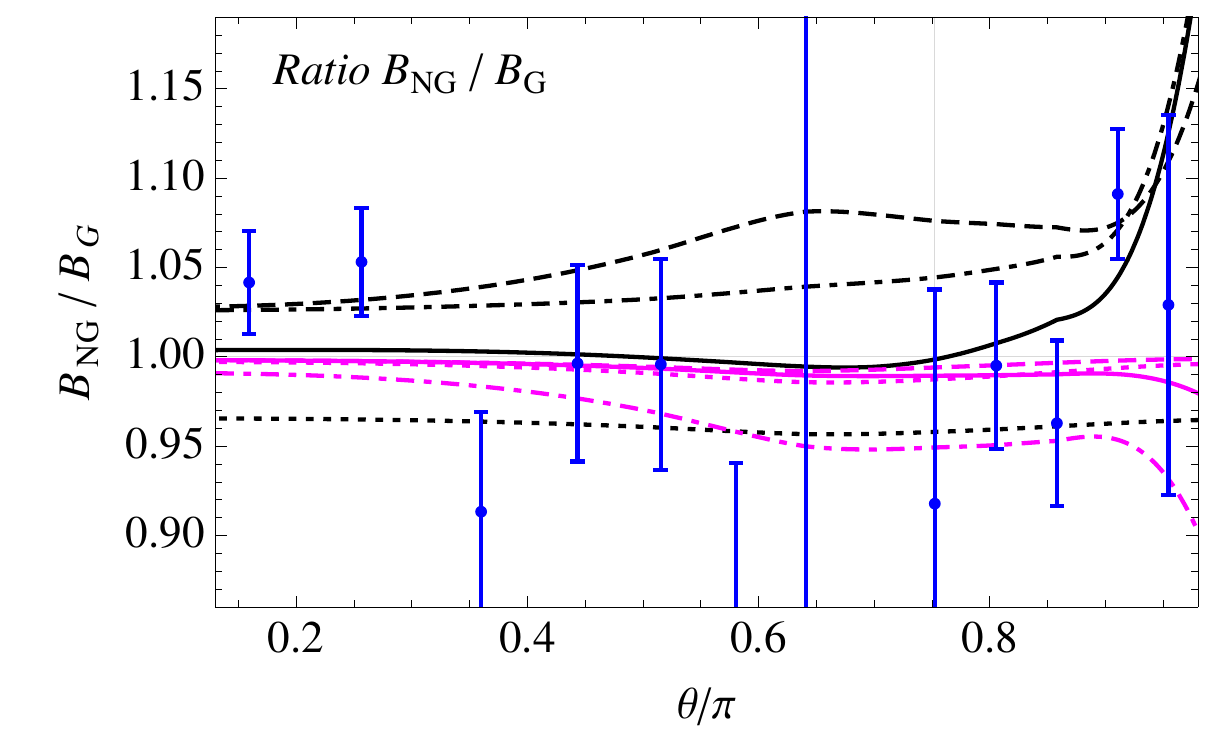}}
{\includegraphics[width=0.48\textwidth]{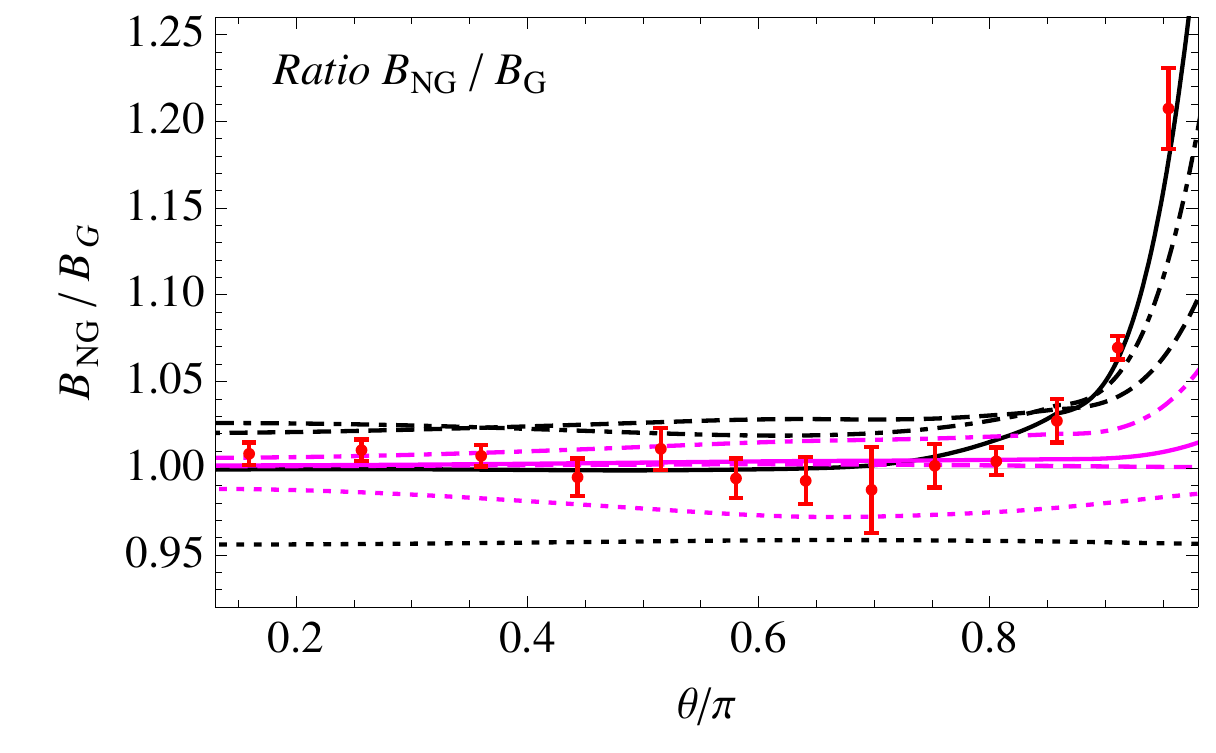}}
{\includegraphics[width=0.48\textwidth]{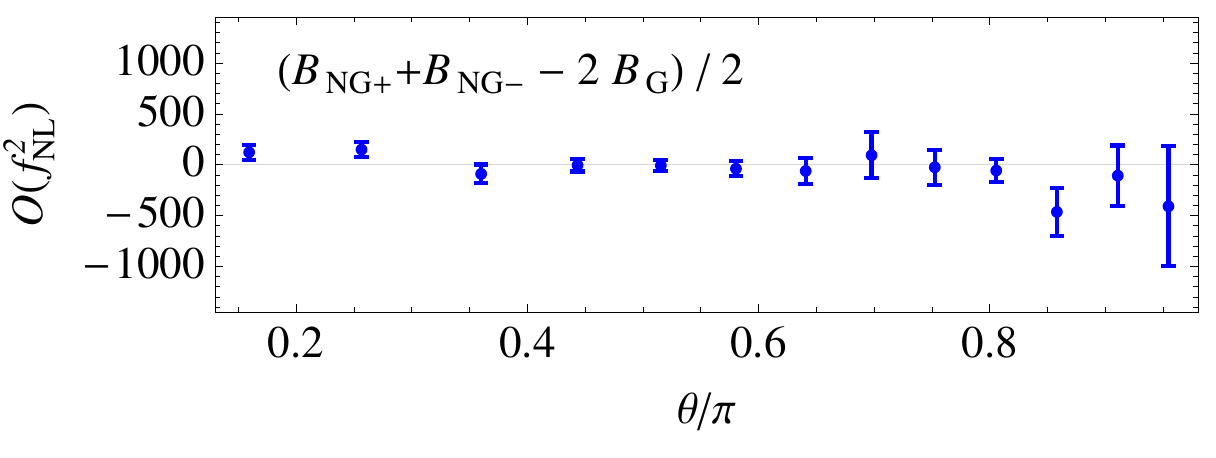}}
{\includegraphics[width=0.48\textwidth]{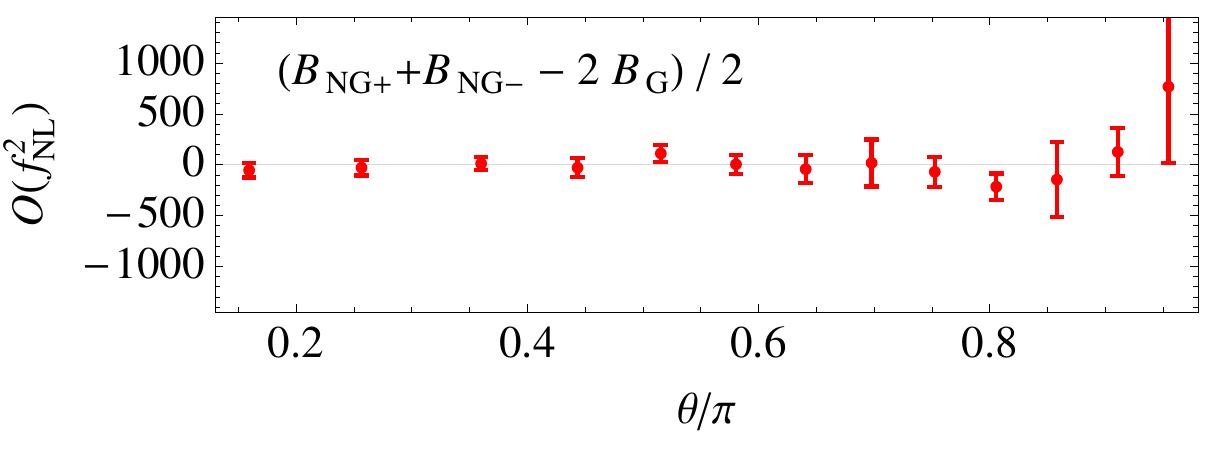}}
\caption{Equilateral configurations of the halo bispectrum, $B_{h}(k,k,k)$, compared with the theoretical prediction assuming the best fit values for the bias parameters $b_1$ and $b_2$.}
\label{fig:bhGg1}
\end{center}
\end{figure}

\begin{figure}[!p]
\begin{center}
\begin{center}{\bf Generic configurations (II)}, $B_{h}(k_1,k_2,\theta)$, $k_1=0.05\kMpc$, $k_2=0.07\kMpc$\end{center}\vspace{0.2cm}
{\includegraphics[width=0.48\textwidth]{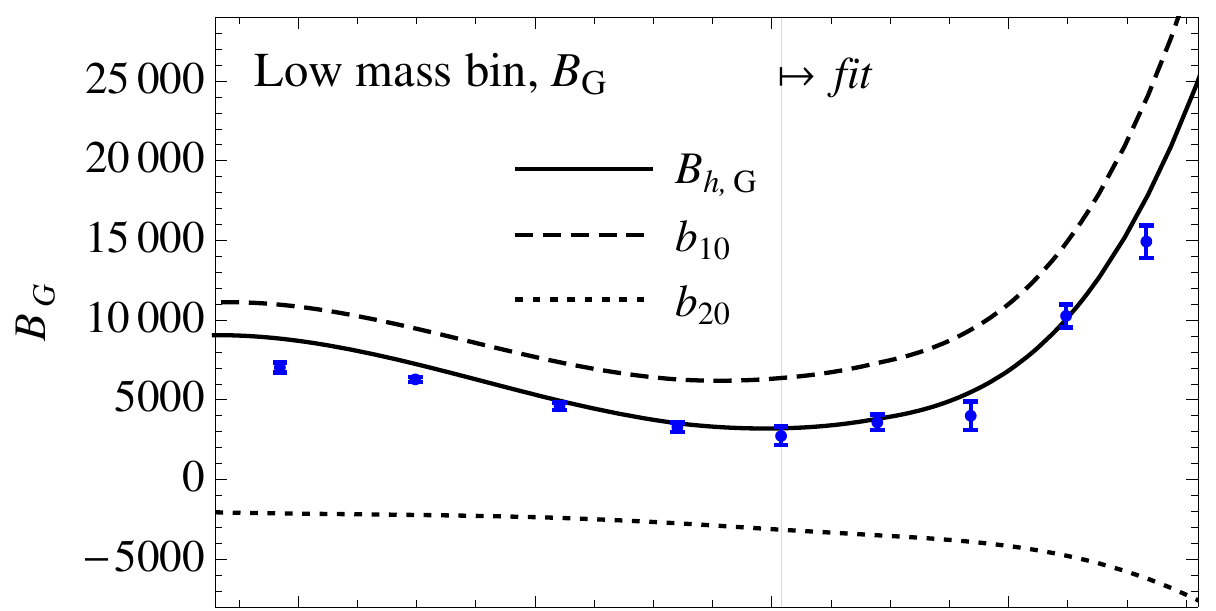}}
{\includegraphics[width=0.48\textwidth]{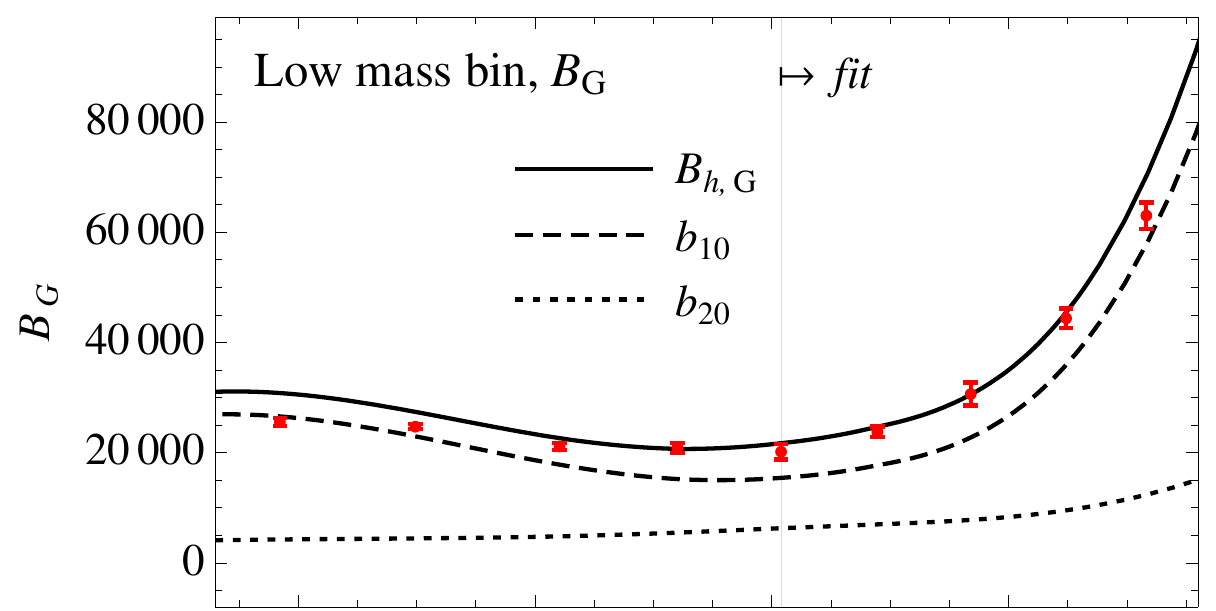}}
{\includegraphics[width=0.48\textwidth]{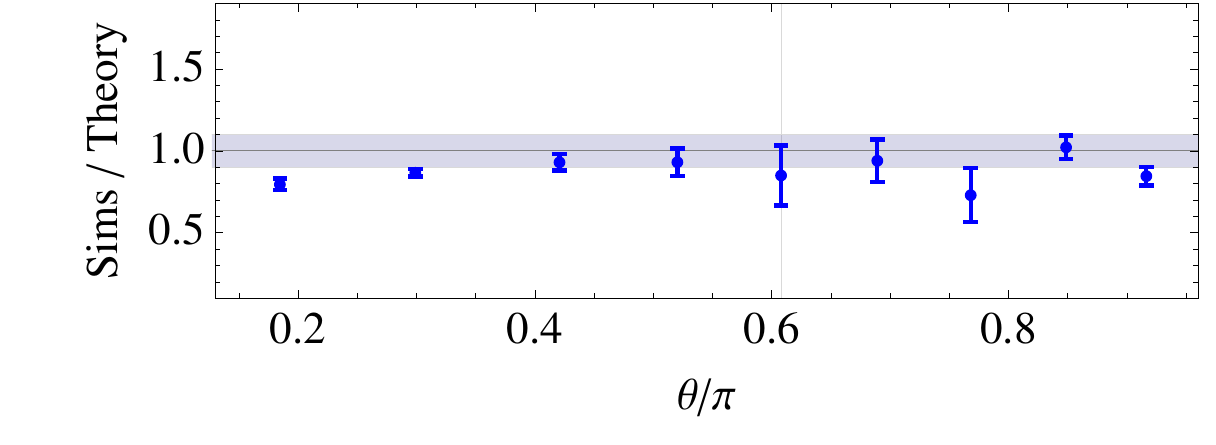}}
{\includegraphics[width=0.48\textwidth]{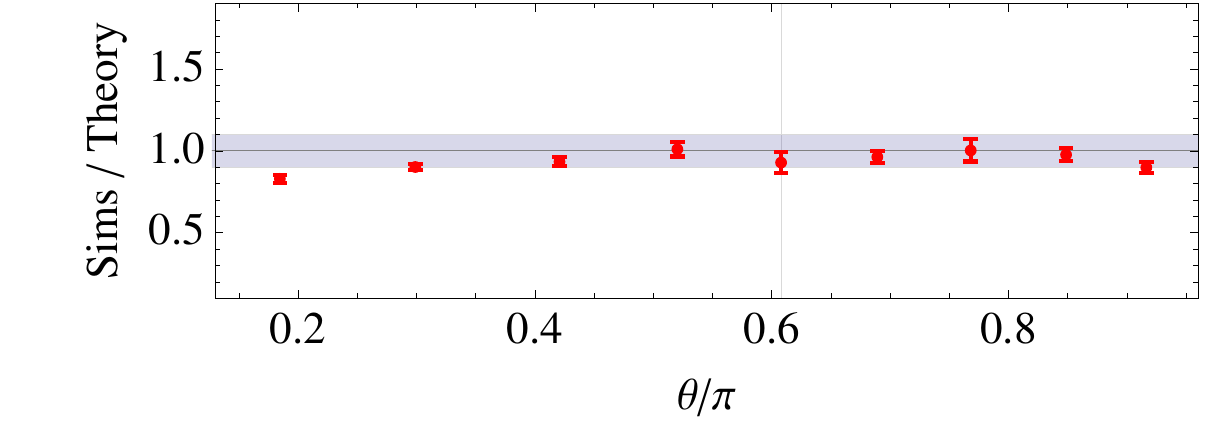}}
{\includegraphics[width=0.48\textwidth]{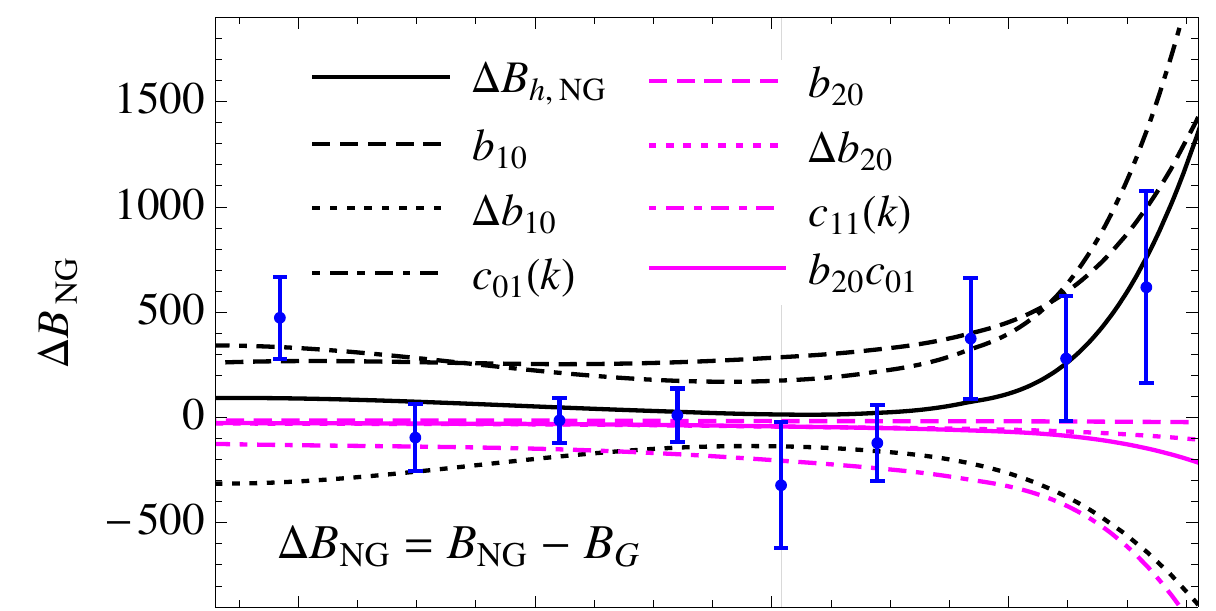}}
{\includegraphics[width=0.48\textwidth]{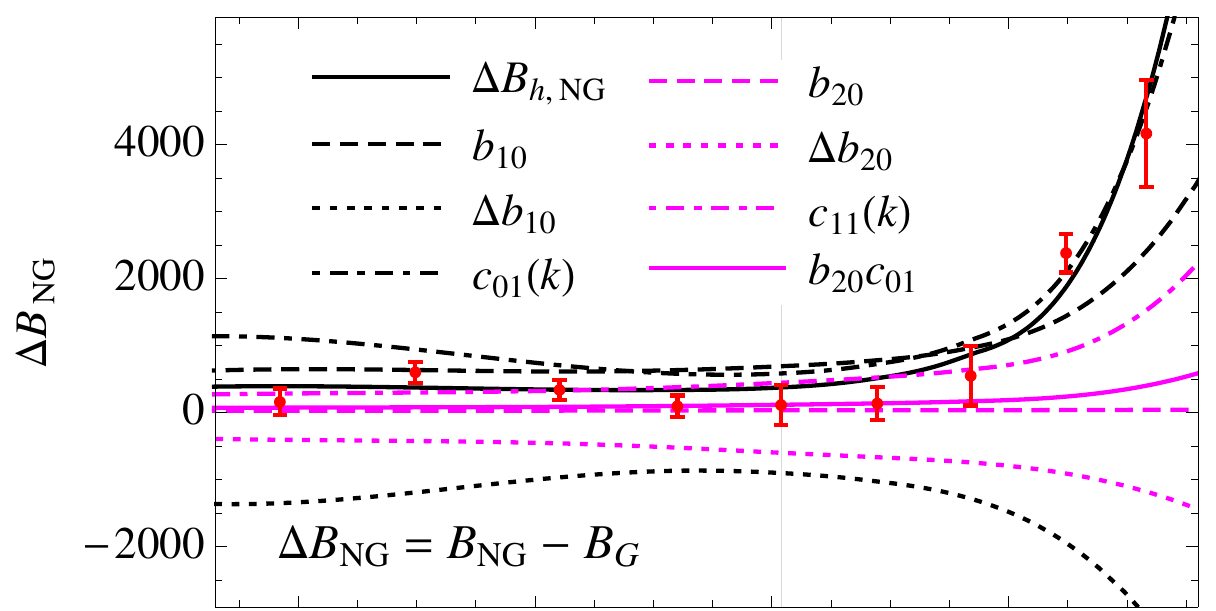}}
{\includegraphics[width=0.48\textwidth]{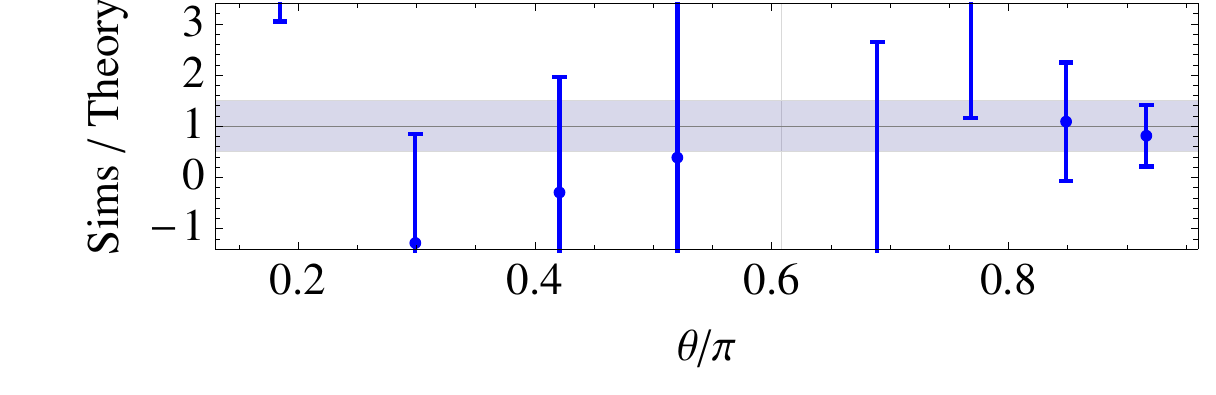}}
{\includegraphics[width=0.48\textwidth]{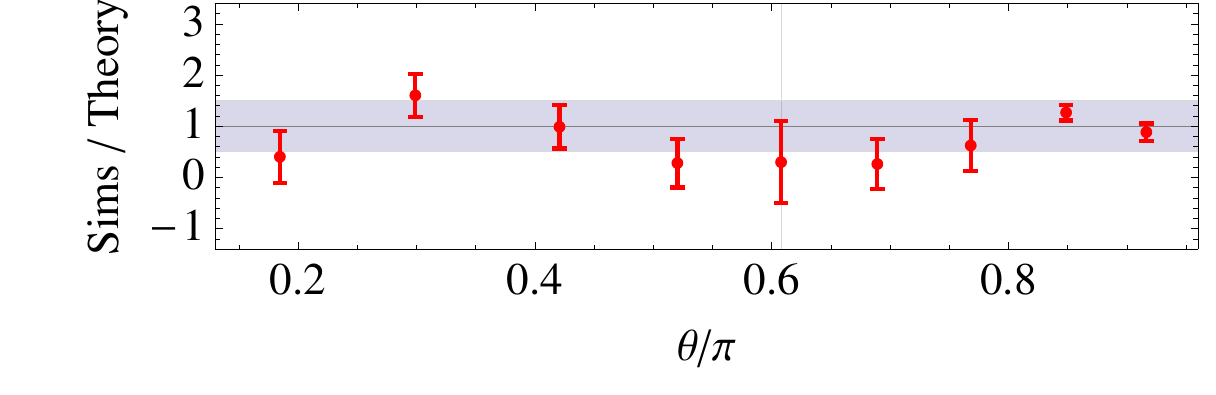}}
{\includegraphics[width=0.48\textwidth]{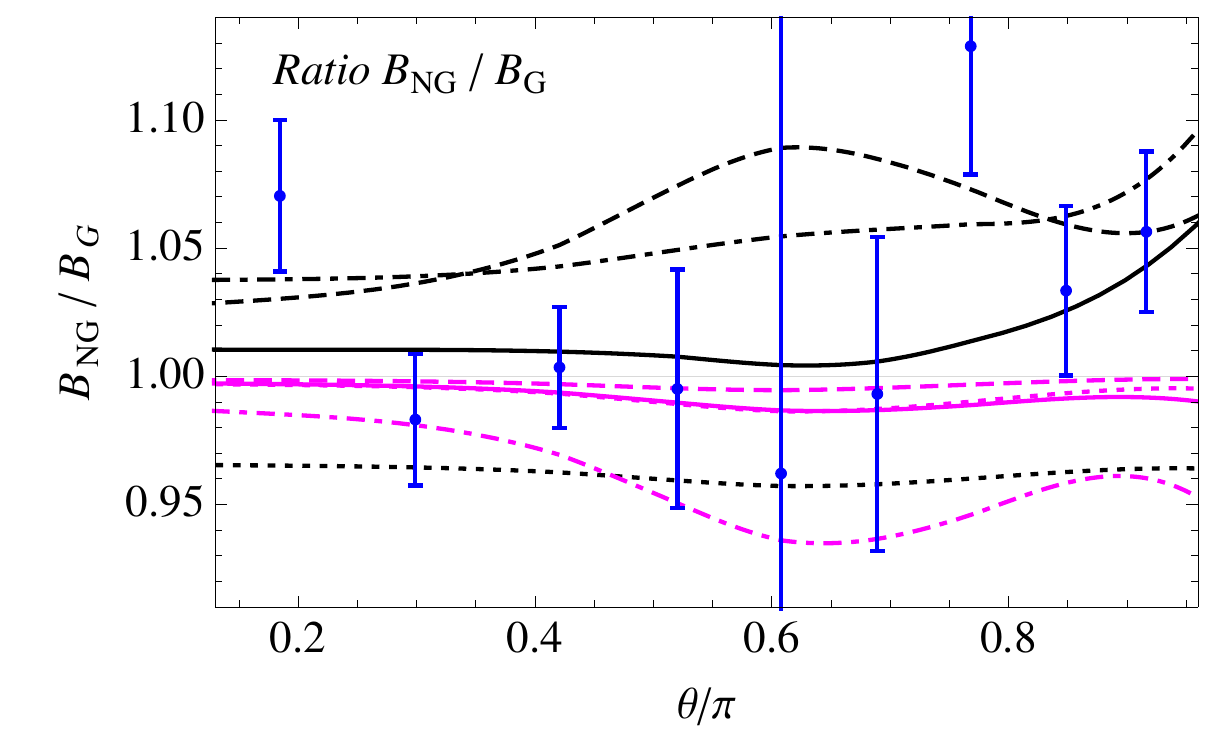}}
{\includegraphics[width=0.48\textwidth]{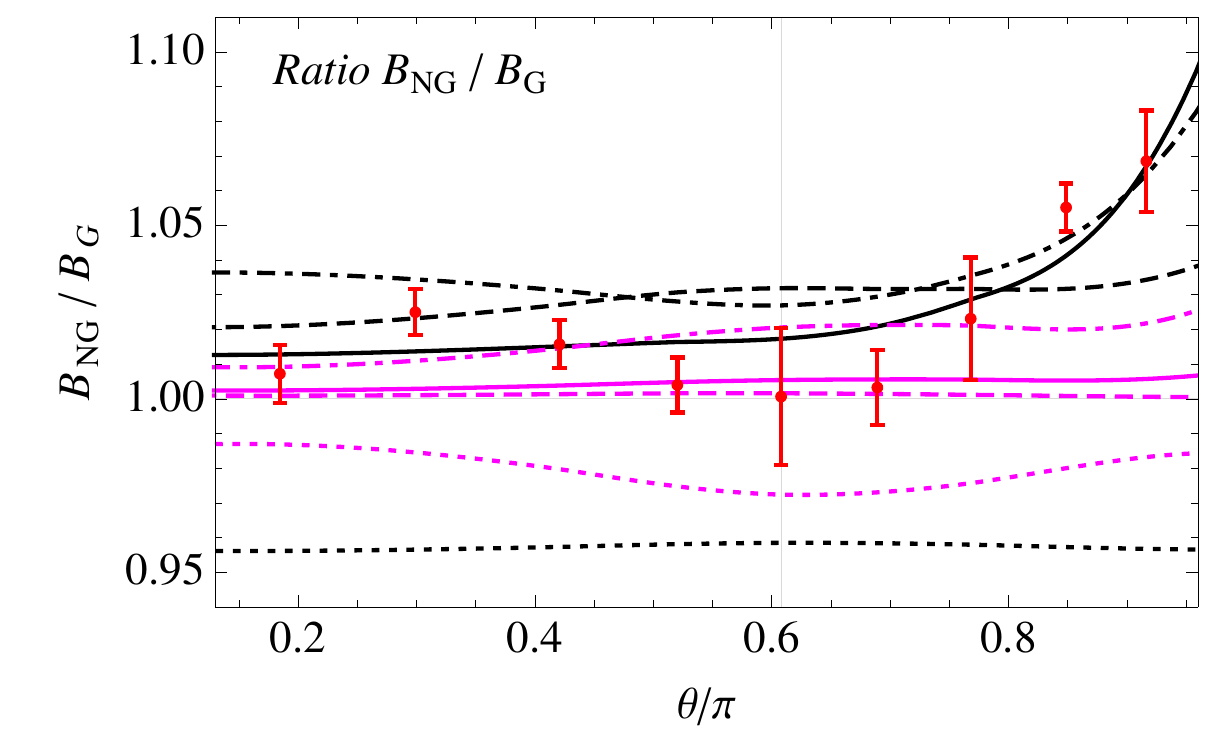}}
{\includegraphics[width=0.48\textwidth]{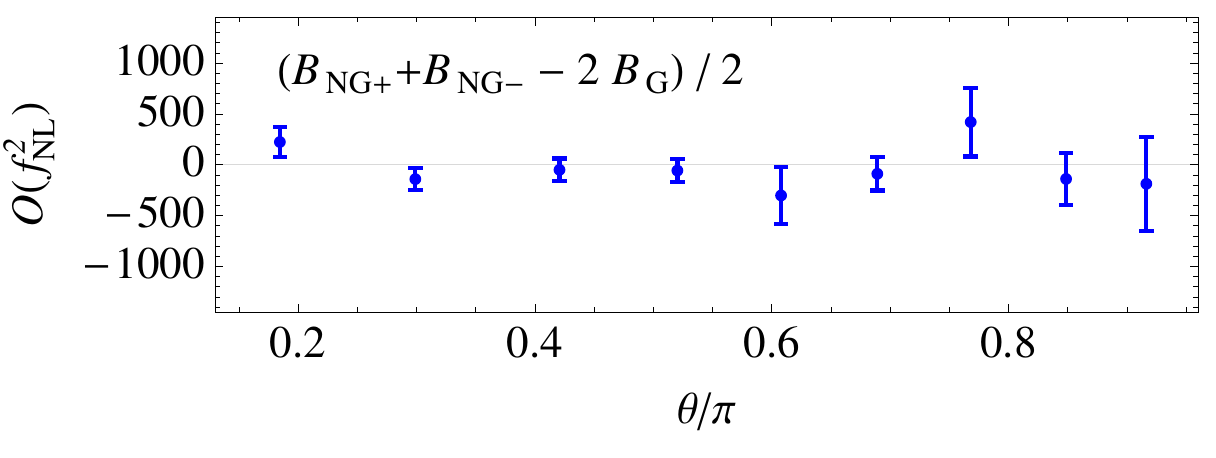}}
{\includegraphics[width=0.48\textwidth]{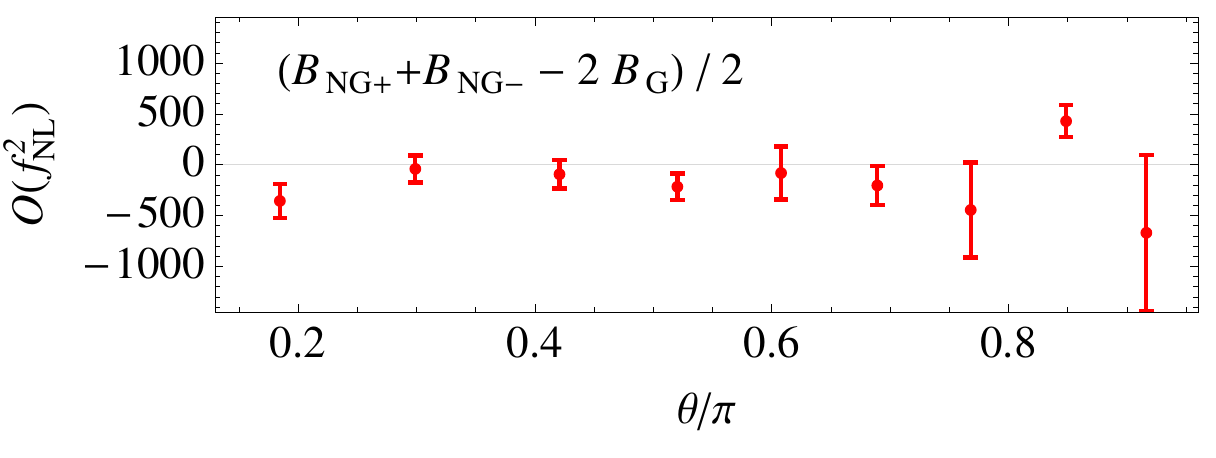}}
\caption{Equilateral configurations of the halo bispectrum, $B_{h}(k,k,k)$, compared with the theoretical prediction assuming the best fit values for the bias parameters $b_1$ and $b_2$.}
\label{fig:bhGg2}
\end{center}
\end{figure}

\begin{figure}[!p]
\begin{center}
\begin{center}{\bf Equilateral configurations}, $B_{h}(k,k,k)$\end{center}\vspace{0.2cm}
{\includegraphics[width=0.48\textwidth]{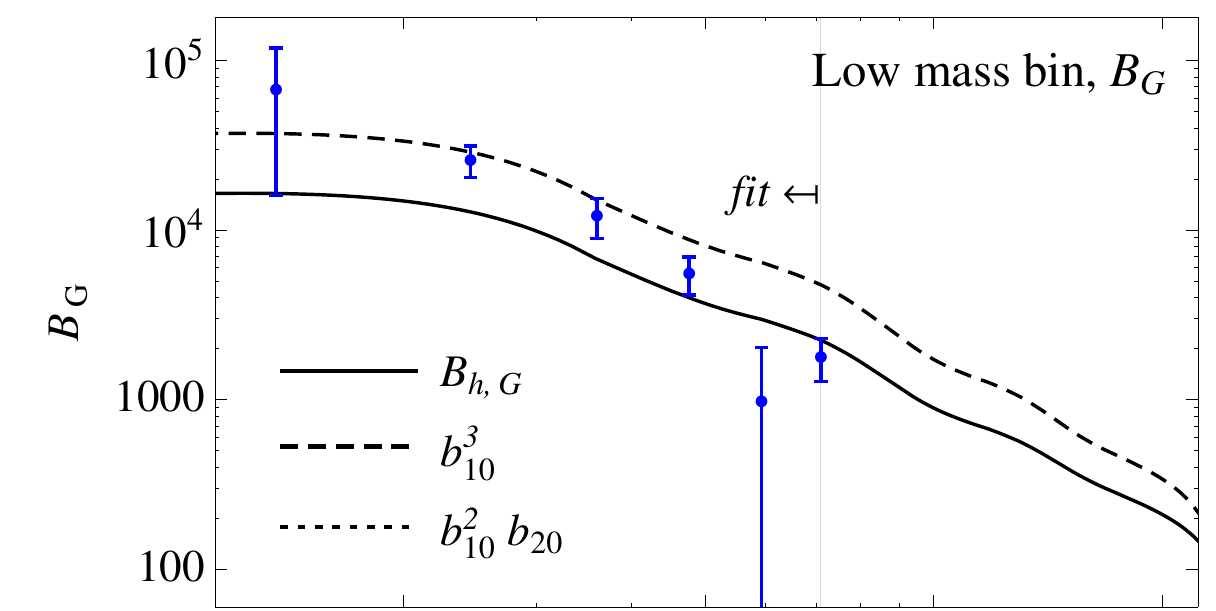}}
{\includegraphics[width=0.48\textwidth]{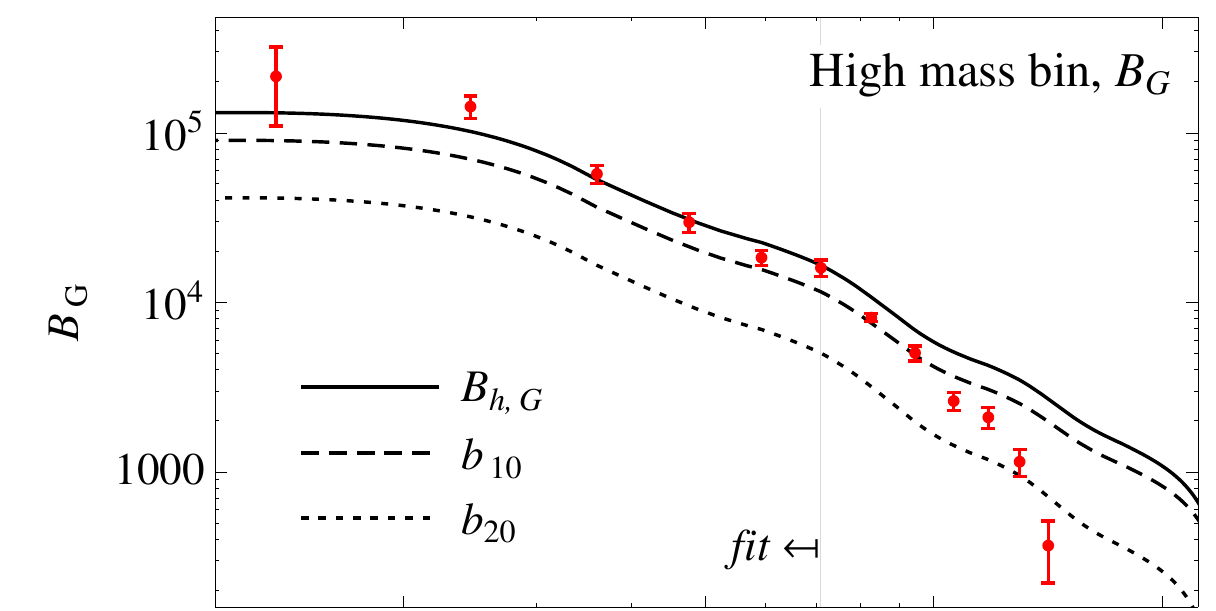}}
{\includegraphics[width=0.48\textwidth]{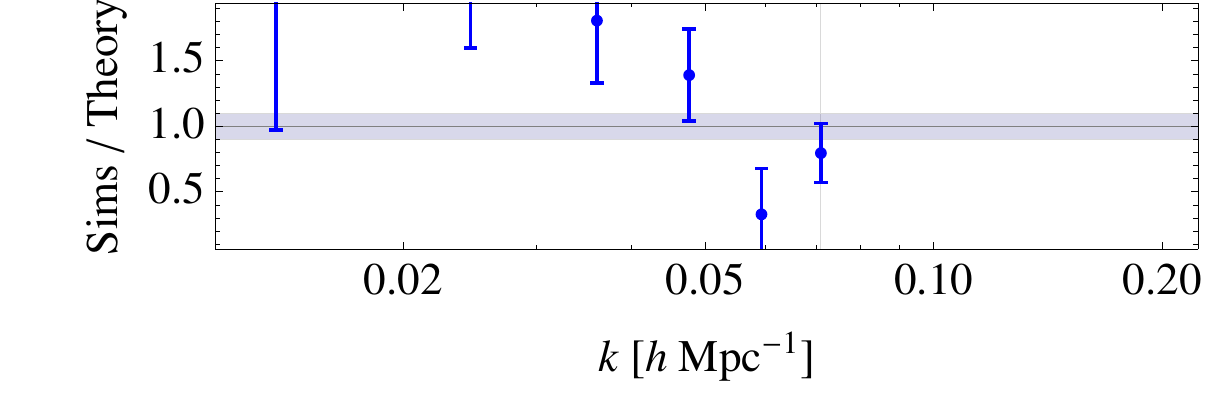}}
{\includegraphics[width=0.48\textwidth]{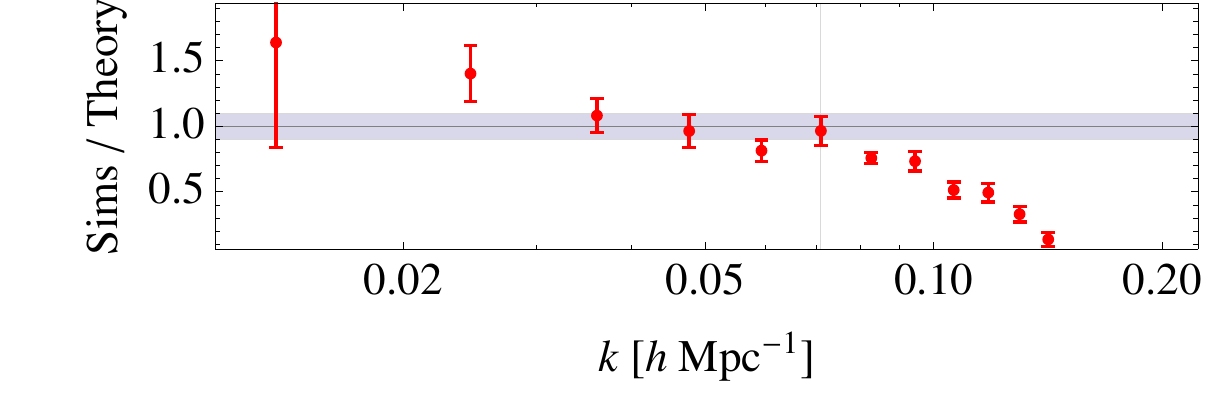}}
{\includegraphics[width=0.48\textwidth]{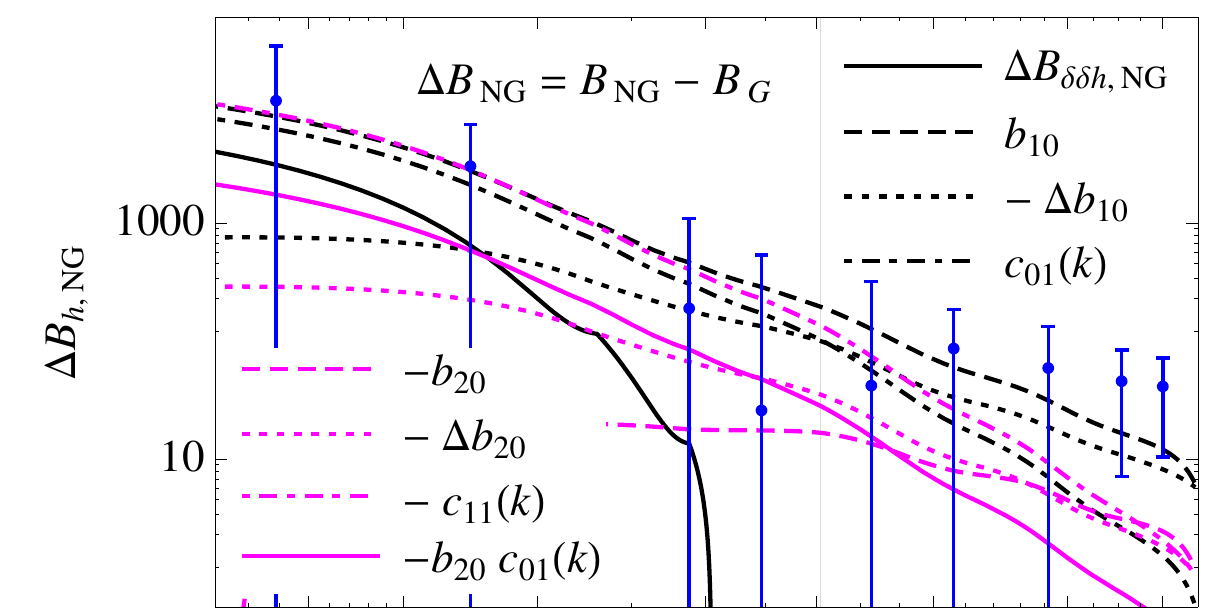}}
{\includegraphics[width=0.48\textwidth]{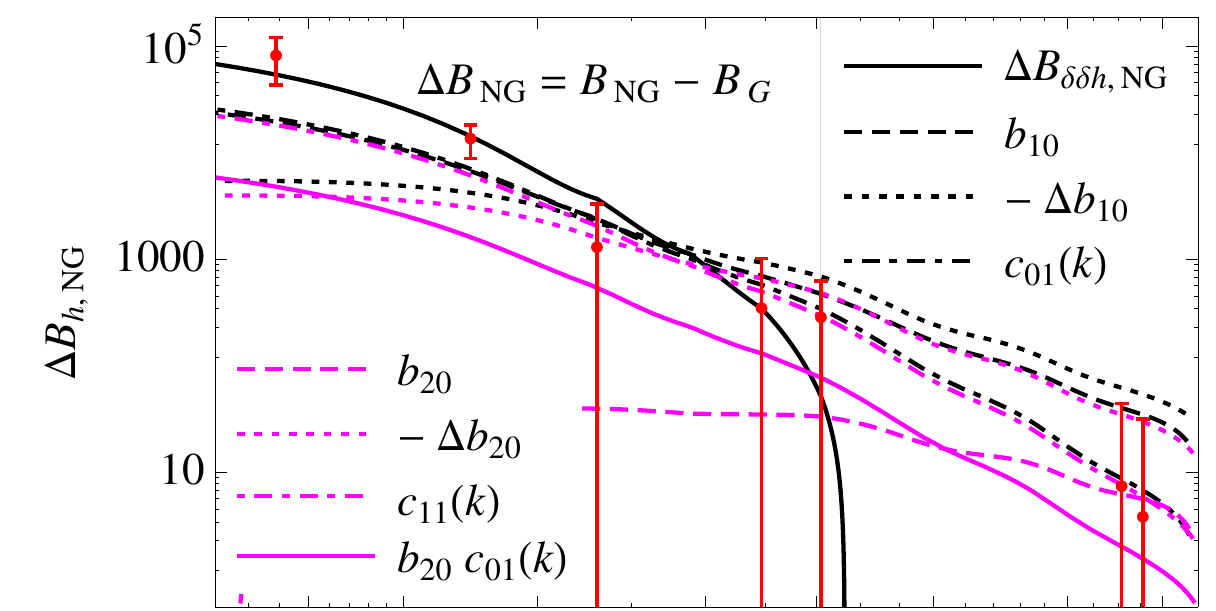}}
{\includegraphics[width=0.48\textwidth]{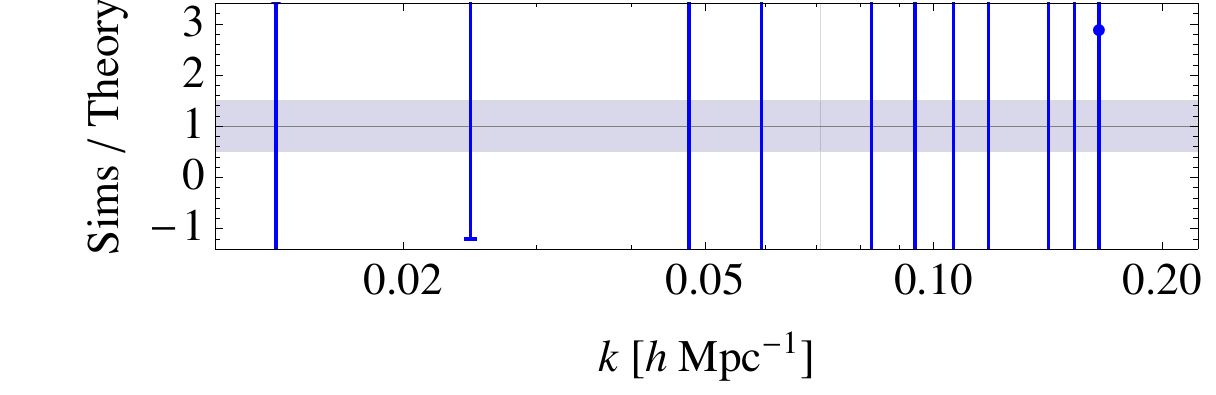}}
{\includegraphics[width=0.48\textwidth]{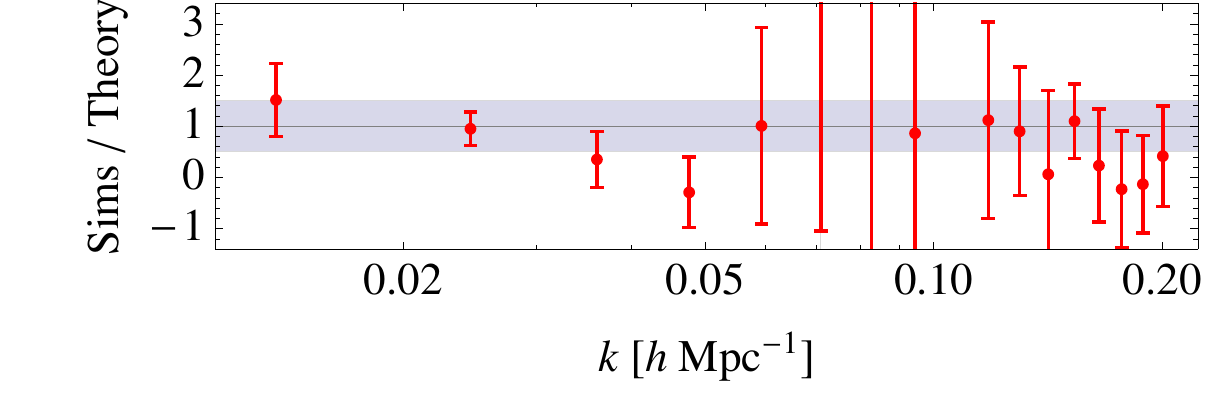}}
{\includegraphics[width=0.48\textwidth]{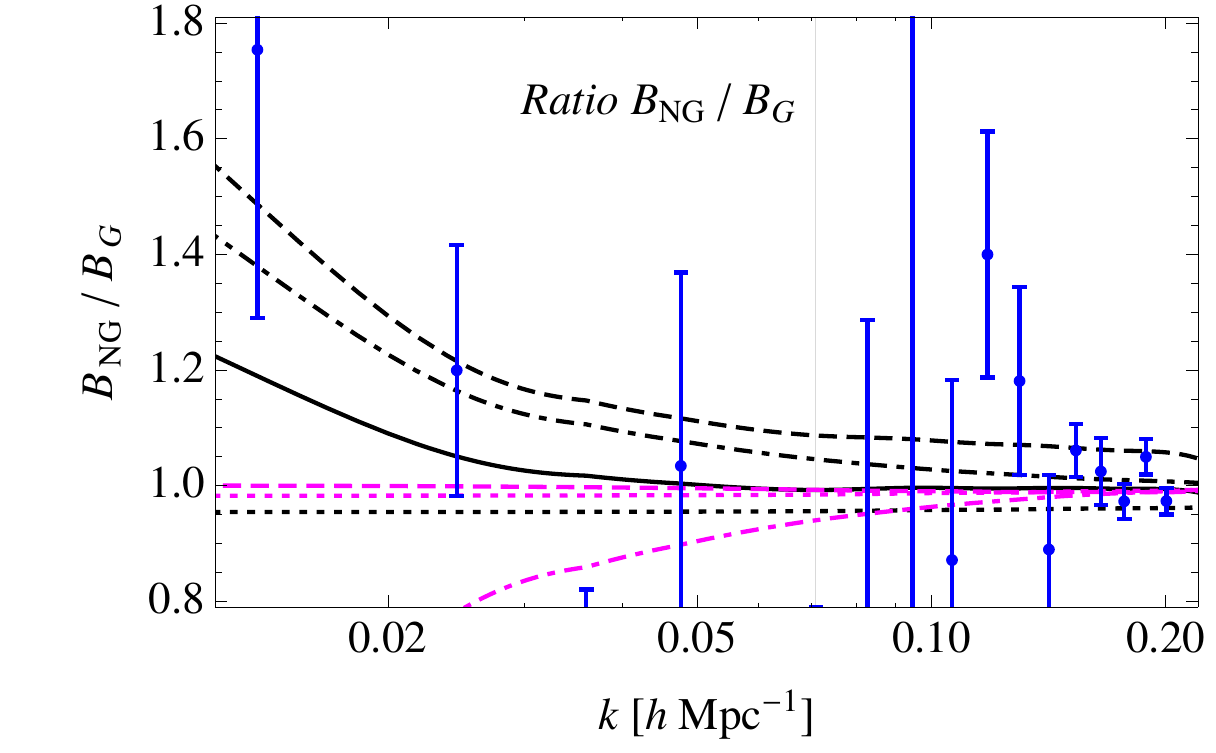}}
{\includegraphics[width=0.48\textwidth]{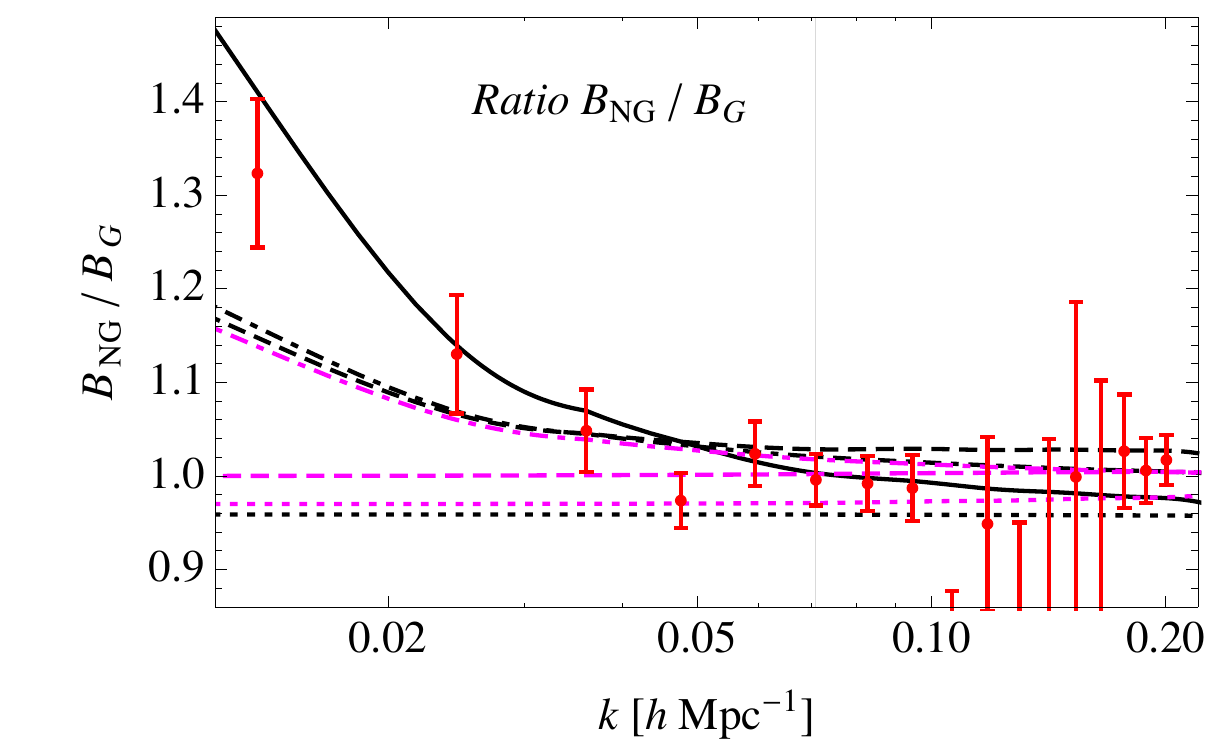}}
{\includegraphics[width=0.48\textwidth]{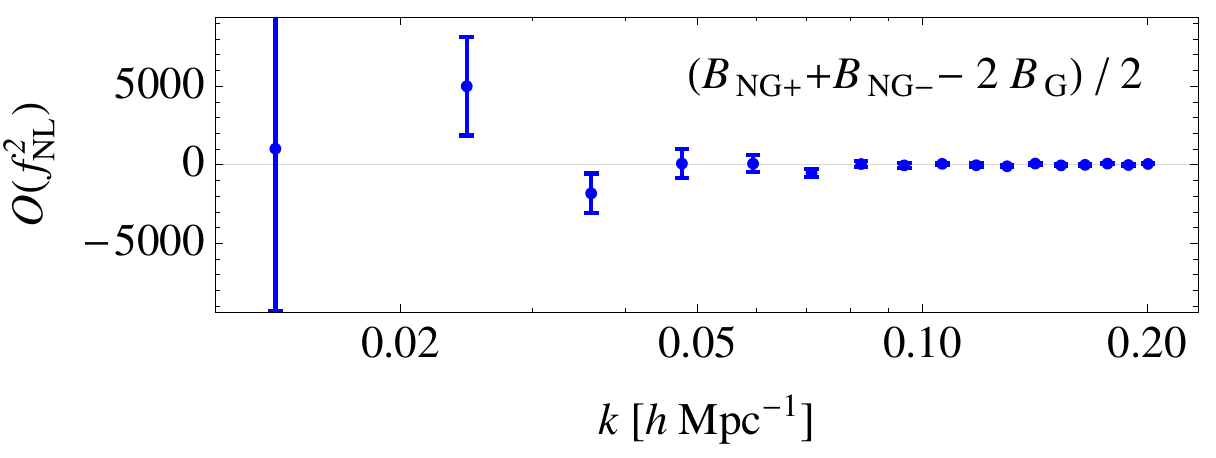}}
{\includegraphics[width=0.48\textwidth]{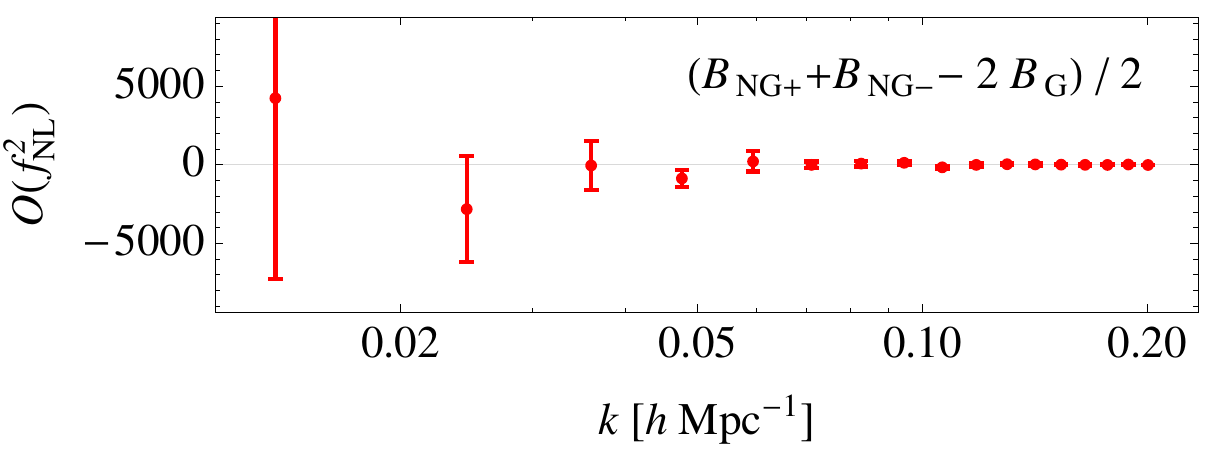}}
\caption{Equilateral configurations of the halo bispectrum, $B_{h}(k,k,k)$, compared with the theoretical prediction assuming the best fit values for the bias parameters $b_1$ and $b_2$.}
\label{fig:bhGeq}
\end{center}
\end{figure}

Figures \ref{fig:bhGsq}, \ref{fig:bhGg1}, \ref{fig:bhGg2} and \ref{fig:bhGeq} show the same subsets of triangular configurations for the halo bispectrum $B_h$. Most of the comments made for the cross-bispectrum $B_{\d\d h}$ can be repeated here. For Gaussian initial conditions, the simple tree-level, local bias prescription provides a model accurate at the 10\% level for generic triangles, the largest deviations occurring for equilateral configurations. In the halo bias case, however, the measurements are considerably noisier than those of the cross matter-matter-halo bispectrum.

The notation adopted to identify each term in the expressions for the Gaussian and non-Gaussian components on the plots is the same as in the previous figures, although additional dependencies on the Gaussian linear bias parameter $b_{10}$ are now present. For Gaussian initial conditions for instance, the $b_{10,G}^2\,b_{20,G}\,P_{\d,G}(k_1)\,P_{\d,G}(k_2)+{\rm cyc.}$ contribution is simply denoted by $b_{20}$ in the legend of the plot. There is one non-trivial additional contribution in the non-Gaussian correction $\D B_{h,NG}$, given by the sixth term on the r.h.s.~of Eq.~(\ref{eq:dBhNG}). It is shown as a continuous magenta (or light gray) curve and labelled $b_{20}\,c_{01}(k)$ in the legends, as it depends both on the scale-dependent linear bias correction {\em and} the quadratic bias.

We remark in the first place that, in our model, the overall non-Gaussian correction to the halo bias $\D B_{h,NG}$ is the sum of several contribution of nearly equal importance. This is particularly evident in the case of squeezed triangles and for the low mass halos (Fig.~\ref{fig:bhGsq}, left column), where all quadratic bias terms are negative and the overall signal thus is the result of large cancellations (with the caveat that for these configurations and mass bin in particular the model shows a relatively larger discrepancy w.r.t. simulations than for other triangles). For the same triangular configurations and for the high mass halos, only the $\D b_{20}$ term is negative and the largest corrections are due to the matter bispectrum and the scale-dependent linear bias correction, the quadratic term proportional to $c_{11}(k)$ is also quite relevant, particularly at the largest scales probed. We notice that the model under-predicts the measurements, in the large mass bin, by about 20\%, as is also evident from the non-Gaussian to Gaussian ratio plots. At the same time, however, it over-predicts by roughly the same amount the measured halo bispectrum for triangles of sides $\left\{\D k, k-\D k, k\right\}$ (not shown), very close to the squeezed configurations considered here. These differences might be due to binning effects to be explored in future works. A better agreement between the model and simulations is found for nearly all generic configurations as can be seen in Fig.~\ref{fig:bhGg1} and \ref{fig:bhGg2}. The plot of the non-Gaussian correction $\D B_{h,NG}$ for the large mass halos ({\em right column, third row of Fig.}~\ref{fig:bhGg2}) is particularly illustrative of the situation, where almost all contributions have a similar absolute values in the $\theta\rightarrow \pi$ squeezed limit and several have a negative sign. These results suggest that it is quite difficult to describe the effects of PNG in terms of few basic corrections, or in terms of the scale-dependent corrections $c_{01}(k)$ and $c_{11}(k)$ alone.

Furthermore, the measurements of the halo bispectrum at low masses present a significant noise, particularly evident for equilateral configurations. We checked that such scatter is not due to any individual faulty realization, but seems to be proper to such halo population. 

Finally, note that $\O(\fNL^2)$ effects are present in the halo bispectrum squeezed configurations. In the high mass bin, a positive $\O(\fNL^2)$ correction is measured for triangles given by $\left\{\D k,k,k\right\}$ with $k\simeq 0.025\kMpc$, accounting for about 20\% the overall non-Gaussian effect $\D B_{h,NG}$. Such a correction is, however, not present in the small mass bin for the same triangles.

\section{Bias parameters}
\label{sec:bias}

\begin{figure}[!t]
\begin{center}
\begin{center} Gaussian halo bias\end{center}\vspace{0.2cm}
{\includegraphics[width=0.48\textwidth]{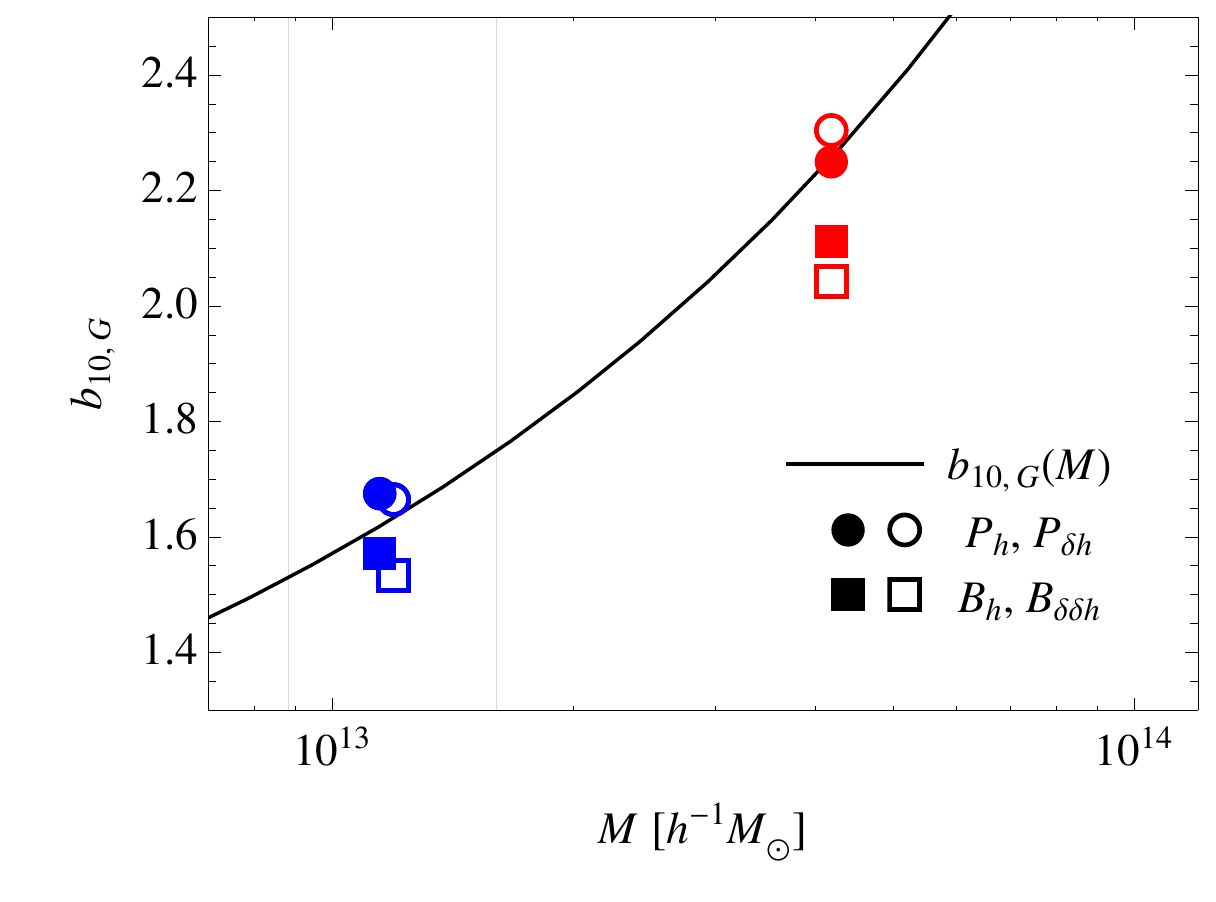}}
{\includegraphics[width=0.48\textwidth]{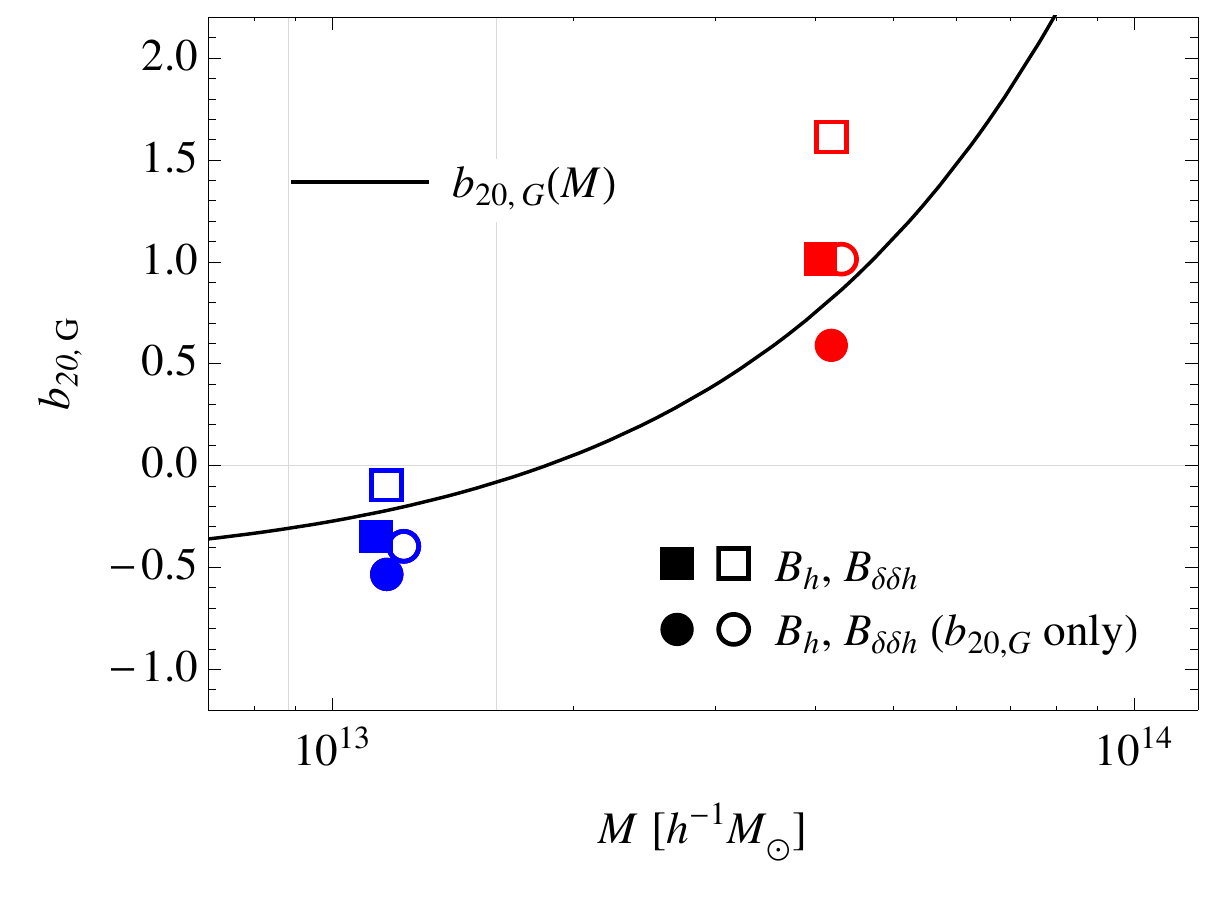}}
\begin{center} Non-Gaussian, scale-independent, halo bias corrections\end{center}\vspace{0.2cm}
{\includegraphics[width=0.48\textwidth]{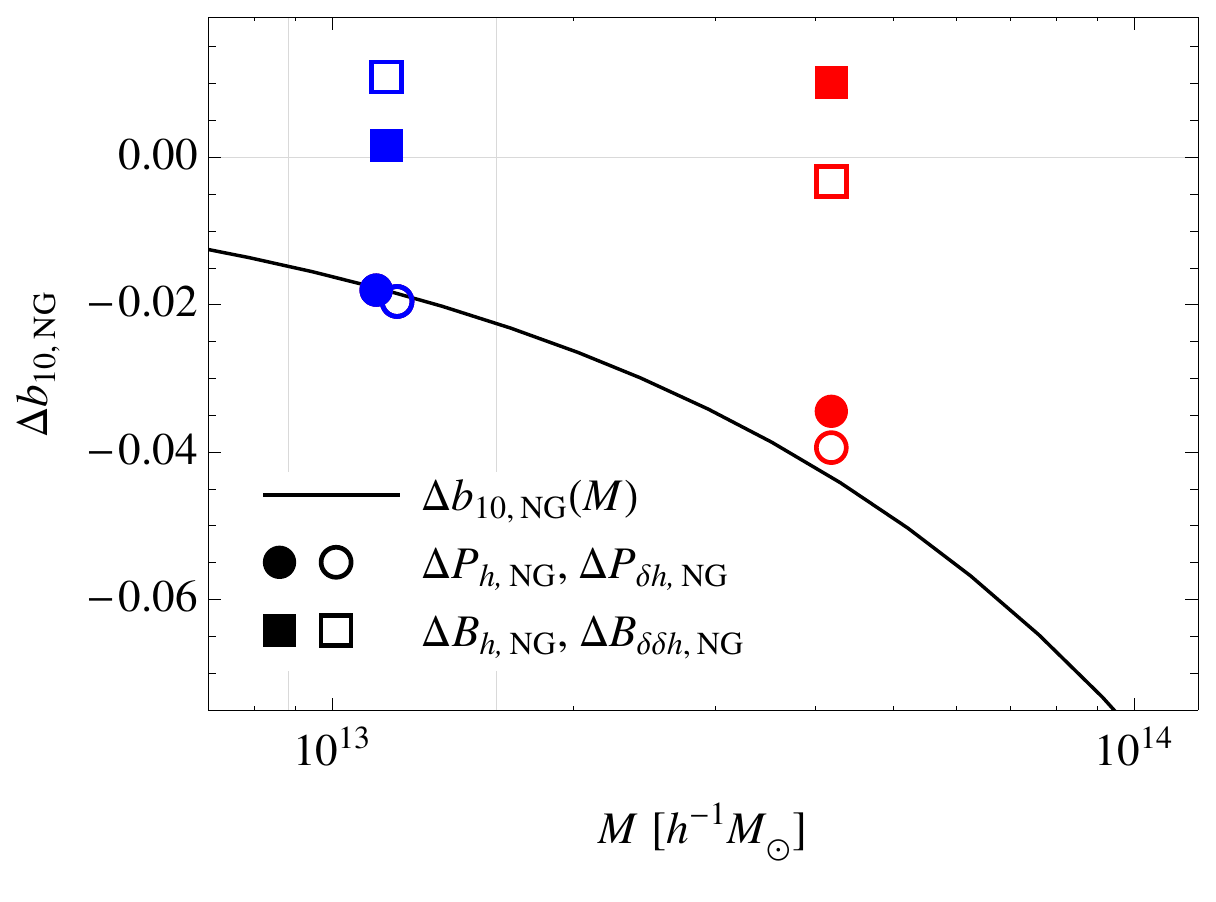}}
{\includegraphics[width=0.48\textwidth]{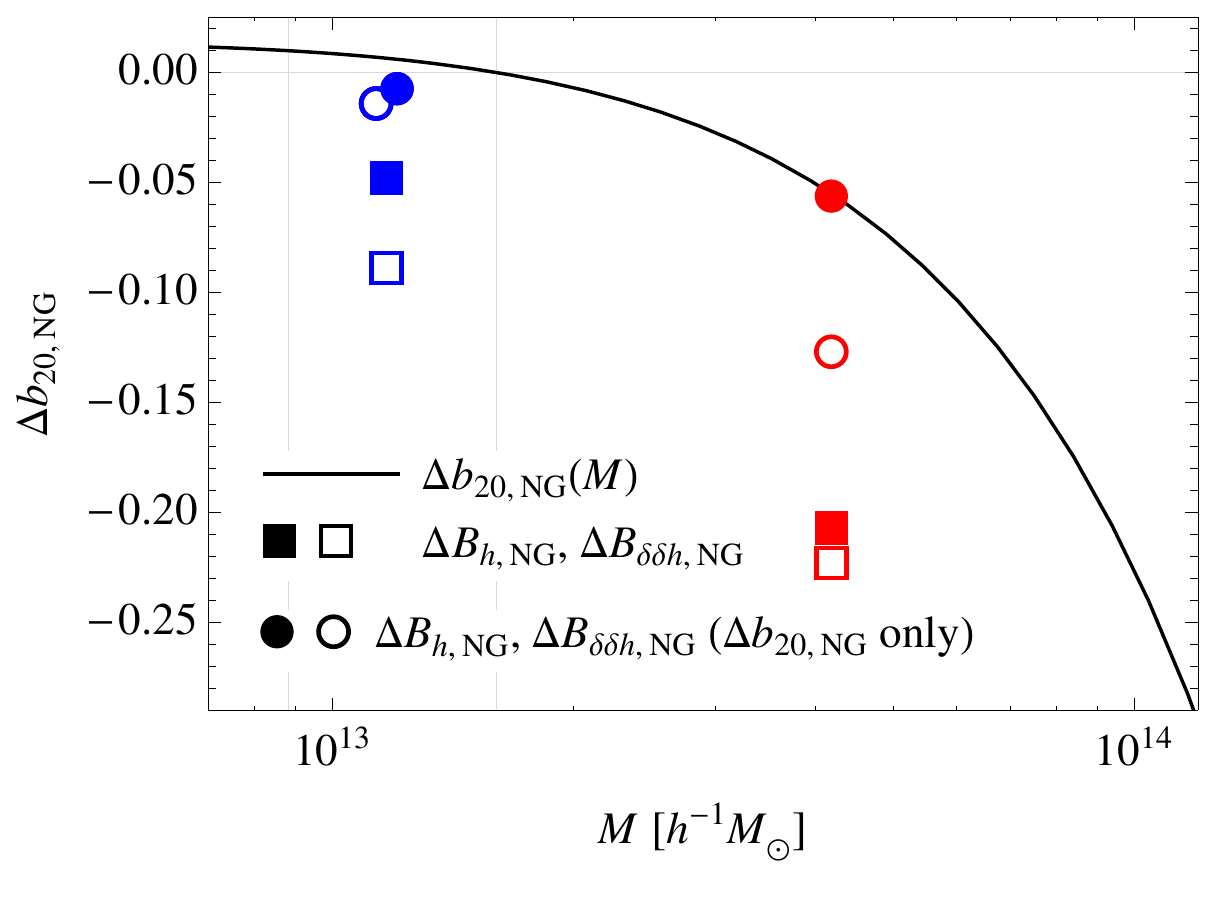}}
\caption{Best-fit bias parameters versus theoretical predictions. In all panels the continuous curve shows the predictions assuming a Sheth-Tormen mass function with the original parameters and the non-Gaussian correction to the mass function with the form proposed by \citep{LoVerdeEtal2008}. The circles correspond to the best-fit bias parameters where the Gaussian linear bias $b_{10,G}$ and its non-Gaussian, scale-independent correction $\D b_{10,NG}$ are determined from power spectrum measurements while only the Gaussian quadratic bias $b_{20,G}$ and its non-Gaussian correction $\D b_{20,NG}$ are determined from the bispectrum. The square data points correspond instead to the same bias parameters determined exclusively from bispectrum measurements. Filled symbols are derived from halo correlators, empty symbols from matter-halo cross-correlators. Data points are plotted at the mean mass value for the corresponding mass bin and are slightly displaced for clarity when needed. Vertical thin gray lines correspond to the thresholds defining the two mass bins.}
\label{fig:bias}
\end{center}
\end{figure}

So far we have not discussed the best-fit values obtained for the Gaussian linear and quadratic bias parameters $b_{10,G}$ and $b_{20,G}$ and their non-Gaussian, scale-independent corrections $\D b_{10,NG}$ and  $\D b_{20,NG}$. 
As explained in Section~\ref{ssec:Analysis}, the linear bias parameter $b_{10,G}$ is determined from the power spectrum measurements in simulations with Gaussian initial conditions while its correction $\D b_{10,NG}$ is obtained from the extra contribution to the power spectrum induced by non-Gaussianity. The quadratic parameter $b_{20,G}$ is then given by fitting the bispectrum with Gaussian initial conditions, while the best-fit value of $\D b_{20,NG}$ is obtained from the non-Gaussian correction to the bispectrum. The whole procedure is applied independently to the matter-halo cross correlators and to the halo correlators (power spectrum and bispectrum).
  
While being likely the most ``predictive'' procedure which does not involve a direct evaluation of the bias parameters, this is by no means the only possible one. We did also consider alternative determinations entirely based on bispectrum measurements. The outcome of these different procedures is shown in Fig.~\ref{fig:bias} and compared to the theoretical predictions of the peak-background split approach \citep{ColeKaiser1989, MoWhite1996, ShethTormen1999, ScoccimarroEtal2001A}. 

In the upper left panel of Fig.~\ref{fig:bias} the best-fit values of $b_{10,G}$ are shown as circles when they are obtained from the power spectrum measurements, and as squares when they are obtained from the bispectrum. These values are plotted at the mean mass for each of the two mass bin considered. Here, like in the other panels, filled symbols refer to halo correlators while empty symbols to matter-halo cross-correlators. Errors on the bias parameters are not shown, because an analysis including the power spectrum and bispectrum variance alone and neglecting covariances underestimate them significantly. Clearly, the best-fit values obtained from the power spectrum are about 10\% larger than those obtained from the bispectrum. A similar discrepancy has been recently reported in \citep{PollackSmithPorciani2011} (who studied Fourier space correlators) and in \citep{ManeraGaztanaga2011} (who studied configuration space correlation functions). Overall, our findings are consistent with the results of these studies. While effects due to smoothing, particularly in relation to the scatter between $\d_h$ and $\d$ in position space, and the limitations due to the tree-level model assumed could plausibly be invoked to explain such a discrepancy, the most likely explanation is the existence of nonlocal terms usually neglected in the halo bispectrum expression, present also for Gaussian initial conditions \citep{Scoccimarro2011talk}. Properly addressing this issue is, however, beyond the scope of this work. The figures also display the predicted value of the linear bias parameter obtained from the peak-background split according to Eq.s~(\ref{eq:biasLGa}) and (\ref{eq:biasEUa}) assuming the Sheth-Tormen (ST) \citep{ShethTormen1999} unconditional mass function. 

The upper right panel of Fig.~\ref{fig:bias} shows instead the best-fit values of the quadratic bias parameter $b_{20,G}$, obtained from the bispectrum measurements. We consider here as well two procedures. In the first one, which is used throughout this paper, the value of $b_{10,G}$ is determined from the power spectrum while the bispectrum only provides $b_{20,G}$; the results are denoted by circles in the plot. In the second one, both $b_{10,G}$ and $b_{20,G}$ are determined from the bispectrum; the results are denoted by squares in the plot. Clearly, the discrepancy discussed above in the determination of $b_{10,G}$ induces different values of $b_{20,G}$. Computing again the quadratic halo bias from the mass function by means of Eq.s~(\ref{eq:biasLGb}) and (\ref{eq:biasEUc}), the values obtained from the ST mass function are in good qualitative agreement with the best-fit values.

In the lower left panel of Fig.~\ref{fig:bias} we show the non-Gaussian correction to the linear bias $\D b_{10,NG}$. Circles indicate the values obtained from $\D P_{\d h,NG}$ and $\D P_{h,NG}$ ({\em empty and filled, respectively}). Squares indicate instead the values determined from $\D B_{\d\d h,NG}$ and $\D B_{h,NG}$ ({\em empty and filled}). The value for the Gaussian component $b_{10,G}$ is provided by measurements of the Gaussian power spectrum in the first case, and measurements of the bispectrum in the second case. The values obtained from the power spectrum are in relatively good agreement with the theoretical prediction of Eq.~(\ref{eq:dngbias1}), which is obtained from the ratio $R_{NG}(\nu)\equiv f_{NG}(\nu)/f_G(\nu)$ of the non-Gaussian to Gaussian mass function (see Section~\ref{sec:nonlocal}). For this quantity, we use the expression of \citep{LoVerdeEtal2008} based on an Edgeworth expansion of the Press-Schechter Gaussian mass function, \citep{PressSchechter1974}. In what follows, we limit the expansion to include linear corrections in $\fNL$ only. We consider an expression given by $R_{NG}(q\,\nu)$ where we fit for the shift parameter $q$ comparing the prediction with the ratio $R_{NG}$ measured in our simulations, finding the best-fit value $q\simeq 0.91$ (and very close values at different redshift). The non-Gaussian correction to the linear bias obtained from the bispectrum alone is in stark disagreement with this prediction. This, in part, motivates our choice to assume the values of $\D b_{10,NG}$ determined from power spectrum measurements. Clearly, in principle, we would obtained the same results using directly the theoretical predictions. 

Finally, the lower right panel of Fig.~\ref{fig:bias} shows the non-Gaussian correction to the quadratic bias $\D b_{20,NG}$. Here as well we consider two different fits. The first one, shown by the circles in the plot and assumed in the previous section, consists in fitting the measurements of the non-Gaussian correction to the bispectrum only for $\D b_{20,NG}$ where the values of $b_{10,G}$ and $\D b_{10,NG}$ are determined from the power spectrum and $b_{20,G}$ from the bispectrum with Gaussian initial conditions. In the second case $\Delta b_{10,NG}$ and $\Delta b_{20,NG}$ are both determined by the non-Gaussian correction of the bispectrum, and shown by squares. We find again that the results in the first case are closer to the theoretical prediction, Eq.~(\ref{eq:dngbias2}), with the exception of the result for the large mass bin from the matter-matter-halo cross-bispectrum.  

Lastly, we note that the $\chi^2$ as a function of $k_{max}$ for the comparison of the model to the bispectrum measurements shown in Fig.~\ref{fig:Chi2} does not significantly change for the different choices of parameters fitting discussed in this section.

\section{Fisher Matrix}
\label{sec:fisher}

The model we have tested can be used to make a preliminary forecast for the ability of galaxy bispectrum measurements in future redshift surveys to constrain a non-Gaussian component in the initial conditions. Here, we perform a Fisher matrix analysis in terms of the non-Gaussian parameter $\fNL$ and of the Gaussian bias parameters $b_{10,G}$ and $b_{20,G}$. In this section we denote the Gaussian parts of the {\em galaxy} bias parameters as $b_1$ and $b_2$, and their non-Gaussian scale-independent corrections as $\D b_1$ and $\D b_2$.

This analysis is purely for illustrative purposes. It is intended to compare constraints from power spectrum and bispectrum measurements, but is not meant to provide specific forecasts for any future survey. A more detailed and realistic study will be considered elsewhere. The Fisher matrix for the galaxy bispectrum is thus simply defined as
\beq
F_{\alpha\beta}\equiv \sum_{k_1,k_2,k_3\ge k_{min}}^{k_{max}}\frac{\partial B_g(\kall)}{\partial p_\alpha}\frac{\partial B_g(\kall)}{\partial p_\beta}\frac{1}{\D B^2_g(\kall)}\,,
\eeq
where the indices $\alpha$ and $\beta$ run over the three parameters $\fNL$, $b_{1}$, $b_{2}$ while, for simplicity, we fix the cosmology to be that of the simulations (see Sec.~\ref{sec:simulations}). The Fisher Matrix for the galaxy power spectrum is defined in an analogous way. The fiducial values of the galaxy bias factors are obtained by computing integrals of the halo bias functions times the mass functions above a certain mass threshold chosen to provide a given galaxy number density of $\bar{n}_g$. This is equivalent to an Halo Occupation Distribution assigning one galaxy per halo above the threshold. In addition, we assume that the scale-independent bias corrections $\D b_{i}$ are functions of the Gaussian parameters themselves, \ie $\D b_{i}(b_{i})$, obtained by varying the galaxy number density. This assumption is partially justified by the strong correlation expected between these parameters as we vary the characteristics of the galaxy population. 

Notice that we account only for the variance of the galaxy bispectrum, given by
\beq
\D B_g^2(k_1,k_2,k_3)=\frac{s_B}{8\pi^2 k_1 k_2 k_3}P_{tot}(k_1)P_{tot}(k_2)P_{tot}(k_3)\,,
\eeq
with $P_{tot}(k)=[P_g(k)+1/[(2\pi)^3\bar{n}_g]$ is the total galaxy power spectrum, including shot noise. The expression for the galaxy power spectrum is given by Eq.s~(\ref{eq:PhG}) and (\ref{eq:dPhNG}) while for the galaxy bispectrum by Eq.s~(\ref{eq:BhG}) and (\ref{eq:dBhNG}) where the bias parameters are now to be interpreted as galaxy bias. For simplicity, we evaluate all the matter correlators at linear and tree-level for the power spectrum and bispectrum, respectively. In addition, the computation of the galaxy bispectrum variance is linearized with respect to $\fNL$.

\begin{figure}[!t]
\begin{center}
{\includegraphics[width=0.48\textwidth]{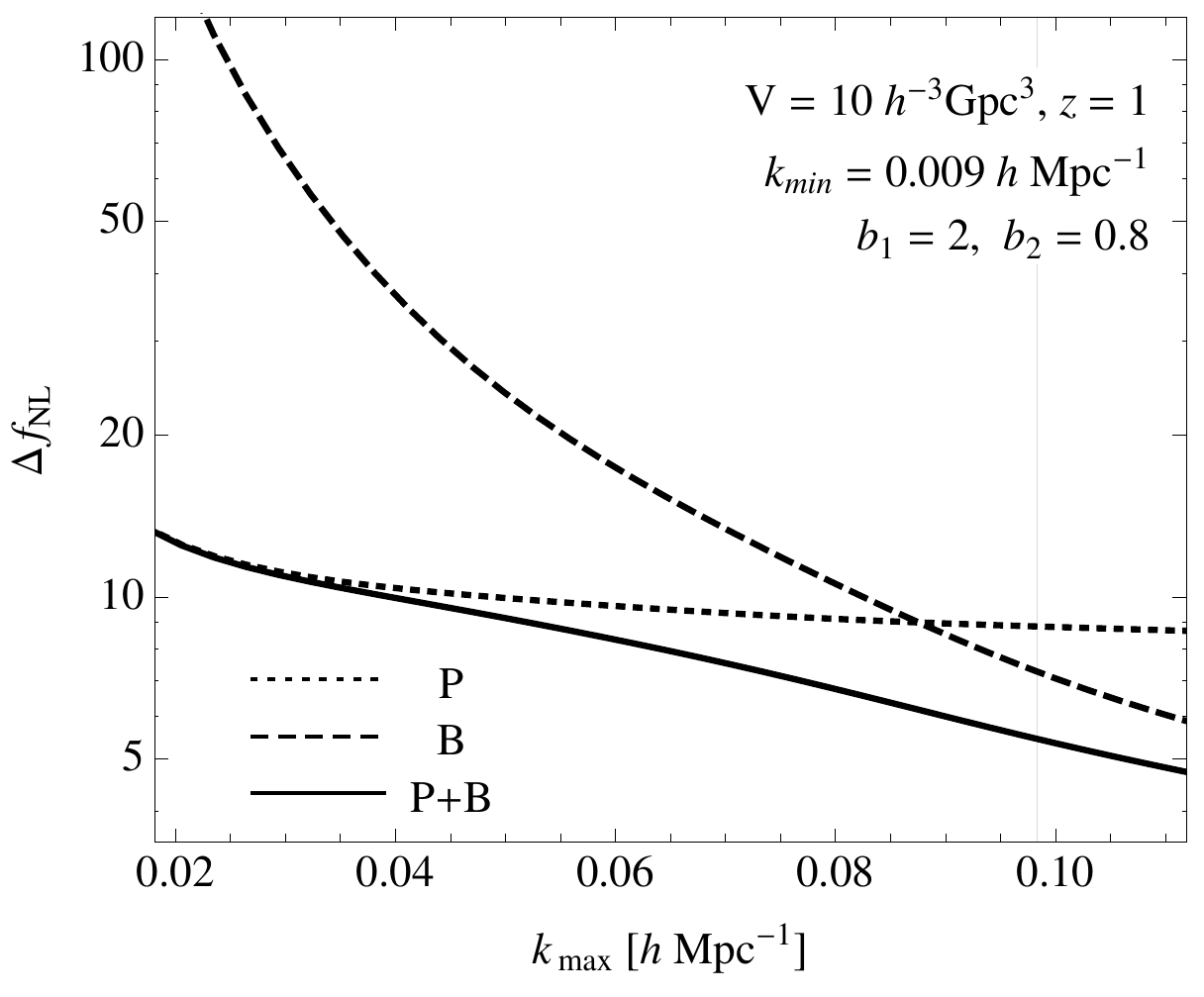}}
{\includegraphics[width=0.48\textwidth]{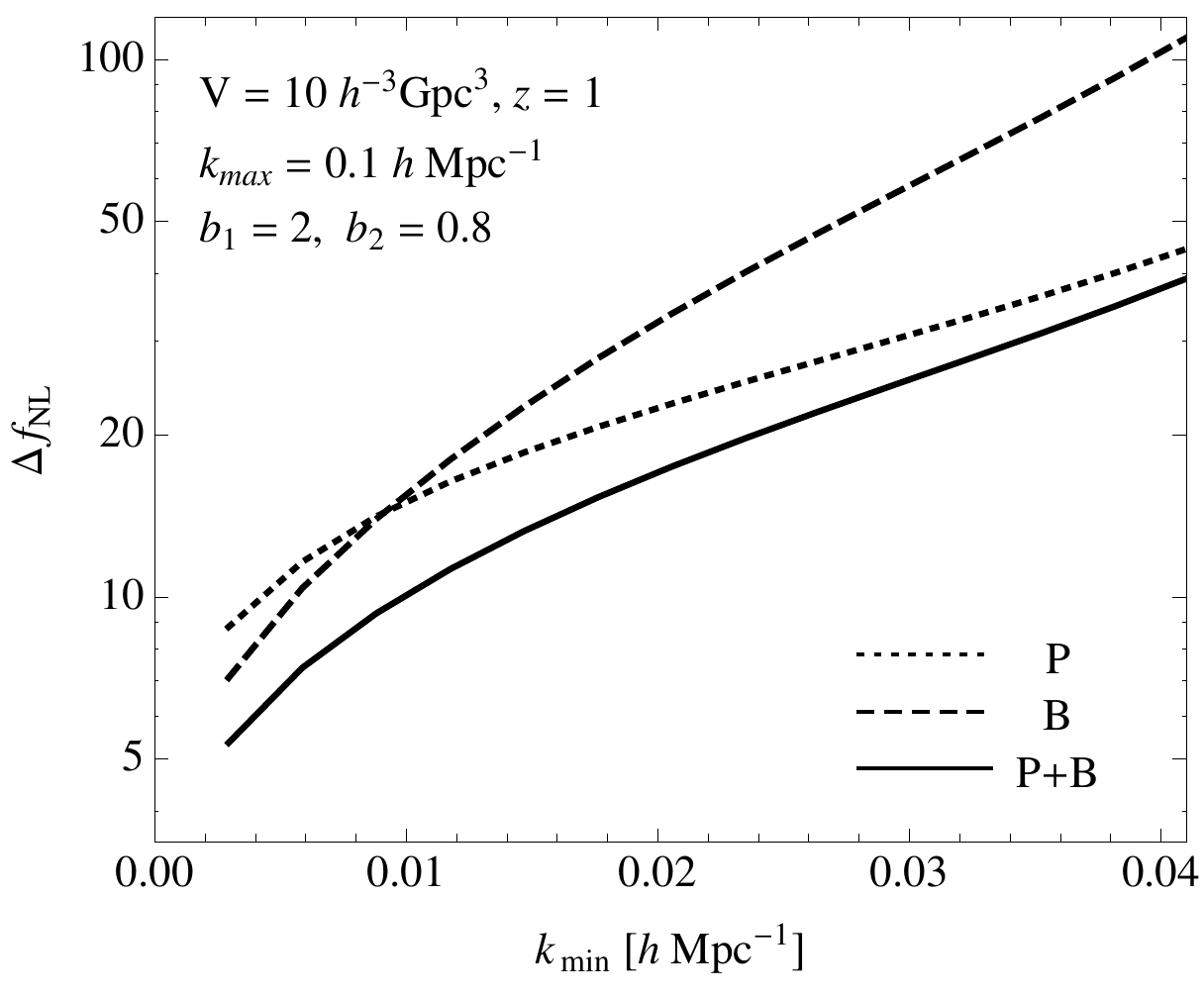}}
\caption{One-$\sigma$ uncertainty on the $\fNL$ parameter, marginalized over the Gaussian bias parameters, obtained from the Fisher matrix analysis of the power spectrum ({\em dotted curve}), bispectrum ({\em dashed curve}) and combined ({\em continuous curve}) for an ideal survey of $10\cGpc$ at redshift $z=1$, assuming a fiducial values for the non-Gaussian and bias parameters given by $\fNL=10$, $b_{10,G}=2$ and $b_{20,G}=0.8$. The left panel shows $\Delta\fNL$ as a function of the maximum wavenumber included, $k_{max}$, while the right panel assumes $k_{max}=0.1\kMpc$ and limits the smallest wavenumber included by $k_{min}$.}
\label{fig:Fisher}
\end{center}
\end{figure}

The results of the Fisher analysis can be read off in Fig.~\ref{fig:Fisher}, which shows the one-$\sigma$ error on the non-Gaussian parameter $\fNL$ obtained upon marginalizing over the two bias parameter. Errors are plotted as a function of the maximum wavenumber $k_{max}$ for an ideal survey of $10\cGpc$ at redshift $z=1$. We consider a galaxy population characterized by number density $\bar{n}_g=10^{-3}\icMpc$ and by the Gaussian bias parameters $b_1=2$ and $b_2=0.8$. The dotted curve represents the error obtained from an analysis of the galaxy power spectrum only, the dashed curve corresponds to the galaxy bispectrum only, and the continuous curve is the constraint from a combined analysis of the galaxy power spectrum and bispectrum.  As expected from signal-to-noise considerations for the effect of primordial non-Gaussianity on halo correlators (see Section~\ref{sec:signal}), the determination of $\fNL$ from power spectrum measurements does not improve significantly as $k_{max}$ increases beyond the largest scales accessible. However, due to the increase in the number of triangles included in the analysis, the bispectrum provides a comparable error $\D \fNL$ for relative small values of $k_{max}$, even before the mildly nonlinear regime. More interestingly, the {\em combined} power spectrum and bispectrum analysis improves that based on the power spectrum alone already at very large scales, even after the marginalization over the bias parameters.  We can compare these results with those of \citep{SefusattiKomatsu2007}, where the sole effect of non-Gaussian initial conditions is on the matter bispectrum and the non-Gaussian galaxy bias is not taken into account. We find a difference between the two analysis of a factor slightly larger than three, essentially due to the effect of PNG on halo bias. 

As we have seen, most of the signal in power spectrum measurements resides in the smallest wavenumber available. In the analysis of observational data, it is thus crucial to avoid any systematic error that may arise from an improper determination of the galaxy selection function. Regarding the bispectrum, since the signal is distributed over a large number of triangular configurations, we can ask ourselves how the error on $\fNL$ depends on the largest scale included in the analysis, defined by the value of $k_{min}$. On the right panel of Fig.~\ref{fig:Fisher} we present the same quantities as the left panel, now as a function of $k_{min}$ and a fixed $k_{max}\simeq 0.1\kMpc$. Clearly, combining power spectrum and bispectrum can provide errors comparable to those obtained from an analysis of the power spectrum alone even if the lowest wavemodes are excluded. This indicates that the galaxy bispectrum can provide, at the very least, a crucial cross-check to any power spectrum results.

We emphasize that the Fisher matrix results presented in this section do not account for several important issues affecting analyses of galaxy survey data. In particular, we are neglecting the effects of the survey selection function {\em and} of the covariance properties of power spectrum, bispectrum and the cross-covariance between the two correlators. These effects are indeed responsible for a significant degradation of the available signal (see for instance the analysis of \citep{SefusattiEtal2006} in the context of cosmological parameters). Their inclusion is essential to provide realistic forecasts for any upcoming mission. This will be the subject of future work.

\section{Conclusions}
\label{sec:conclusions}

In this work, we have presented the first detailed analysis of the effects of non-Gaussian initial conditions of the local kind on the bispectrum of halos extracted from numerical simulations. We have measured all triangular configurations at large scales for two different halo populations. 

We have shown that the cumulative signal-to-noise in the bispectrum exceeds the signal-to-noise in the power spectrum when all triangles down to mildly nonlinear scales are taken into account. The effects of local non-Gaussianity on the halo power spectrum are mainly due to the scale-dependent corrections to the linear halo bias, concentrating the signal in the smallest wavenumbers (i.e. largest scales) accessible in the simulations. On the other hand, non-Gaussian initial conditions of the local kind have more complex effects on the halo bispectrum. This leads to a non-Gaussian signal distributed across a wide variety of triangular configurations (with very different scales and shapes). In fact, the halo bispectrum includes, in the first place, the linearly evolved primordial contribution to the matter bispectrum, which is significant at large scales for generic models of non-Gaussianity. Furthermore, for models of PNG characterized by a large primordial bispectrum in the squeezed limit, scale-dependent corrections to the bias are present both at linear and quadratic level. As we have seen, for local non-Gaussianity these are as important as the primordial contribution at large scales. In general however, even scale-independent corrections are relevant and must be properly modeled in order to reproduce the simulations. 

We have compared our measurements with the theoretical model derived in \citep{BaldaufSeljakSenatore2011} from the multivariate halo bias expansion of \citep{GiannantonioPorciani2010}, Eq.~(\ref{eq:dhpos}). At large scales, the confirmation of the validity of the tree-level approximation for the halo bispectrum obtained from this perturbative expansion of the halo density is one of the main results of this work. We have studied both the halo bispectrum and the matter-matter-halo cross bispectrum, where the lower shot-noise allows for a more accurate comparison between predictions and measurements. The value of the constant linear and quadratic bias parameters both for Gaussian and non-Gaussian initial conditions are fitted to the measured halo power spectra and simultaneously to all, large-scale triangular configurations of the bispectra and later compared to their theoretical expectations from the peak-background split approach.

We have found that the model discussed in Section~\ref{sec:model} provides a quite accurate description, at the 10\% level, at large scales, \ie $k\lesssim 0.07\kMpc$, for almost all triangles of any shape, both for the halo bispectrum measured in simulations with  Gaussian initial conditions and for the correction to the halo bispectrum due to local primordial non-Gaussianity. Since, as a first step, we fit for the constant bias parameters and their non-Gaussian corrections, such results signifies that {\em the model presents all relevant functional dependencies on the triangular configurations necessary to describe the specific shape dependence of the halo bispectrum resulting from nonlinearities in the gravitational evolution and in the bias relation between the halo and matter distributions and from the peculiar correlations induced by local non-Gaussianity}. While the large-scale agreement between the simple tree-level, local bias model and numerical results for the halo bispectrum is an established fact for Gaussian initial conditions, recently confirmed for instance by \citep{SmithShethScoccimarro2008, PollackSmithPorciani2011}, the agreement of the model with the additional contribution to the halo bispectrum due to local non-Gaussianity, is, on the other hand, not trivial. In fact, the model describes such contribution alone by means of up to eight distinct terms, see Eq.~(\ref{eq:dBhNG}), each characterized by different scale and shape dependences. We show that for generic triangles none of these terms can provide on its own an accurate description of the non-Gaussian effects on the halo bispectrum which is rather given by the {\em sum of several different contributions}, and, for negative values of quadratic bias, also by relevant cancellations between them. This is true, in particular, for squeezed triangular configurations, where most of the signal from local models of PNG is concentrated. Interestingly, for such triangles the validity of model can be extended to the mildly nonlinear regime, $0.1\kMpc<k<0.2\kMpc$. 

The specific choice adopted for fitting procedure of the bias parameters (where the linear bias parameters are determined from power spectrum measurements) does not allow for a great freedom to adapt to the data, proving to a large extent the predictivity of the model. We compare as well the best-fit value for the linear and quadratic bias parameters and their non-Gaussian, scale-independent corrections, to their predictions in the context of the peak-background split approach finding a broad, qualitative agreements in their mass dependence. 

Finally, we perform a Fisher matrix analysis to compare the ability of power spectrum and bispectrum measurements to constrain the non-Gaussian parameter $\fNL$, and, more importantly, to quantify the possibilities given by their combined analysis, a necessary step toward a full exploitation of the data available in future large-scale structure surveys. Under the strong assumptions of neglecting the effects of covariance and window functions, we show that a combined power spectrum and bispectrum analysis can improve over the power spectrum alone by a factor of a few for a very large-volume redshift survey. At the same time, the bispectrum can provide a fundamental confirmation of any power spectrum result leading, in perspective, to the large-scale structure as a robust test for the initial conditions, with expected constraints of the order of those achievable by CMB observations, that is $\D \fNL\sim$ few. 

For this to be possible a substantial amount of work is still in order. This paper represents the first detailed study and test of a viable model for the halo bispectrum at large-scale. To achieve the accuracy necessary to place large-scale structure and CMB observations on the same footing with respect to their ability to constrain non-Gaussian initial conditions further investigations are required. In the first place, the discrepancies observed between the values of the bias parameters determined from power spectrum and bispectrum measurements in \citep{ManeraGaztanaga2011, PollackSmithPorciani2011} might hint at additional contribution, maybe due to nonlocal effects, relevant even for Gaussian initial conditions: in general a more accurate determination of the properties of halo and galaxy bias will be needed. In the second place, only a comprehensive study of the covariance properties of the halo bispectrum in combination with selection function effects can provide robust forecasts for the constraints on primordial non-Gaussianity expected from galaxy bispectrum observations. Our work will hopefully provide strong motivations for future investigations along these directions.

\begin{acknowledgements}

We thank T. Baldauf and R. Scoccimarro for useful discussions and T.~Giannantonio for comments on the draft. E.S. acknowledges support by the Marie Curie IEF program. M.C. acknowledges support by the Spanish Ministerio de Ciencia e Innovacion (MICINN), project AYA2009-13936, Consolider-Ingenio CSD2007- 00060, European Commission Marie Curie Initial Training Network CosmoComp (PITN-GA-2009-238356), research project 2009-SGR-1398 from Generalitat de Catalunya and the Juan de la Cierva MICINN program. V.D. is supported by the Swiss National Science Foundation under contracts No.  $200021-116696/1$ and PP$00$P$2\_1133577$.

\end{acknowledgements}
 
\appendix

\section{Evaluation of matter correlators in Perturbation Theory}
\label{app:PT}

The matter correlators appearing in the halo power spectrum and bispectrum expressions are computed, respectively, at 4th and at 6th order in the linear matter density $\d_0$ in Eulerian Perturbation Theory (EPT). We refer to the review \citep{BernardeauEtal2002} and references therein for an introduction to cosmological perturbation theory. More recent reviews focusing on non-Gaussian initial conditions and higher-order correlators of the Large-Scale Structure can be found in \citep{LiguoriEtal2010, DesjacquesSeljak2010B}. We notice that promising resummation approaches in EPT such as those of \citep{CrocceScoccimarro2006A, BernardeauCrocceScoccimarro2008, Pietroni2008} can be extended to non-Gaussian initial conditions \citep{BernardeauCrocceSefusatti2010, BartoloEtal2010}, leading to more accurate predictions than those considered here. 

In addition to the standard loop-corrections to the matter power spectrum and bispectrum we explain here in detail the corresponding nonlinear correction for the cross-power spectrum and bispectrum $P_{\d\d_0}$ and $B_{\d\d\d_0}$ involving both the nonlinear and non-Gaussian density contrast $\d$ and its linear and Gaussian counterpart $\d_0$. We adopt the notation of \citep{Sefusatti2009} for the EPT contributions.

In Fourier space, the perturbative solution for the nonlinear matter overdensity $\d_\kv$ is expressed by the series \citep{BernardeauEtal2002}
\beq\label{eq:PT}
\d_\kv=\d_\kv^{(1)}+\d_\kv^{(2)}+\d_\kv^{(3)}+\ldots,
\eeq
where $\d_\kv^{(1)}\equiv\d_0$ is the linear solution (here non-Gaussian) and  
\beq\label{eq:PTterm}
\d_\kv^{(n)}\equiv\int d^3q_1\ldots d^3q_nF_n(\qv_1,\ldots,\qv_n)\d_{\qv_1}^{(1)}\ldots\d_{\qv_n}^{(1)},
\eeq
where $F_n(\qv_1,...,\qv_n)$ is the symmetrized kernel for the $n$-th order solution. From this expansion one can derived in turns perturbative solutions for matter correlators, once the initial conditions, \ie the initial correlators are specified.

\subsection{Matter power spectra}

Up to 4-th order in $\d_0$ and at linear level in $\fNL$, the nonlinear matter power spectrum $P_{\d}$ defined as $\la\d_{\kv_1}\d_{\kv_2}\rangle\equiv \d_D(\kv_{12})P_{\d}$ is given by \citep{JainBertschinger1994, MakinoSasakiSuto1992, TaruyaKoyamaMatsubara2008}
\beq
P_\d = P_{11}+P_{12}+P_{22}+P_{13}+ \O(\d_0^5,\fNL^2),
\eeq 
where, $P_{11}\equiv  P_0$ is the linear matter power spectrum, while the other terms correspond to one-loop corrections given by
\bea
P_{12}(k)
& = &
2\intq \,F_2(\qv,\kv-\qv)~B_0(k,q,|\kv-\qv|),\\
P_{13}(k)
& = & 
6~P_0(k)\intq F_3(\kv,\qv,-\qv)~P_0(q),\\
P_{22}(k)
& = & 
2\intq\, F_2^2(\qv,\kv-\qv)~P_0(q)~P_0(|\kv-\qv|).
\eea
The Gaussian component is
\beq
P_{\d,G} = P_{11}+P_{22}+P_{13}+ \O(\d_0^5),
\eeq 
while the non-Gaussian correction is simply
\beq
\Delta \D P_{\d,NG} = P_{12}+ \O(\d_0^5,\fNL^2).
\eeq 

The cross-power spectrum $P_{\d\d_0}$ is defined as $\la\d_{{\kv_1}}\d_{0,{\kv_2}}\rangle\equiv \d_D(\kv_{12})P_{\d\d_0}$ so the perturbative expansion applies only to one field in the expectation value. We denote the perturbative contributions separating the order of correction in $\d$ from the linear term so that $\la\d_{{\kv_1}}^{(i)}\d_{0,{\kv_2}}\rangle\equiv \d_D(\kv_{12})P_{i,1}$. The EPT expansion for $P_{\d\d_0}$ is therefore given by
\bea
P_{\d\d_0} & = & P_{1,1}+P_{2,1}+P_{3,1}+ \O(\d_0^5,\fNL^2),
\eea 
where $P_{1,1}=P_{11}\equiv  P_0$ is the linear matter power spectrum, while the 1-loop corrections are given by
\bea
P_{2,1}(k)
& = &
\intq F_2(\qv,\kv-\qv)~B_0(k,q,|\kv-\qv|)=\frac12 P_{12},
\\
P_{3,1}(k)
& = & 
3~P_0(k)\intq F_3(\kv,\qv,-\qv)~P_0(q)=\frac12 P_{13}.
\eea
The Gaussian component is simply
\bea
P_{\d\d_0,G} & = & P_{1,1}+P_{3,1}+ \O(\d_0^5).
\eea
while the non-Gaussian correction is $\D P_{\d\d_0,NG}=P_{2,1}+ \O(\d_0^5,\fNL^2)$ which is neglected since it enters at second order in $\fNL$ in the halo power spectra. 
 
The evaluation of the matter-matter-halo bispectrum $B_{\d\d h}$ and of the halo bispectrum $B_h$, involves products of power spectra that, for consistency with the matter bispectra calculation, need to be calculated up to 6th-order in the linear density field. In particular
\bea
P_{\d,G}(k_1)P_{\d,G}(k_2) & = & P_{11}(k_1)P_{11}(k_2)+
\left[P_{11}(k_1)P_{22}(k_2)+P_{22}(k_1)P_{11}(k_2)\right]+
\nonumber\\ & &
\left[P_{11}(k_1)P_{13}(k_2)+P_{13}(k_1)P_{11}(k_2)\right]+
\O(\d_0^7)\,,
\eea
and 
\beq
\Delta P_{\d,NG}(k_1)P_G(k_2)  =  P_{12}(k_1)P_{11}(k_2)+\O(\d_0^7,\fNL^2)\,.
\eeq
Finally, for the products involving $P_{\d\d_0}$ we have
\bea
P_{\d\d_0,G}(k_1)P_{\d,G}(k_2) & = & P_{1,1}(k_1)P_{11}(k_2)+P_{1,1}(k_1)P_{22}(k_2)+
\nonumber\\ & &
\left[P_{1,1}(k_1)P_{13}(k_2)+P_{3,1}(k_1)P_{11}(k_2)\right]+
\O(\d_0^7)\,.
\eea

\subsection{Matter bispectra}

We evaluate all bispectra (and products of power spectra) up to 6-th order in the linear density field, $\d_0$. For non-Gaussian initial conditions this implies \citep{Scoccimarro1997, Sefusatti2009},
\bea
\label{eq:Bexp}
B_\d & = &  B_{111}+
B_{112}^I+B_{122}^I+B_{122}^{II}+B_{113}^I+B_{113}^{II}+
B_{222}^I+B_{123}^{I}+B_{123}^{II}+B_{114}^{I}+\O(\fNL^2,\d_0^7),
\eea
where $B_{111}\equiv B_0$ is the initial bispectrum and 
\beq\label{eq:B112I}
B_{112}^I  =  
2~F_2(\kv_1,\kv_2)~P_0(k_1)~P_0(k_2)+{\rm 2~perm.},
\eeq
is the other tree-level contribution, while the 1-loop corrections are given by
\bea
B_{122}^{I}
&=&
2 ~P_0(k_1)\left[F_2(\kv_1,\kv_3)\intq~F_2(\qv,\kv_3\!-\!\qv)~
B_0(k_3,q,|\kv_3-\qv|)+(k_3\leftrightarrow k_2)\right]+{\rm 2~perm.}
\nonumber\\
& = &
F_2(\kv_1,\kv_2)\left[P_0(k_1)~P_{12}(k_2)+P_0(k_2)~P_{12}(k_1)\right]+
{\rm 2~perm.},
\\
B_{122}^{II}
&=&
4 \intq~F_2(\qv,\kv_2\!-\!\qv)~F_2(\kv_1\!+\!\qv,\kv_2\!-\!\qv)~
B_0(k_1,q,|\kv_1\!+\!\qv|)~P_0(|\kv_2\!-\!\qv|)
 +{\rm 2~perm.},
\\
B_{113}^I
&=&
3B_0(k_1,k_2,k_3)\intq~F_3(\kv_3,\qv,-\qv)P_0(q)+
{\rm 2~perm.},
\\
B_{113}^{II}
&\!=\!&
3 P_0(k_1)\!\!\intq~F_3(\kv_1,\qv,\kv_2\!-\!\qv)B_0(k_2,q,|\kv_2\!-\!\qv|)+
(k_1\leftrightarrow k_2)+{\rm 2~perm.},
\\
B_{222}^I
\!&\!=\!&\!
8 \!\!\intq F_2(-\qv,\qv\!+\!\kv_1)F_2(-\!\qv\!-\!\kv_1,\qv\!-\!\kv_2)
F_2(\kv_2\!-\!\qv,\qv)P_0(q)P_0(|\kv_1\!+\!\qv|)P_0(|\kv_2\!-\!\qv|),
\\
B_{123}^{I}
\!&\!=\!&\!
6~P_0(k_1)\!\! \intq ~F_3(\kv_1,\kv_2\!-\!\qv,\qv)~F_2(\kv_2\!-\!\qv,\qv)~
\!P_0(|\kv_2\!-\!\qv|)~P_0(q)+{\rm 5~perm.},
\\
B_{123}^{II}
\!&\!=\!&\!
6~P_0(k_1)~P_0(k_2)~F_2(\kv_1,\kv_2)
\int d^3q~ F_3(\kv_1, \qv,-\!\qv)~P_0(q)+{\rm 5~perm.}
\nonumber\\
&=&
F_2(\kv_1,\kv_2)\left[P_0(k_1)~P_{13}(k_2)+P_0(k_2)~P_{13}(k_1)\right]+
{\rm 2~perm.},
\\
B_{114}^I
&\!=\!&\!
12\,P_0(k_1)\,P_0(k_2)\!\! \intq \,F_4(\!\qv,\!-\qv,\!-\kv_1,\!-\kv_2)\,P_0(q)+
 {\rm 2~perm.}.
\eea 
Specifically, the one-loop contributions present because of non-Gaussian initial conditions are all the fifth-order terms $B_{122}^I$, $B_{122}^{II}$, $B_{113}^I$ and $B_{113}^{II}$, which depend on the initial bispectrum $B_0$. The Gaussian component to the matter bispectrum is therefore given by
\beq
\label{eq:BexpG}
B_{\d,G} = B_{112}^I+B_{222}^I+B_{123}^{I}+B_{123}^{II}+B_{114}^{I}+\O(\d_0^7),
\eeq
while the non-Gaussian correction is
\beq
\label{eq:BexpNG}
\D B_{\d,NG} = B_{111}+B_{122}^I+B_{122}^{II}+B_{113}^I+B_{113}^{II}+\O(\d_0^7,\fNL^2),
\eeq

Similarly to the cross power spectrum $P_{\d\d_0}$, for the cross-bispectrum $B_{\d\d\d_0}$, defined as $\la\d_{{\kv_1}}\d_{{\kv_2}}\d_{0,{\kv_3}}\rangle\equiv \d_D(\kv_{123})B_{\d\d\d_0}$ we denote the perturbative contributions separating the order of correction in $\d$ from the linear term so that $\la\d_{{\kv_1}}^{(i)}\d_{{\kv_2}}^{(j)}\d_{0,{\kv_3}}\rangle\equiv \d_D(\kv_{123})B_{ij,1}$. Here the comma indicates that permutations with the third index are not considered. We have
\bea
B_{\d\d\d_0} & = &  B_{11,1}+
B_{12,1}^I+B_{22,1}^I+B_{22,1}^{II}+B_{13,1}^I+B_{13,1}^{II}+
+B_{23,1}^{I}+B_{23,1}^{II}+B_{14,1}^{I}+\O(\fNL^2,\d_0^7),
\eea
where $B_{11,1}=B_{111}\equiv B_0$ is the initial bispectrum and 
\beq
B_{12,1}^I  =  
2~F_2(\kv_1,\kv_3)~P_0(k_1)~P_0(k_3)+2~F_2(\kv_2,\kv_3)~P_0(k_2)~P_0(k_3),
\eeq
is the other tree-level contribution, while the 1-loop corrections are given by
\bea
B_{22,1}^{I}
&=&
P_0(k_3)\left[F_2(\kv_1,\kv_3)P_{12}(k_1)+F_2(\kv_2,\kv_3)P_{12}(k_2)\right],
\\
B_{22,1}^{II}
&=&
4 \intq~F_2(\qv,\kv_2\!-\!\qv)~F_2(\kv_3\!+\!\qv,\kv_2\!-\!\qv)
B_0(k_3,|\kv_3\!+\!\qv|,q)~P_0(|\kv_2\!-\!\qv|),
\\
B_{13,1}^I
&=&
3\, B_0(k_1,k_2,k_3)\left[\intq~F_3(\kv_1,\qv,-\qv)P_0(q)+\intq~F_3(\kv_2,\qv,-\qv)P_0(q)\right],
\\
B_{13,1}^{II}
&\!=\!&
3\, P_0(k_1)\!\!\intq~F_3(\kv_1,-\qv,\kv_3\!+\!\qv)B_0(k_3,q,|\kv_3\!+\!\qv|)+
 \nonumber\\ & & 
3\, P_0(k_3)\!\!\intq~F_3(\kv_3,-\qv,\kv_1\!+\!\qv)B_0(k_1,q,|\kv_1\!+\!\qv|)+
(k_1\leftrightarrow k_2),
\\
B_{23,1}^{I}
\!&\!=\!&\!
6~P_0(k_3)\!\! \intq ~F_3(\kv_3,\kv_1\!-\!\qv,\qv)~F_2(\kv_1\!-\!\qv,\qv)~
\!P_0(|\kv_1\!-\!\qv|)~P_0(q)+(k_1\leftrightarrow k_2),
\\
B_{23,1}^{II}
\!&\!=\!&\!
6~P_0(k_1)~P_0(k_3)~F_2(\kv_1,\kv_3)
\int d^3q~ F_3(\kv_3, \qv,-\!\qv)~P_0(q)+(k_1\leftrightarrow k_2)
\nonumber\\
&=&
P_{13}(k_3)\left[F_2(\kv_1,\kv_3)P_0(k_1)+F_2(\kv_2,\kv_3)P_0(k_2)\right],
\\
B_{14,1}^I
&\!=\!&\!
12\,P_0(k_3)\left[P_0(k_1)\!\! \intq \,F_4(\!\qv,\!-\qv,\!-\kv_1,\!-\kv_3)\,P_0(q)+P_0(k_2)\!\! \intq \,F_4(\!\qv,\!-\qv,\!-\kv_2,\!-\kv_3)\,P_0(q)\right].
\eea 
In our approximation, we are only interested in the Gaussian component of $B_{\d\d\d_0}$ given by
\bea
\label{eq:Bddd0expG}
B_{\d\d\d_0,G} & = & B_{12,1}^I+B_{23,1}^{I}+B_{23,1}^{II}+B_{14,1}^{I}+\O(\fNL^2,\d_0^7).
\eea
We finally notice that the evaluation of the bispectrum contributions for a given triangle takes into account in part the effect of the finite bin $\D k$ defining the wavenumbers $k_1$, $k_2$ and $k_3$. A detailed explanation of the procedure is given in Section 3.1 of 
\citep{SefusattiCrocceDesjacques2010}.

\bibliography{Bibliography}

\end{document}